\pdfoutput=1
\documentclass[usenatbib,useAMS,twocolumn]{mn2e}
\usepackage{color}

\usepackage{float} 
\usepackage[fleqn]{amsmath}
\usepackage{epsfig,floatflt}
\usepackage{natbib}
\usepackage{subfigure}
\usepackage{amssymb}
\usepackage{multirow}
\usepackage{ulem}

\newcommand{\LCDM}{ $\Lambda$CDM }
\newcommand{\be}{\begin{equation}} \newcommand{\ee}{\end{equation}}
\newcommand{\ba}{\begin{eqnarray}} \newcommand{\ea}{\end{eqnarray}}
\newcommand{\brr}{\begin{array}} \newcommand{\err}{\end{array}}
\newcommand{\kms}{km s$^{-1}$}
\newcommand{\kmsMpc}{km s$^{-1}$ Mpc$^{-1}$}
\newcommand{\Mpch}{$h^{-1}$Mpc}
\newcommand{\correction}[1]{#1}
\begin{document}

\title{Observational biases in Lagrangian reconstructions of cosmic velocity fields}

\author[Lavaux et al.]{ G. Lavaux$^{1}$, R. Mohayaee$^{1}$, 
 S. Colombi$^{1}$, R. B. Tully$^{2}$,
 F. Bernardeau$^{3}$, J. Silk$^{1,4}$\\
$^{1}$ Institut d'Astrophysique de Paris -- UMR 7095, 98bis bd Arago,
France, CNRS/Universit\'e Pierre et Marie Curie \\
$^{2}$ Institute for Astronomy, Univ. of Hawaii, Honolulu, USA \\
$^{3}$ Service de Physique Th\'eorique, CEA/DSM/SPhT, Unit\'e de
recherche associ\'ee au CNRS, CEA/Saclay 91191 Gif-sur-Yvette c\'edex \\
$^{4}$ Department of Astrophysics, University of Oxford, Keble Road, Oxford OX1 3RH}

\maketitle

\begin{abstract}
  Lagrangian reconstruction of large-scale peculiar velocity fields
  can be strongly affected by observational biases.
  We develop
  a thorough analysis of these systematic effects by relying on specially selected mock
  catalogues.
  For the purpose of this paper, we use the
  Monge-Amp\`ere-Kantorovitch (MAK) reconstruction method, although any
  other Lagrangian reconstruction method should be sensitive to the same
  problems. We extensively study the uncertainty in the
  mass-to-light assignment due to incompleteness (missing luminous mass tracers),
  and the poorly-determined relation between mass and luminosity. The impact of redshift distortion
  corrections is analyzed in the context of MAK and we check the
  importance of edge and finite-volume effects on the reconstructed
  velocities. Using three mock catalogues with different average
  densities, we also study the effect of cosmic variance. In particular,
  one of them presents the same global
  features as found in observational catalogues that extend to  
  80~$h^{-1}$Mpc scales.
We give recipes, checked using the aforementioned mock catalogues, to handle
  these particular observational effects, after having introduced them
  into the mock catalogues so as to quantitatively mimic the
  most densely sampled currently available galaxy catalogue of the nearby universe. Once biases have been taken care
  of, the typical resulting error in reconstructed velocities is
  typically about a {\it quarter} of the overall velocity dispersion, and
  without significant bias.
We finally model our reconstruction errors to propose an improved Bayesian approach to measure 
  $\Omega_\text{m}$    in an unbiased way  by comparing the reconstructed velocities to the
  measured ones in distance space, even though they may be plagued by large
   errors. We show that, in the context of observational data, it is possible to build a nearly unbiased
  estimator of $\Omega_\text{m}$ using MAK
  reconstruction.

\end{abstract}



\label{firstpage}

\section*{Introduction}
\label{sec:introduction}

Galaxy redshift catalogues provide us with the radial velocities of
the galaxies, 
\begin{equation}
  cz=H_0\,r+v_r,
  \label{eq:cz}
\end{equation}
which are partly due to the global Hubble expansion ($H_0\,r$ with
$H_0$ the present value of the Hubble parameter) and partly due to the
line-of-sight components of the peculiar velocities ($v_r$). Peculiar
velocities are the deviations of galaxy velocities from the uniform
Hubble expansion, due to the non-homogeneous distribution of matter in
the Universe. The peculiar velocities are thus tracers of mass
distribution in the Universe and can have far-reaching implications
for cosmology.
As tracers of
dark matter, peculiar velocities can be used to determine the local and
global distribution of dark matter. 
From expression (\ref{eq:cz}), it is evident that  observations of
galaxy redshifts ($z$) supplemented by measure of radial distances
($r$), would yield the peculiar velocities.  However, measuring
distances is a non-trivial exercise.  The Tully-Fisher relation, surface
brightness fluctuations, the Faber-Jackson relation for ellipticals (and
their siblings, including the fundamental plane and the ${\rm
D}_{\rm n}-\sigma$ methods, the
Tip of the Red Giant Branch, Cepheids,  and SNIa are the most usual methods for obtaining
distances.  The data gathered is however rather sparse: out of about a
million galaxies whose redshifts are presently known with surveys such
as 2dF and SDSS, the distances to only a few thousand have measured distances.
Moreover, distances for most of these galaxies have 
too large peculiar velocity errors (due essentially to errors
in distance measurements) to be useful in  studying dynamics.
For instance, distance indicators such as the Tully-Fisher relation suffer from 20\% relative
distance errors and thus produce quite noisy measurements at relatively moderate
redshifts ({\it i.e.} $c z \ga 3000$~\kms). The data also suffers
from selection biases \citep{StraussWillick, Tully00}.
One way of
reducing the error bars on distances is to average over many distance
measurements for galaxies in clusters or groups and also by combining
the results from different distance estimators. This treatment decreases the
error bars on distances to about $8\%$ relative distance errors \citep{TullyVoid07}.
Even though all these difficulties can be surmounted, one can finally
hope to only have a sparse sample (as compared to redshift samples) of radial
components of peculiar velocities. Fortunately, we now have
Lagrangian velocity reconstruction schemes that are based solely on
current redshift positions of mass tracers. The reconstructed
velocities depend on cosmological parameters. Thus, comparing
predictions obtained through Lagrangian reconstruction algorithms and
the measured velocities may give estimations of these parameters. 

This brings us to the {\it main
point} that this paper tries to address: developing a {\it robust} and {\it unbiased} method of
Lagrangian peculiar velocity reconstruction using redshift
catalogues, in particular when observational effects
distort most of the required data needed for the reconstruction of the
dynamics. The reconstructed velocities are then compared to the
measured ones using an {\it ad hoc} algorithm to yield a measurement
of $\Omega_\text{m}$, the mean matter density of the Universe.

Throughout the paper, we will try to mimic observational effects as
they appear in the most densely sampled currently available galaxy catalogue of the
nearby universe which has been compiled by one of the authors
(R. B. Tully). This galaxy catalogue is built from different
sources such as ZCAT \citep{ZCAT92} and SSRS \citep{SSRS88}. Only
galaxies for which $c z \le 8000$~\kms{} have been introduced in the
catalogue.
This catalogue is named NBG-8k, standing for NearBy Galaxy
catalogue with a depth of 8000~\kms. 
Although selection criteria for this catalogue are not well defined, it
will prove to be useful for the study of smaller  galaxy
catalogues such as NBG-3k \citep{TullyVoid07}. 

For the purpose of this paper, we use a recently developed technique, called the
Monge-Amp\`ere-Kantorovitch reconstruction method (MAK 
hereafter), which is an approximation to the full non-linear dynamics to
trace orbits back in time. This is a Lagrangian method, such as PIZA
\citep{Piza97} or the Least-Action method \citep{Peebles89}, and not a Eulerian
technique such as, e.g., POTENT \citep{POTENT}. One must note that the results of this
paper are also valid for the other Lagrangian reconstruction methods
as all the effects we are going to analyze are explainable in terms of
gravitational dynamics. The MAK
reconstruction has already been largely discussed when applied on
numerical simulations \citep{moh2005,Brenier2002}. It is based on
assuming that the dark matter displacement field is convex and potential,
{\it i.e.} irrotational. In doing so, we exclude displacement fields
which include multistreaming regions. The main result is
that it is then possible to reconstruct accurately and uniquely the displacement
field of dark matter particles between their original position and
their current position. Practically, to solve the MAK problem, one must
minimize a cost function for the assignment of a dark matter particle
at the present comoving position ${\bf x}_i$ and its initial
comoving position ${\bf q}_j$:
\begin{equation}
  S_{\sigma}=\sum_{i=1}^N\left({\bf x}_i - {\bf q}_{\sigma(i)}\right)^2\;.
  \label{eq:assign}
\end{equation}
If the Universe is assumed to be initially homogeneous, which is a
fair hypothesis supported by CMB data \citep[e.g. WMAP first year
  in][]{WMAP1map},\footnote{\cite{Brenier2002} actually shows
  the uniformity is
  even required to prevent singularities in the solution of the
  Euler-Poisson system of equations.} then ${\bf q}_j$ must be distributed on a uniform
grid and the solution to the MAK
problem is unique and given by the assignment $\sigma$ which minimizes
$S_\sigma$. The derived solution is then necessarily irrotational and
derives from a convex potential. To solve this problem, we have implemented a parallel
version of the so-called ``auction'' algorithm proposed
by \cite{Bertsekas79}.\footnote{We implemented a parallel version for
  shared-memory supercomputers and MPI clusters. On the Magique2 cluster, it
  needs 50 minutes on 2 processors to solve the assignment of 74000
  particles. The algorithm is already sparse, {\it i.e.} it only looks
  for candidates for assignment in a limited region of the
  catalogue. The MPI efficiency is here optimal using 2 processors. It must
  be noted that the time complexity depends highly on the catalogue
  that is being reconstructed. For a given
  catalogue, the time needed to solve the assigment problem increase as $N^{2.25}$ with $N$ the number of
  particles. }  Of course, as we are using an approximation to the
dynamics, the solution to the problem will be only valid above some
scale (typically a few $h^{-1}$Mpc). Once the solution is found, the immediate
output of MAK reconstruction is the nonlinear displacement field ${\bf
  \Psi}({\bf q}) = {\bf x}({\bf{q}}) - {\bf q}$, which can be used to
find the peculiar velocity field ${\bf v}$ using the first-order
Zel'dovich approximation: 
\begin{equation}
  {\bf v}_i = \beta {\bf \Psi}_i\;,
\end{equation}
where the subscript $i$ indicates the comparison is achieved on the
corresponding field averaged over the object $i$ ({\it i.e.} in a
Lagrangian way), and the linear growth factor $\beta
\simeq \Omega_{\text{m}}^{5/9}$ \citep{Bouchet95}. 
This best fit for
$\beta$ is valid as soon as $\Omega_\text{m} + \Omega_\Lambda = 1$,
$\Omega_\Lambda$ being the present dark energy density.
It appears
then that a direct comparison of ${\bf \Psi}_i$ against  ${\bf v}_i$
should in principle give us $\beta$ and thus $\Omega_\text{m}$. Though
naive measurements \citep{MohTu2005} and preliminary studies
\citep{Branchini02,Phelps2006} on mock redshift catalogues have already been tried, the
observational biases and systematic errors in the velocity-velocity comparison
have never been studied thoroughly.

This paper is organized as follows. 
In Section~\ref{sec:mock_building_main}, we describe the simulation and
the basic mock catalogues that are used in the rest of this
paper. Subsequent mock catalogues integrate more and more observational
features but are still based on the same original basic mock
catalogues presented in this section. Section~\ref{sec:mak_error}
gives a model for the error
distribution on MAK velocities and discuss the first problematic
features of the comparison between MAK and measured velocities. This
error distribution helps us in particular to establish the
likelihood analysis in Section~\ref{sec:malmquist}. We go
then to the first main topic of this paper in Section~\ref{sec:ml_assign}
by studying the systematic errors introduced by arbitrary
mass-to-light assignments in redshift catalogues. This section includes
a study of missing mass correction (\S~\ref{sec:diffuse_mass}),
unknown $M/L$ function (\S~\ref{sec:ml_ratio}) and
incompleteness effects (\S~\ref{sec:incompleteness}; technical details
are given in Appendix~\ref{app:magnitude_limit}). In
Section~\ref{sec:redshift_distortion}, we 
discuss the problem of redshift distortions and the way to account for
it during the MAK reconstruction. Section~\ref{sec:edge_effects} is devoted
to the handling of finite volume and edge effects, i.e. issues related
to the zone of avoidance 
(\S~\ref{sec:zoa}), the choice of the Lagrangian volume of the reconstruction
(\S~\ref{sec:lag_volume}), and finally the so-called cosmic variance 
(\S~\ref{sec:cosmic_var}). The last section
(\S~\ref{sec:malmquist}) of this paper investigates the effect of
distance measurement errors on the comparison between reconstructed
and measured velocities, and  proposes a maximum likelihood
estimator (\S~\ref{sec:likelihood}) to account for them in the
measurement of $\Omega_\text{m}$. Results given by this estimator are
then discussed in \S~\ref{sec:lik_result}.

\section{Mock catalogues}
\label{sec:mock_building_main}

To study various effects and systematic biases on
the MAK reconstructed velocity field, we generated a number of mock
catalogues extracted from a $N$-body simulation (\S~\ref{subsec:nbody}). Although
many recipes will be employed later to address various observational
biases, we will always start from the same three\footnote{The
  computationally high cost of the reconstruction considerably limits 
  the number of possible realisations.} ``main'' halo catalogues as
described in \S~\ref{sec:mock_building:basic_mock}. The first
catalogue aims to reproduce to some extent the main features of the
local universe, in particular the presence of  a large cluster at about
40~$h^{-1}$Mpc and a super-cluster at about 70~$h^{-1}$Mpc.  The
second and the third catalogues have less salient features but
represent locally overdense and underdense realisations in order to
address the problem of cosmic variance.  

\subsection{The $N$-body sample}
\label{subsec:nbody}

Our $128^3$ particles $N$-body sample \citep{moh2005} was generated with the public version of the $N$-body code {\sc HYDRA}
\citep{CouHydra95} to simulate collisionless structure formation in a standard
\LCDM cosmology.
\correction{The sample covers a comoving volume of 200$^3$$h^{-3}$~Mpc$^3$.}
The mean matter density is $\Omega_{\text{m}}=0.30$ and the cosmological constant $\Omega_\Lambda=0.70$. The Hubble constant
is $H_0=65$~\kmsMpc. The normalisation of the density
fluctuations in a sphere of radius 8~$h^{-1}$~Mpc, is
$\sigma_8=0.99$. We note that this value  of $\sigma_8$ is
significantly larger than the value suggested by present WMAP data which sets
$\sigma_8=0.74$ \citep{SpergelCosmo2006}, but this should not affect
significantly the results presented in this paper. In fact, a lower $\sigma_8$ compared to $0.99$ would reduce both
non-linearities and cosmic variance effects, hence improving the
quality of the measurements.

\correction{As the velocity field presents significant fluctuations on a larger
  scale than for the density field, one may worry about the small size
  of the simulation volume. 
  We have checked, using linear theory, that the velocity dispersion in
  $200^3 h^{-3}$~Mpc$^3$, for our cosmology, is $40$~\kms. This value
  has to be compared to the typical errors appearing while doing
  velocity reconstructions to ensure that cosmic variance effects are
  negligible for our purpose.}

\subsection{The basic mock catalogues}
\label{sec:mock_building:basic_mock}

\begin{figure}
  \begin{center}
    \includegraphics*[width=\linewidth]{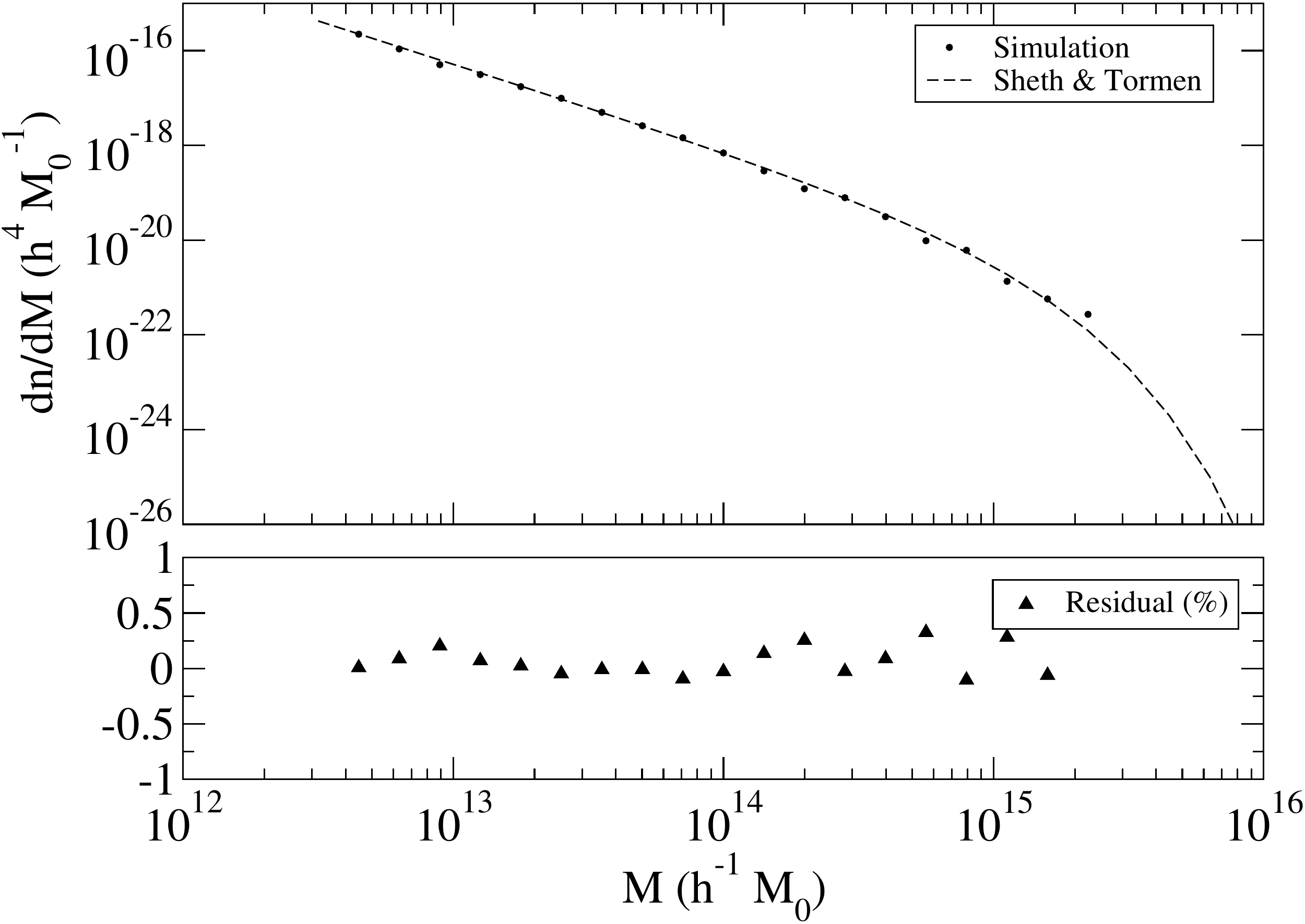}
  \end{center}
  \caption{ \label{fig:press_schechter}
   {\it Sheth \& Tormen mass function / Diffuse mass} -- The top
   panel of this plot gives the number density of haloes in a mass bin
   as a function of. the mass. The round points give the measurement
   of this function in the halo catalog whereas the dashed line is
   obtained using the \citet{ShethTormen02} theory. The residuals
   between the prediction and the measurement are given in the lower
   panel (relative differences). Most of the time, the points are
   within a few percent of the theoretical prediction.}
\end{figure}

To build mock catalogues, we have selected haloes from the $N$-body
experiment using a standard Friend-Of-Friend algorithm with a
traditional value of the linking parameter given by $l=0.2$
\citep{EFWD}. Haloes with less than 5 particles, {\it i.e.} with mass smaller
than $M_\text{min} = 1.62 \times 10^{12}\; h^{-1}\text{ M}_\odot$, were discarded.
Fig.~\ref{fig:press_schechter} shows the good agreement between the
measured halo mass function
and the \cite{ShethTormen02} model for haloes with $M \ga M_\text{min}$.
However about 63\% of the mass is not clumped in these haloes and is
distributed in the {\it background field}. In realistic galaxy
samples such as the NBG-8k or the 2MASS catalogue the lower mass cut-off is of the
order of $10^{11}\text{ M}_\odot$, a value much smaller
than our $M_\text{min}$. 
To mimic galaxies with mass smaller than
$M_\text{min}$, as will be required in the following, we just use
dark matter particles unassigned to any halo as tracers. The catalogue
containing all the haloes and all the field particles will be called
{\it FullMock}. \correction{One could here worry that the $N$-body sample that we are using
  has a too low resolution as the spatial distribution of small halos is
  biased but not the particles of the background field. We have actually checked
  that using a $512^3$ $N$-body sample with
  nearly the same cosmology [the simulation is described in
    \protect\cite{VelGrav07}] does not change any measurements
  presented in \S~\ref{sec:mak_error}. }

Out of {\it FullMock}, we have extracted three spherical cuts of radius
  40~$h^{-1}$Mpc (hereafter denoted by {\it 4k-mockX}), where the velocity-velocity comparisons are
  conducted, and twice deeper counterparts (hereafter denoted by {\it
    8k-mockX}) are used to give better constraints (\S~\ref{sec:lag_volume}) on the reconstruction 
  within the volume of analyses.  
 Each of these catalogues is centered in a different place in the
 simulation such that:
\begin{itemize}
\item {\it 4k-mock6} is mildly overdense, with an effective mean matter
  density $\Omega_\text{eff}=0.35$, and contains 495 haloes. It is
  designed in such a way that large voids and large concentrations of matter (clusters or super-clusters)
  are present near its boundaries, similarly as found in real redshift catalogues
  of our local neighbourhood, such as the UZC \citep{UZC99}, the
  NBG-3k \citep{Shay95,TullyVoid07} and the NBG-8k. This catalogue and its deeper
  counterpart, {\it 8k-mock6}, are particularly suited to address
  edge effects on the NBG-3k (which terminates at Hydra and
  Centaurus clusters) and the NBG-8k (which stops at the Great Wall), respectively.
\item {\it 4k-mock7} is highly overdense,with $\Omega_\text{eff}=0.50$, and contains 656 haloes.
  Very little mass has come in and out of this volume: it behaves somewhat like an isolated universe,
  with small external tides.
\item {\it 4k-mock12} is underdense, with $\Omega_\text{eff}=0.19$,
  and contains 213 haloes. It presents as well a low level of density
  fluctuations along its boundary. 
\end{itemize}

While there is no ambiguity in setting up a $128^3$ MAK mesh when
using all the haloes and the background particles (such as in {\it
  FullMock}), it is less trivial 
to consider lower resolution meshes that will be used in some of the subsequent
analyses. Indeed, the number of mesh elements
assigned to each tracer is not necessarily an integer
anymore. Appendix~\ref{app:subsampling} details the general procedure used to
associate elements of the MAK mesh to each tracer.

\section{Errors in MAK velocities}
\label{sec:mak_error}

Before going over observational issues, we address errors intrinsic to
MAK reconstruction. First, there is  scatter in the reconstruction of
the displacement field itself which is expected to
be rather small \citep{moh2005}. Second, there is scatter due to the
Zel'dovich approximation one uses to convert a displacement field into a velocity
field and to deal with redshift distortions. An accurate knowledge of
the distribution of errors on the reconstructed velocities is
eventually required
for the likelihood analysis we want to introduce in 
\S~\ref{sec:likelihood}. In this section, we measure such a
distribution in real space while redshift space
will be addressed in \S~\ref{sec:redshift_distortion}. In principle, the
width of such a distribution is expected to increase when
observational biases are taken into account while its shape should not
change significantly.

\begin{figure}
  \begin{center}
    \includegraphics*[width=.9\linewidth]{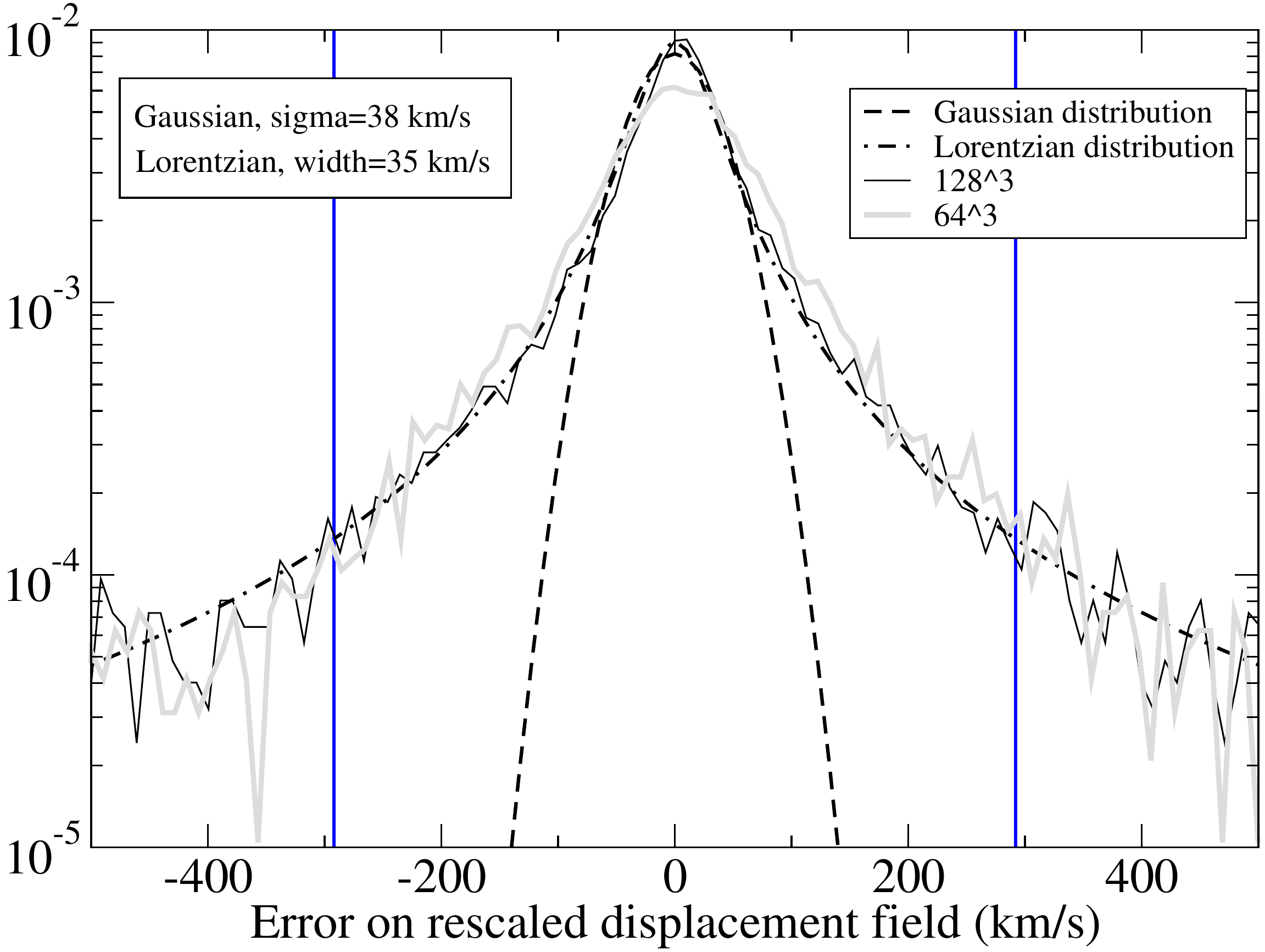}
  \end{center}
  \caption{\label{fig:reconstructed_displacement_stat} {\it Error
      in reconstructed displacements} -- This
    plot displays the probability distribution of the quantity
    $\beta \left(\Psi_\text{r,rec} - \Psi_\text{r,sim}\right)$
    measured in {\it FullMock} (solid
    curve), where $\Psi_\text{r,rec}$ and
    $\Psi_\text{r,sim}$ are the
    line-of-sight component of the reconstructed and simulated
    displacement fields, respectively, after choosing
    an observer at the center of the simulation box. The dashed and
    dot-dashed curves give the best fit of a Gaussian and a Lorentzian
    distribution, respectively.}
\end{figure}

\correction{
We consider, in this section, reconstructions based on the catalog
{\it FullMock}, for which periodic
boundary conditions are applied to avoid edge effect problems.
We also assume that we know the mass of all of described catalog objects (haloes and
individual particles). Our subsequent reconstructions have a resolution
within $64^3$ and $128^3$ mesh elements. We will thus present two
reconstructions obtained on two different initial MAK mesh, $128^3$
and $64^3$, obtained using the procedure presented in
Appendix~\ref{app:subsampling}. The results on the reconstructed
displacement field are given in Fig.~\ref{fig:reconstructed_displacement_stat}. 
These plots give the distribution of differences, $P_\text{DE}$, between 
the line of sight component of the reconstructed
displacement field and the ``exact'' one, given by the simulation. 
}

The dot-dashed and dashed curves correspond to a least-square fit of the function
$P_\text{DE}$ corresponding to the $128^3$ reconstruction
respectively with a Gaussian fit, and a Lorentzian fit given by
\begin{equation}
  P_\text{Lor}(x) = \frac{1}{\pi B} \frac{1}{1 +
    \frac{x^2}{B^2}}\text{ .}
\end{equation}
Examination of Fig.~\ref{fig:reconstructed_displacement_stat} supports
the Lorentzian approximation with $B=35$~\kms, which reproduces better
the long tails of $P_\text{DE}$ than the Gaussian.

The width, $B$, of $P_\text{DE}$ is rather small compared to the
 line-of-sight dispersion, $\langle \beta^2 \Psi_r^2 \rangle^{1/2} \simeq 292$~\kms,
as expected. Naturally, the function $P_\text{DE}$ is slightly flatter and
larger for the $64^3$ case than for the $128^3$ one. However, the far
end tails of $P_\text{DE}$ are the same for $64^3$ and $128^3$. In this
regime, the measurements are not influenced by the resolution
of the grid used to perform the reconstruction but rather by the
inability of MAK to reproduce the internal dynamics of 
massive, relaxed objects \citep{moh2005}.

\begin{figure*}
  \begin{center}
   \begin{tabular}{ccc}
     Reconstruction $128^3$ & Reconstruction $64^3$ & Simulation\\
     \includegraphics[width=.3\linewidth]{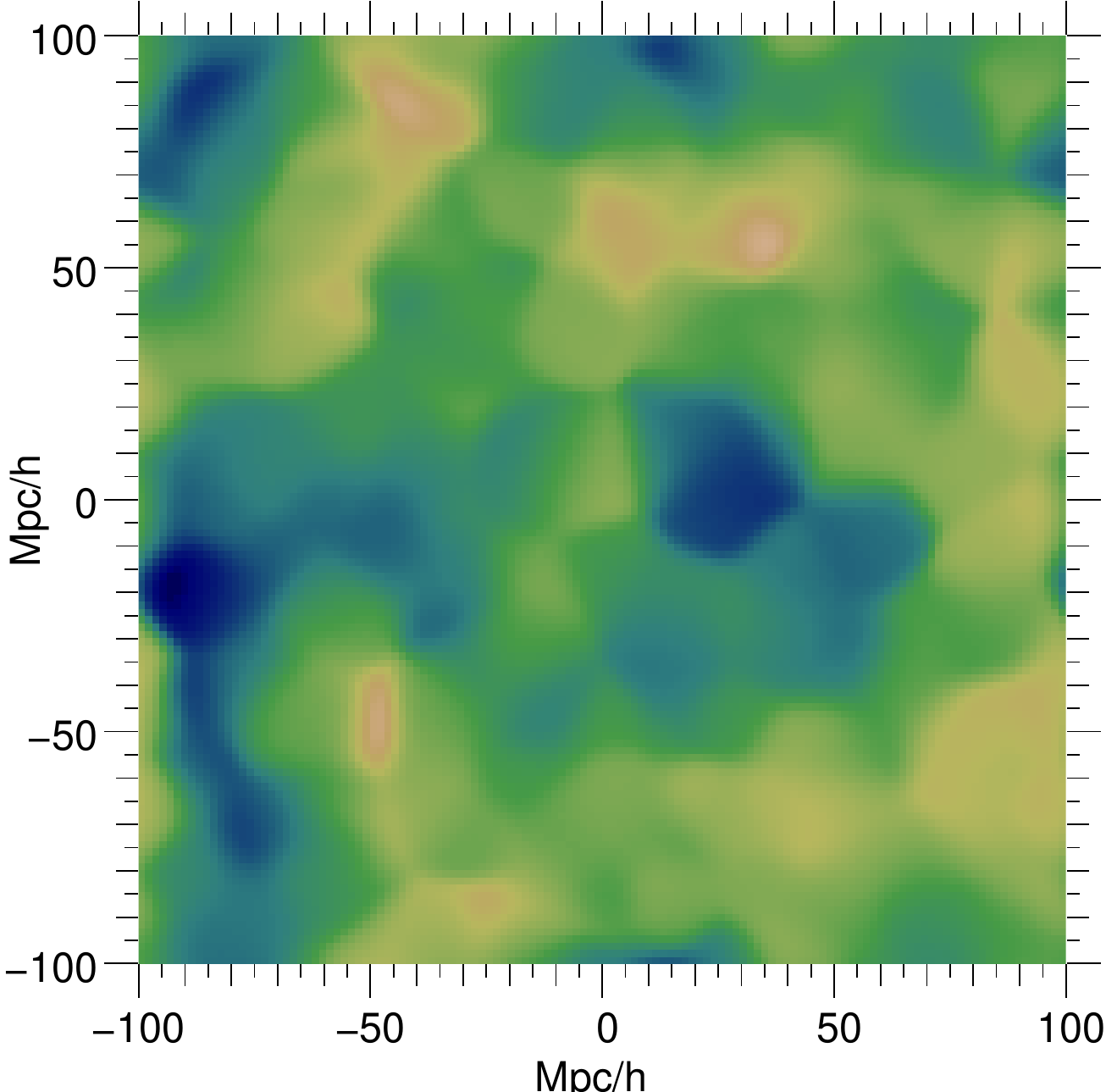} &
     \includegraphics[width=.3\linewidth]{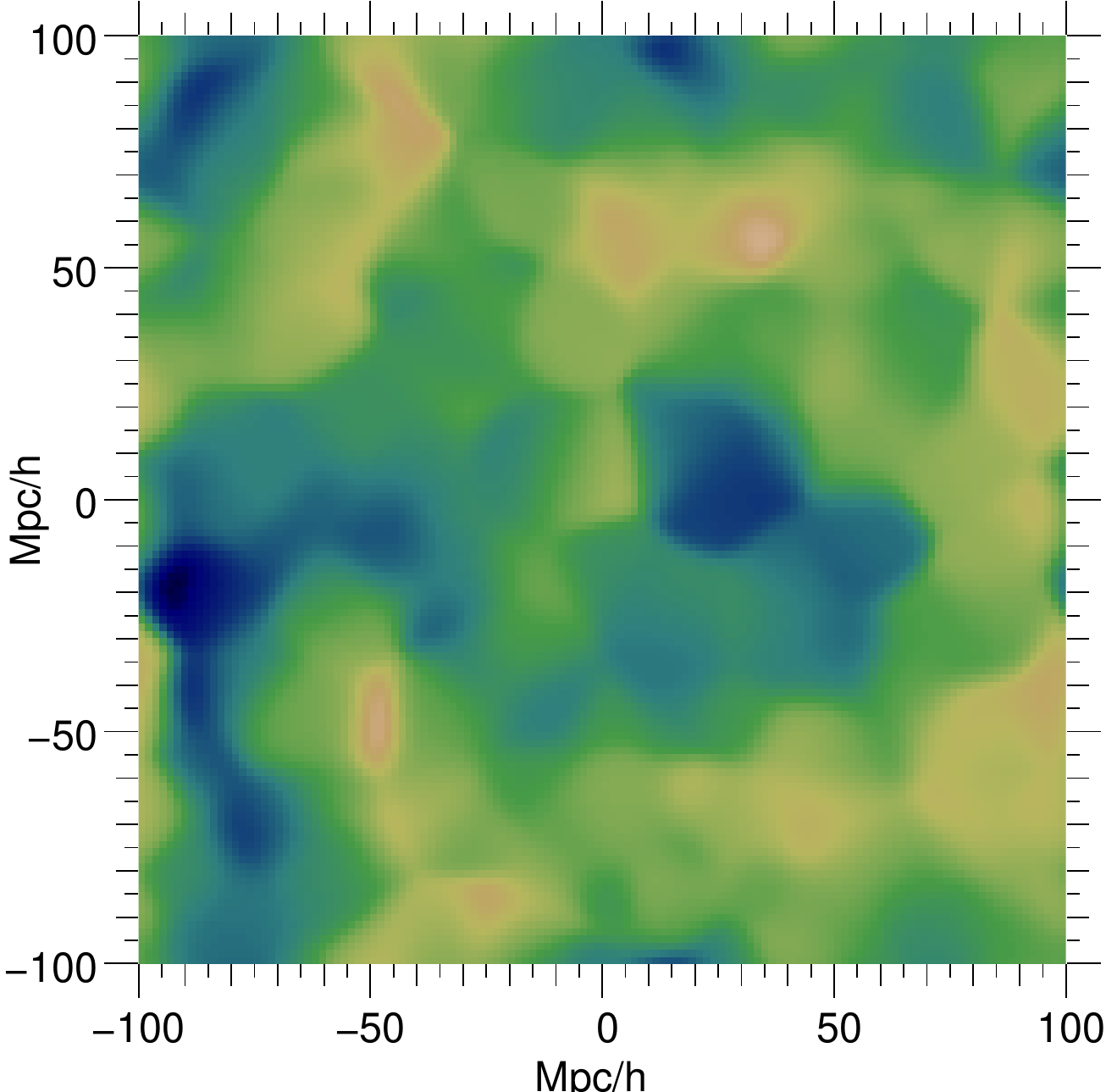} &
     \includegraphics[width=.3\linewidth]{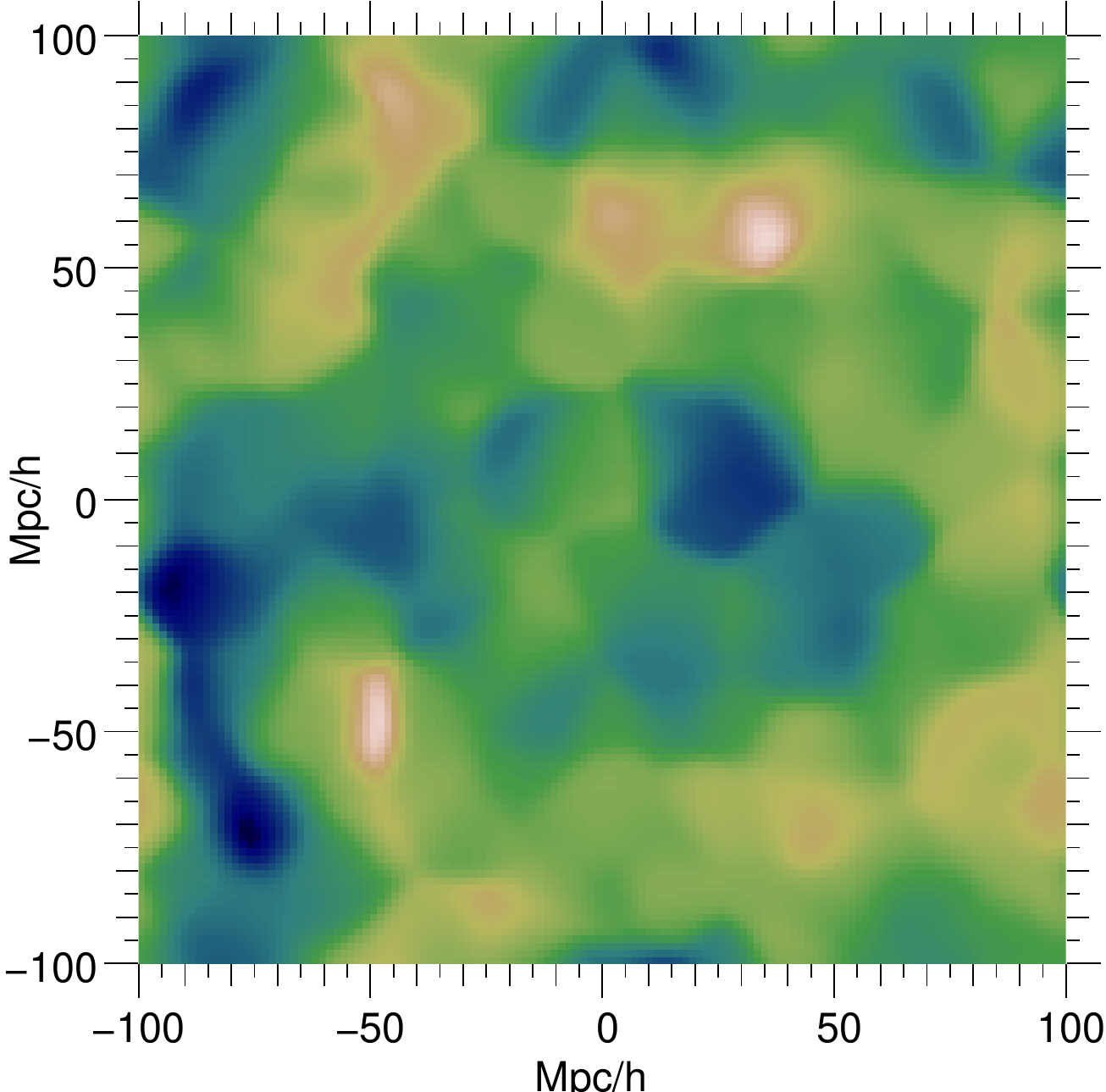} \\
     \includegraphics[width=.3\linewidth]{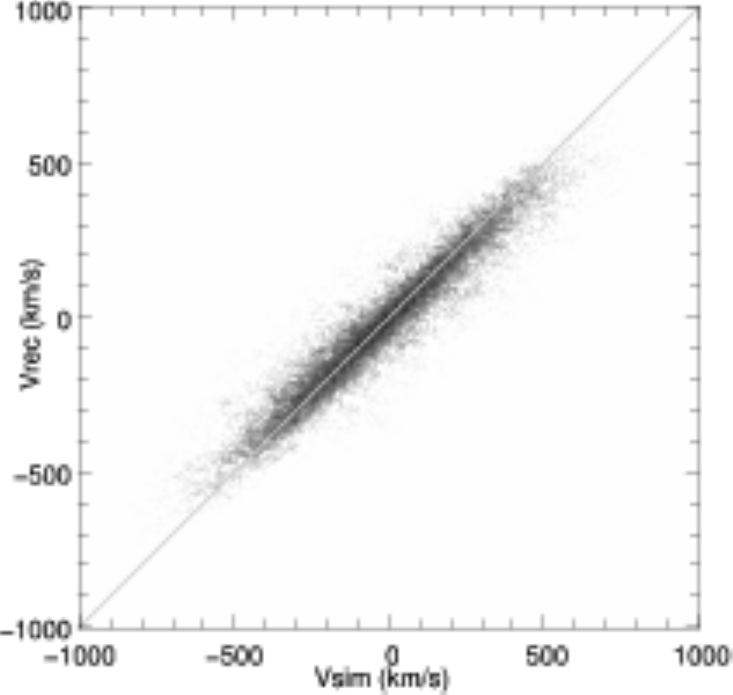}  &
     \includegraphics[width=.3\linewidth]{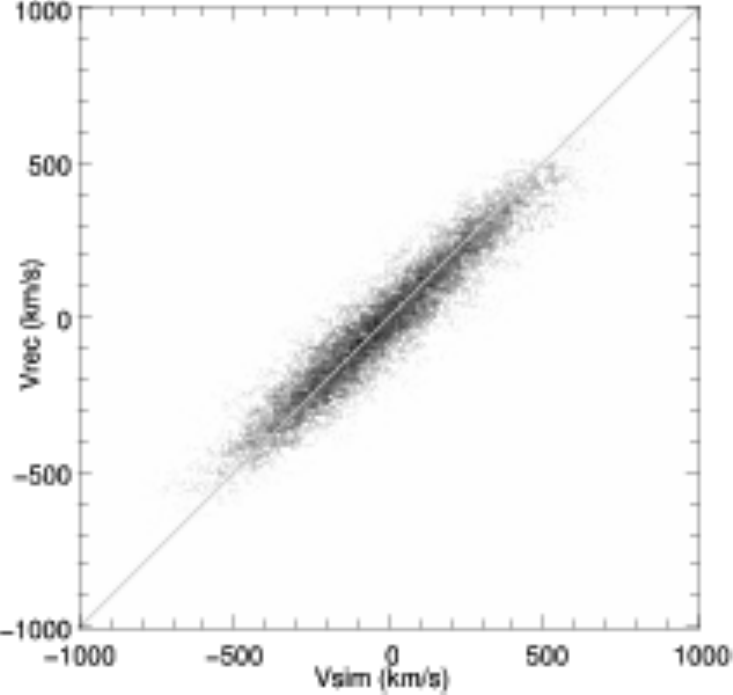} &
     \includegraphics[width=.3\linewidth]{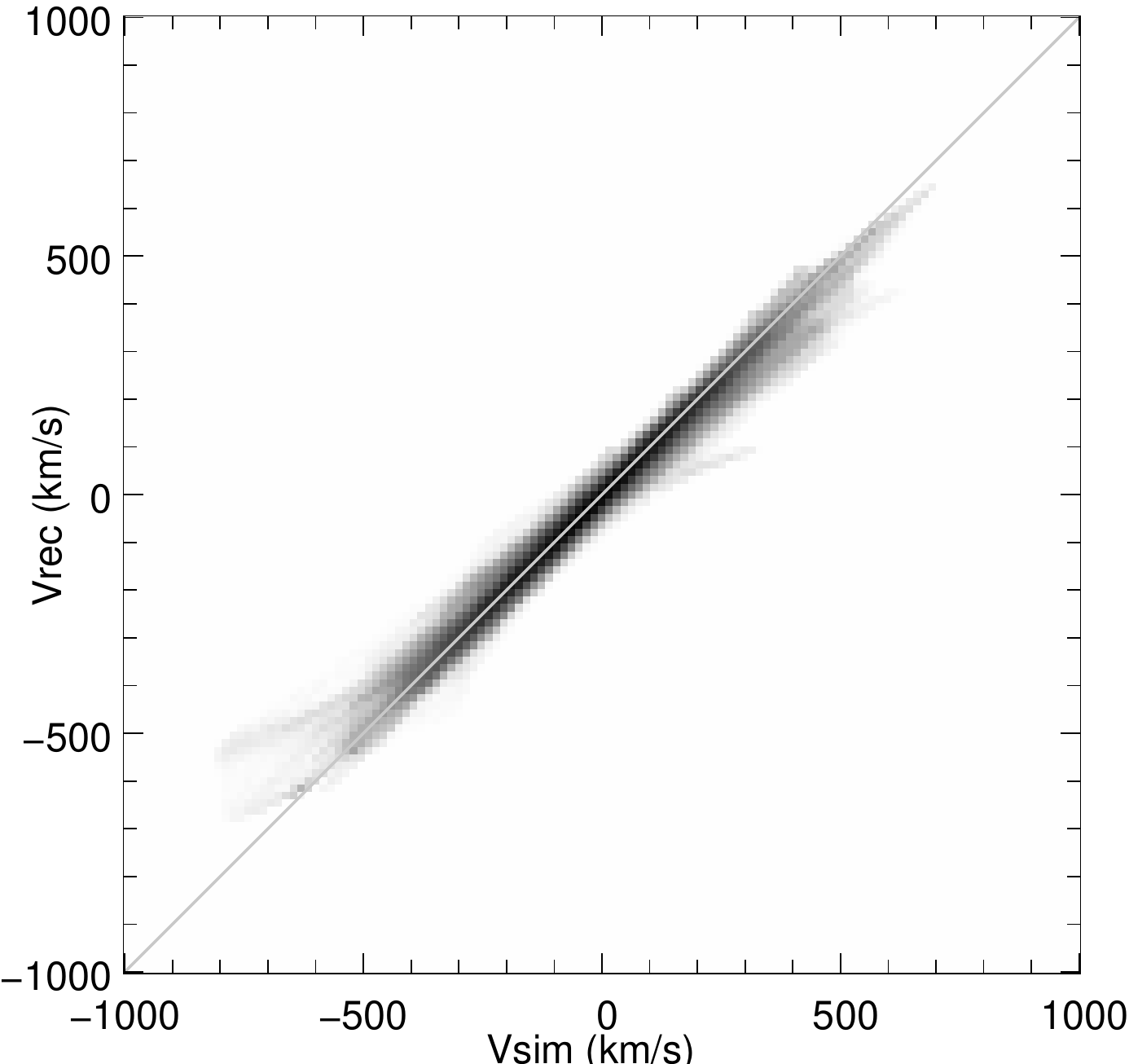}
   \end{tabular}
  \end{center}
  \vspace*{-0.4cm}\caption{\label{fig:scatter_velocity_collapse}    
    {\it Velocity field reconstruction on FullMock} -- 
    Top panels: A slice of the line-of-sight component of the
    simulated velocity field, $v_\text{r,sim}$, and the reconstructed one,
    $v_\text{r,rec}$, after smoothing with a
    5~$h^{-1}$Mpc Gaussian window. The observer is at the center of this
    slice. Bottom panels: Scatter plots between $v_\text{r,sim}$ and
    $v_\text{r,rec}$ for invidual haloes (left) and after smoothing (right). }
\end{figure*}

\begin{figure}
  \begin{center}
    \includegraphics*[width=.9\linewidth]{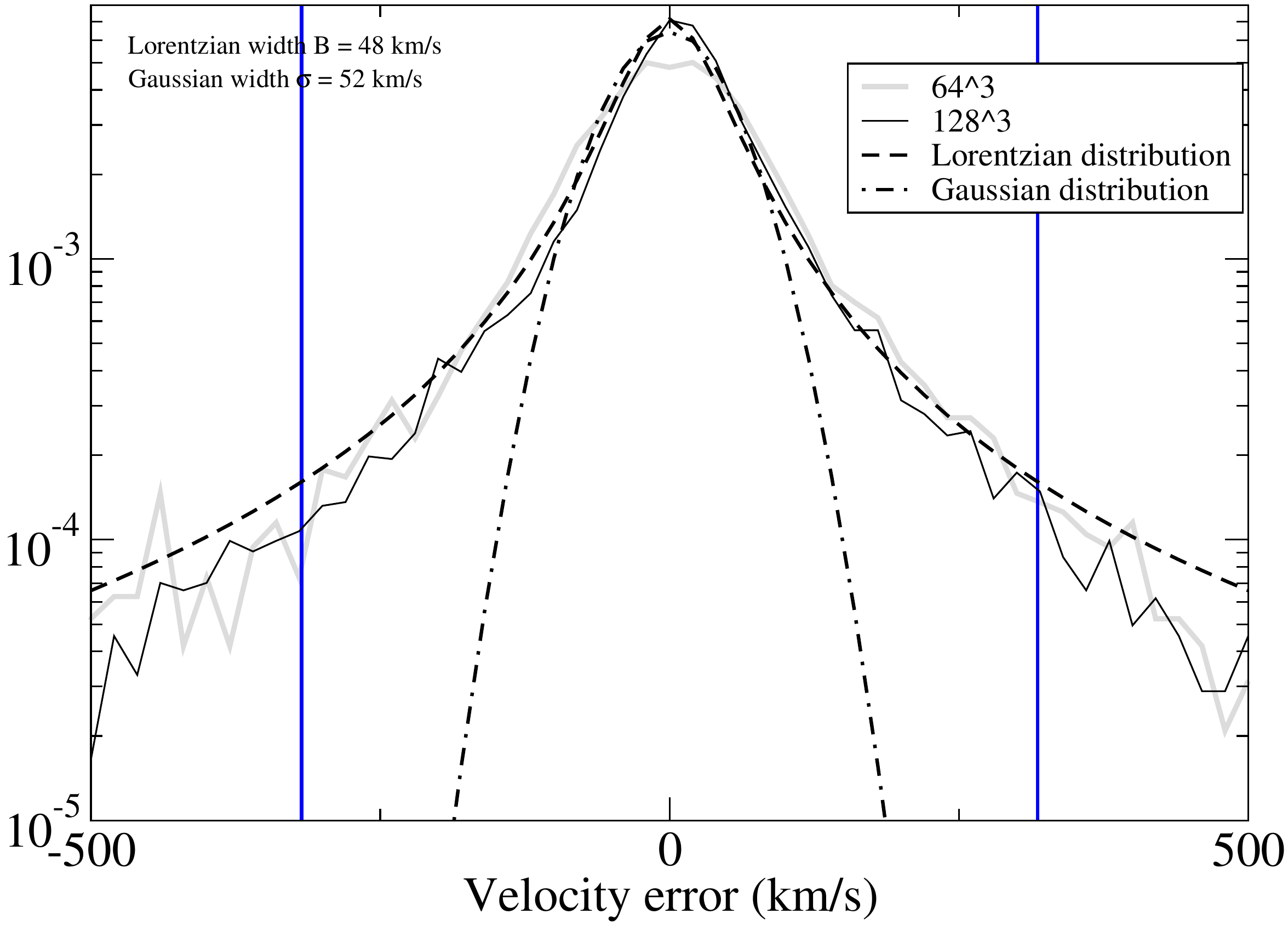}
  \end{center}
  \caption{\label{fig:reconstructed_velocity_stat} {\it Error
      in reconstructed velocities} -- Same as in
    Fig.~\ref{fig:reconstructed_displacement_stat} but the solid curve
    corresponds to the
    probability distribution of the quantity $v_\text{r,rec} -
    v_\text{r,sim}$, where $v_\text{r,rec}$ and $v_\text{r,sim}$ are the line-of-sight
    reconstructed and simulated velocities, respectively.}
\end{figure}

 Fig.~\ref{fig:reconstructed_velocity_stat} is similar to
 Fig.~\ref{fig:reconstructed_displacement_stat} but considers line of
 sight reconstructed velocities vs ``exact'' ones.
Although Zel'dovich approximation introduces extra noise as shown by
a wider width of the distribution, $P_{DE}$ remains roughly Lorentzian with a
small width $B=48$~\kms. \correction{This error variance is
  grossly 25\% higher than
  the expected velocity field variance on the simulation volume
  (\S~\ref{subsec:nbody}). We are thus not affected by cosmic variance
  effects that could have been induced by modes larger than the box size of
  the simulation.} 

These results are fully supported by the examination
of Fig.~\ref{fig:scatter_velocity_collapse}. However, the lower panels
of this figure shows that the joint distribution
$P(v_\text{sim},v_\text{rec})$ presents non-trivial tails above the
diagonal line in the lower left quadrant and below the diagonal line
in the upper right quadrant, respectively. These tails do not
disappear even after smoothing of the velocity field with a 5~$h^{-1}$Mpc
Gaussian window. This is due to non-linear features in the
dynamics not taken into account by our MAK+Zel'dovich prescription, which
produces a slightly smoother velocity field than the real one. As a result,
upper left panel of Fig.~\ref{fig:scatter_velocity_collapse}, which
corresponds to the reconstruction, is less
contrasted than the upper right one, which corresponds to the
simulation.

These non-linear tails give a propeller shape to
$P(v_\text{sim},v_\text{rec})$ which is susceptible to inducing a small bias
on the final velocity-velocity comparison. For instance, one can estimate
the slope of the lower left scatter plot of
Fig.~\ref{fig:scatter_velocity_collapse} using the ratio
$s = \sigma_{v,rec} / \sigma_{v,sim}$, where
$\sigma^2_{v,rec}$ and $\sigma^2_{v,sim}$ are the variances of the
reconstructed and simulated velocity fields, respectively. In this
case, the estimated $\beta$ is biased to higher values by about 7\%. 
However, visually inspecting the scatter shows no measurement bias
should occur if only the central part of the scatter is used for the
computation. To achieve this, we have first applied an adaptive SPH
filter on the scatter plot 
to produce a Probability Density Function (PDF), which is probed
by the scatter in the points, on a
regular mesh grid. We then compute the 1.5$\sigma$ isocontour 
which encloses the region where the integrated PDF is equal to 68\%. This procedure
has already been used in \cite{VelGrav07} for the gravity-velocity
comparison with total success. Only the
points enclosed by the 1.5$\sigma$ isocontour are used to compute the new $s_{med,68}$
coefficient. The $\beta$
parameter deduced from $s_{\text{med},68}$ is now statistically unbiased.
Similarly, we define two other slope estimators $s_{\text{min},68}$ and
$s_{\text{max},68}$ whose relevance is discussed in Appendix~\ref{app:err_analysis}.
In this paper, until \S~\ref{sec:malmquist}, we will only discuss the
measurement of $\Omega_\text{m}$ obtained through the estimation of
$s_{\text{med},68}$. The $\Omega_\text{m}$ obtained by this method is identified
by a ``$1.5\sigma$'' to make a difference with the one obtained through
the likelihood analysis that will be established in
\S~\ref{sec:malmquist} and which is identified by a ``$\mathfrak{L}$''
in the tables and figures. A test of this method on a simulated
scatter distribution, whose shape is built on analysis of
reconstruction errors, is detailed in Appendix~\ref{sec:stat_bias}.

\section{Mass-to-light assignment}
\label{sec:ml_assign}

Most reconstruction methods, including ours, infer the total
matter distribution as a function of the visible matter distribution
traced by galaxies. The fundamental assumption one  usually makes is
that the relation between these two distributions is highly
deterministic. In other words, one assigns to each galaxy of a given
luminosity $L$ a dark matter concentration (a halo) of mass $M=f(L)$.
However, there are several issues in this procedure:
\begin{itemize}
  \item {\it Mass-to-light ratio} -- The choice of a function $f(L)$ influences considerably the
    results and is expected to introduce significant bias on the
    measured $\beta$ if performed unwisely. Now, the function $f(L)$ is
    coarsely determined 
    \citep{TullyML05,Marinoni02} from direct measurements in
    observations. One way to infer this
    function is to rely on semi-analytic models of galaxy formation, but
    this represents a very strong prior on the
    measurements. Furthermore, $f(L)$ remains a mean relation around
    which there can be some significant scatter. This dispersion can as well
    introduce some significant biases.
  \item {\it Missing tracers / Magnitude limitation} -- Even if function $f(L)$ is perfectly known, fainter galaxies are still
    missing in the catalogues due to the  limitations of observational
    instruments. For instance, in magnitude-limited catalogues, the
    number density of detected galaxies decreases with distance from
    the observer. These missing tracers have unknown positions and
    correspond to a part of the dark matter distribution which is
    totally undefined. This missing mass has to be taken into account
    in some way.
\end{itemize}

In what follows, we will first address the second issue in a very
simple way which assumes that the function $f(L)$ is well known (namely
the masses of dark matter haloes themselves) but
there is a fixed low-mass cut-off. The problem then consists in
determining the unknown part of the dark matter distribution (namely the
particles unassigned to any halo). Clearly
it is correlated with the detected mass tracers but less
clustered. There are two extreme ways to locate this missing
mass 
\begin{enumerate} 
  \item[(a)] associate it with the existing tracers as usually done
    with the analysis of real observations 
  \item[(b)] associate it with a uniform background. 
\end{enumerate}
Of course, the real solution is somewhat intermediate between (a) and (b) as will be
shown in \S~\ref{sec:diffuse_mass}.

Then, we turn in \S~\ref{sec:ml_ratio} to the issue of the choice of
$f(L)$. In this paper, we
prefer to be as free as possible from strong priors so we deliberately
do not use results from semi-analytic models of galaxy formation.
Instead, we use determinations of $f(L)$ from observational data but,
unfortunately, there are large uncertainties in these
measurements. The point here is to quantify, quite heuristically
though, the effect of these uncertainties, random or systematic, on
the measurement of $\beta$. Indeed, one is both confronted with a
possibility of a wrong
approximation of $f(L)$ and most
probably a large scatter around this mean relation.
 
In sufficiently deep galaxy catalogues, the effect of the missing
tracers is expected to be negligible close to the observer and, in
general, to increase with the distance from the observer. With
appropriate weighting of the data, one can minimize the bias brought
by the procedure used to infer the missing mass distribution far from the
observer. In \S~\ref{sec:incompleteness}, we shall illustrate
this point by considering the case of a magnitude-limited catalogue where all the
missing mass is associated with the existing tracers [method (a) above].

\subsection{Missing tracers}
\label{sec:diffuse_mass}

\begin{figure*}
  \begin{center}
    \begin{tabular}{ccc}
      & Simulation & \\
      & \includegraphics[width=.3\linewidth]{refVelSim} & \\
      All missing mass in haloes & Optimal compromise & All missing
      mass to background \\
      \includegraphics[width=.3\linewidth]{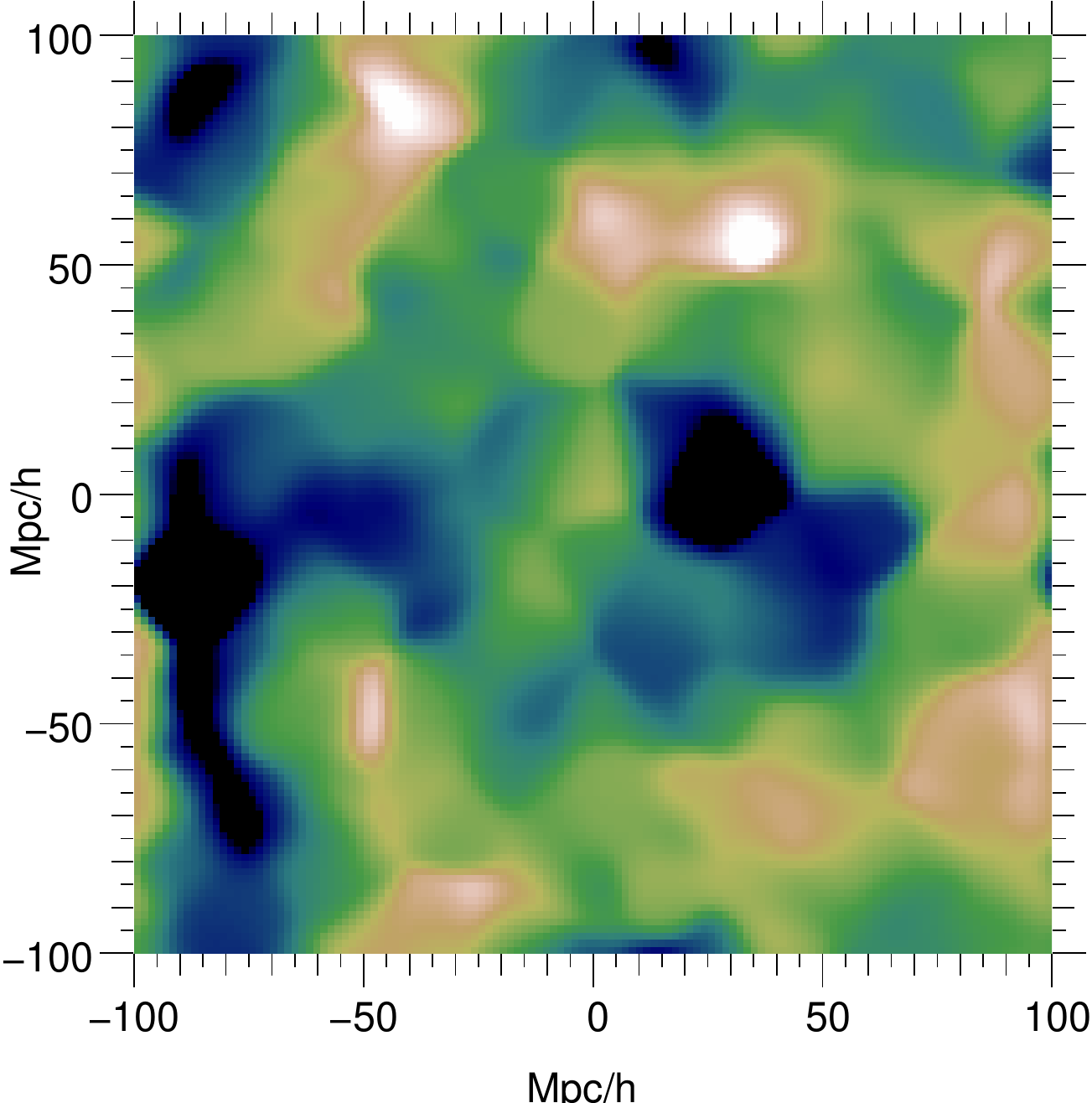} &
      \includegraphics[width=.3\linewidth]{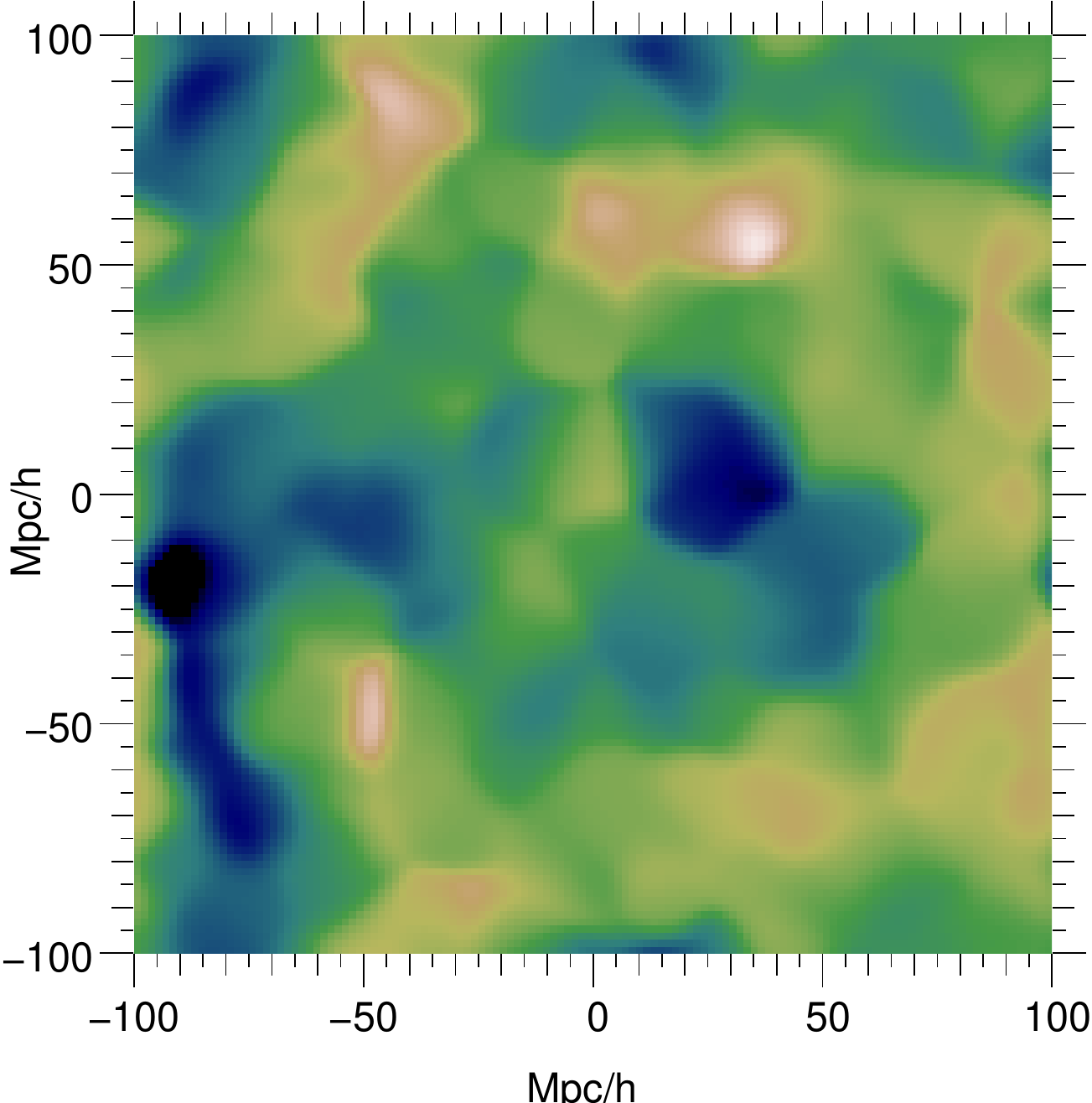} &
      \includegraphics[width=.3\linewidth]{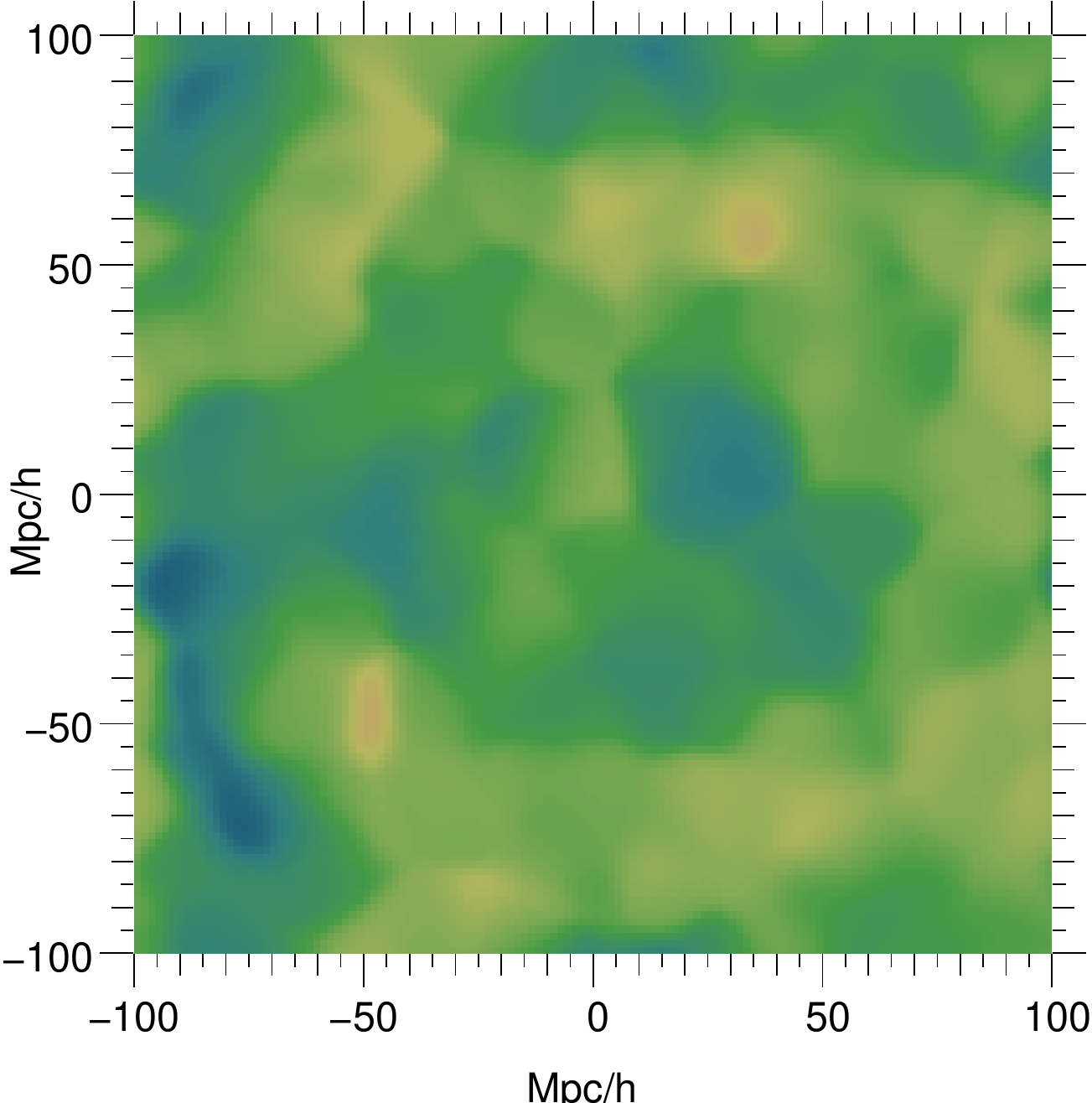}
      \\
      \includegraphics[width=.3\linewidth]{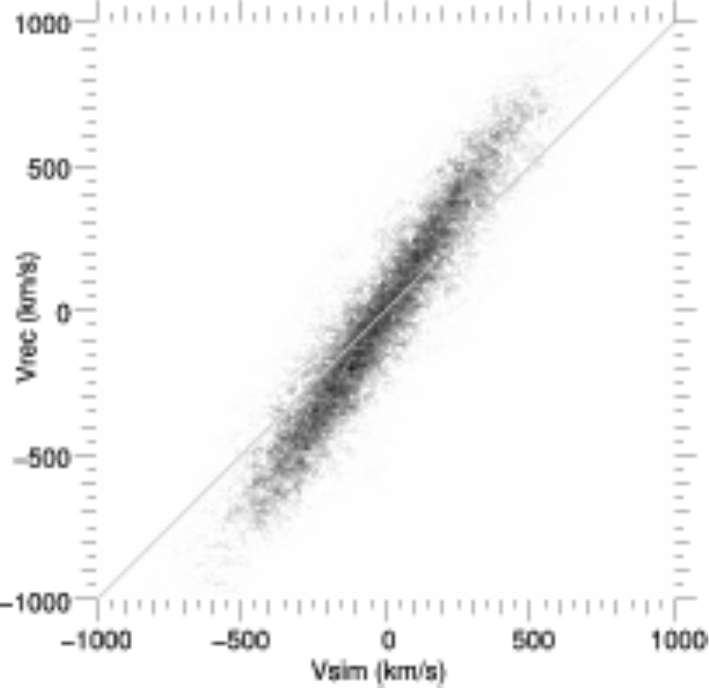} &
      \includegraphics[width=.3\linewidth]{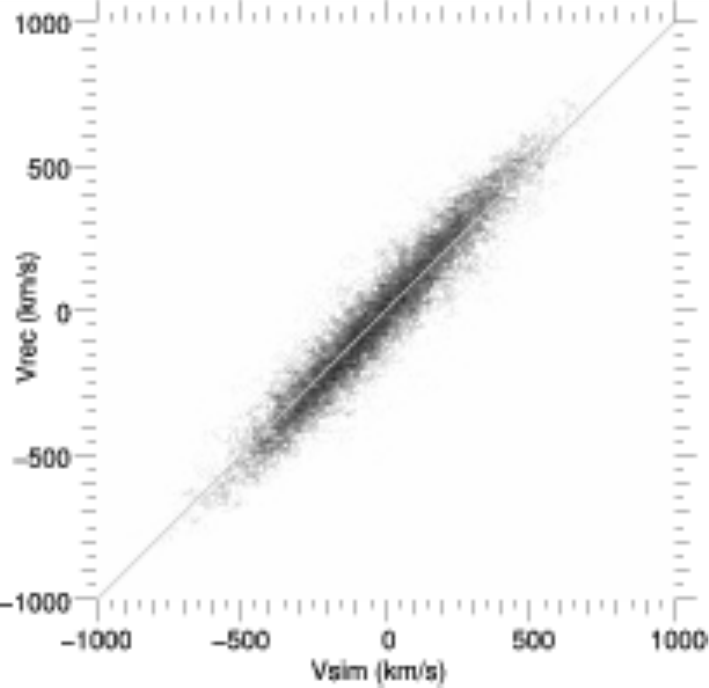} &
      \includegraphics[width=.3\linewidth]{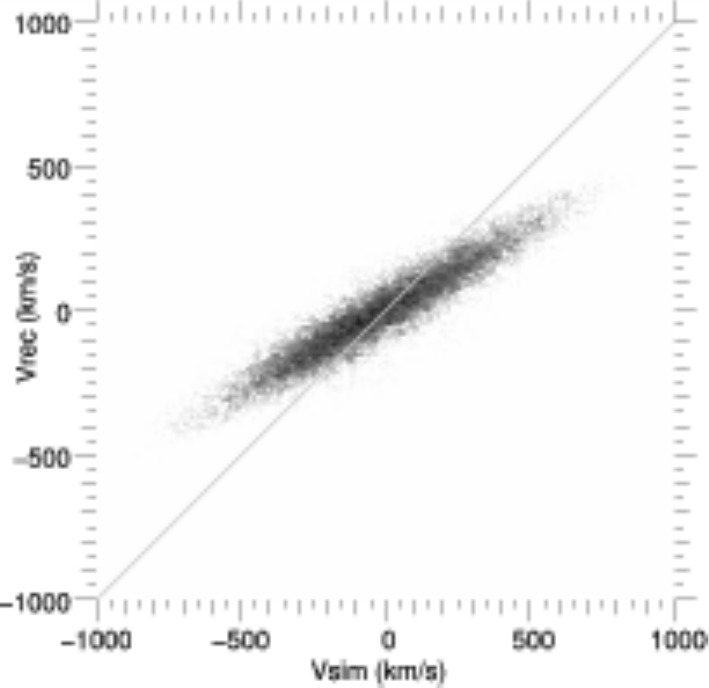}
    \end{tabular}
  \end{center}
  \caption{\label{fig:velocity_random} {\it Diffuse mass correction}
    -- The top panel gives a slice of the line-of-sight
    component of the simulated velocity field, after smoothing with a
    5~$h^{-1}$Mpc Gaussian window. The observer has been put at the
    center of this slice. The second row of panels
    represents the line-of-sight component of the reconstructed
    velocity field, smoothed in the same way, for different
    corrections of the diffuse mass. The third row of panels give the
 scatter distribution of individual reconstructed velocities of haloes {\it vs}
 simulated ones. 
  The left panels give the result of a 
    reconstruction on a mock catalog which only contain the haloes and
    not the background field but at the same time conserves the total
    mass of the catalog by reassigning the missing mass to the
    haloes. The right panels give the result for a reconstruction
    based on a mock catalogue for which the missing diffuse mass is
    represented by 
    a background field composed of particles placed randomly in the
    catalogue. 
  The center panels give the result of a reconstruction on a mock
    catalogue which only contain the haloes and a random background
    field. The mass that have been initially removed from the mock
    catalogue (the background ``galaxies'') is reassigned as follows: 
    60\% to
    haloes and  40\% to the background. }
\end{figure*}

\begin{figure}
  \begin{center}
    \includegraphics*[width=.9\linewidth]{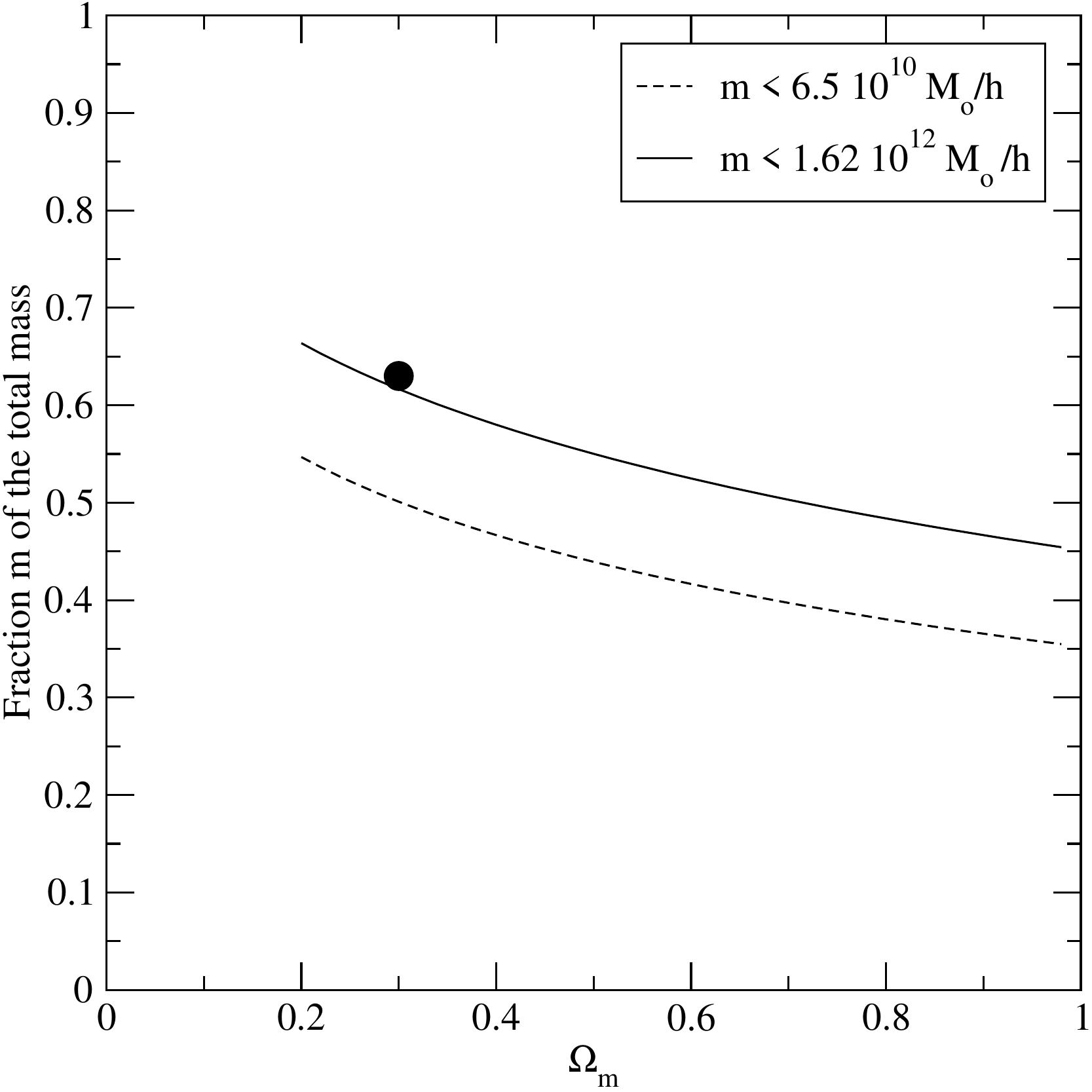}
  \end{center}
  \caption{\label{fig:mass_fraction} {\it Diffuse mass} -- In this
    plot, we represent the fraction of the clustered mass below two
    mass resolutions for a standard \LCDM type cosmology ($h=0.65$,
    $\sigma_8=0.99$). \correction{We used a power spectrum as proposed by \protect\cite{BBKS}}. The
    curvature of the Universe is kept flat while $\Omega_{\text{m}}$
    varies. This fraction is plotted for  mass resolutions:
    $2.5\times 10^{12}\text{ M}_\odot$ (corresponding to the lower
    mass limit of haloes in our simulation) and $10^{11}\text{ M}_\odot$ ($\simeq
    10^{9}\text{ L}_{\odot,B}$). The unclustered fraction in {\it
    FullMock} is given by the back filled circle.
    The fraction of mass below both of these limits is still considerable.}
  \smallskip
\end{figure}

Fig.~\ref{fig:mass_fraction} shows the expected fraction of the total
mass below a fixed threshold as a function of $\Omega_\text{m}$, using
the \cite{ShethTormen02} model (see also
Fig.~\ref{fig:press_schechter}). The solid line corresponds to the
mass cut-off of haloes in {\it FullMock} and agrees, as expected, with the measurement
in the simulation for $\Omega_\text{m} = 0.30$. Here, 63\% of
the mass is outside of the haloes, which represent
our ``galaxies'' with known $M/L$ ratio. The particles not
linked to the haloes
represent the missing mass. In
Fig.~\ref{fig:scatter_velocity_collapse}, their exact location was used
to perform the reconstruction. The only information available now is
the distribution of ``visible galaxies''. The missing mass needs to be
redistributed using only these pieces of information.
We propose two extreme ways to do so:
\begin{enumerate}
  \item[I.] {\it All missing mass to background} -- 
    Prior to the reconstruction, the missing mass is divided into
    particles which are randomly put in the catalog following
    a poissonian distribution. In the example illustrated by the right
    panels of Fig.~\ref{fig:velocity_random} we choose for simplicity particles of
    the same mass as those in the simulation.    
  \item[II.] {\it All missing mass in haloes} --
    The missing mass is attributed to the existing haloes
    in proportion to their masses, as illustrated by left panels of
    Fig.~\ref{fig:velocity_random}. This approach is equivalent, in real
    observations, to multiplying
    the $M/L$ ratio of galaxies or group of galaxies by a constant
    $\alpha > 1$. 
\end{enumerate}

Obviously, in I, the screening effect due to the background is
exagerated, hence the reconstructed velocity is less contrasted and
$\beta$ is over-estimated to compensate for this. In II, on the other
hand, the potential wells are more contrasted than they should be, which
leads to the opposite effect. At this point, it is extremely
tempting to try to find a simple compromise between I and II as
illustrated by middle panels of Fig.~\ref{fig:velocity_random} where
60\% of the missing mass was linked to the tracers and the remaining
to a uniform background. With this particular choice of the
redistribution, the match between the reconstructed and the simulated
velocity fields is spectacular. This result is non-trivial given the
simplicity of the handling of this {\it sixty three} percent
missing mass all the more since the scatter on the middle-lower panel
of Fig.~\ref{fig:velocity_random} is {\it of the same order of} that
of the lower left panel of Fig.~\ref{fig:scatter_velocity_collapse},
where all the tracers contribute optimally.

Although the choice of the optimal
redistribution remains a priori unknown in a real galaxy catalogue one
can at least infer error bars from I and II. In that framework,
Fig.~\ref{fig:velocity_random} unfortunately provides quite a bad constraint on 
$\beta$, $0.36 \la \beta \la 0.85$. However, in real galaxy catalogues, such as
the NBG-3k or the NBG-8k, the minimum luminosity is of the order of
$10^9$~L$_\odot$. This corresponds to a less abrupt mass cut-off, $M_\text{cut} \sim
10^{11}\text{ M}_\odot$, than in Fig.~\ref{fig:velocity_random}, where
$M_\text{cut} = 2.5\times 10^{12} \text{ M}_\odot$. Therefore, one expects
the problem of missing mass to be less 
saliant in real observations, as illustrated by the dashed curve of
Fig.~\ref{fig:mass_fraction}. Furthermore, an
appropriate use of mock catalogues can help at calibrating the
redistribution of mass, as performed in middle panels of
Fig.~\ref{fig:velocity_random}. 

\subsection{Mass-to-light ratio}
\label{sec:ml_ratio}

\newcommand{\mydataflow}[3]{%
\text{\begin{minipage}{1.25cm}\centering Halo catalog\end{minipage}}
\xrightarrow{\text{#1}}\text{\begin{minipage}{1.55cm}\centering
    Luminosity Catalog\end{minipage}} \xrightarrow{\text{#2}}
\text{\begin{minipage}[t]{1cm}\centering MAK\end{minipage}(#3)}
}

\newcommand{\myexpansion}[2]{%
\begin{minipage}{#1}
  \begin{center}
    #2
  \end{center}
\end{minipage}
}

\begin{figure}
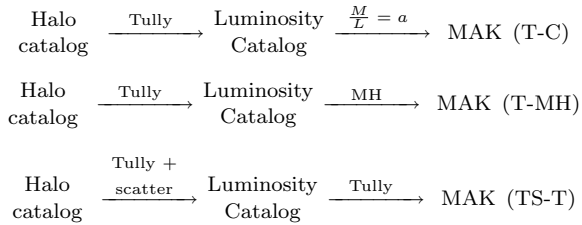

  \begin{center}
    \begin{equation*} \displaystyle
      \begin{array}{c}
        \mydataflow{\myexpansion{1cm}{Tully}}{\myexpansion{1cm}{$\frac{M}{L}=a$}}{T-C} \\
        \\
        \mydataflow{\myexpansion{1cm}{Tully}}{\myexpansion{1cm}{MH}}{T-MH} \\
        \\
        \mydataflow{\myexpansion{1cm}{Tully + scatter}}{\myexpansion{1cm}{Tully}}{TS-T}
      \end{array}
    \end{equation*}
  \end{center}
  \caption{\label{fig:flow_luminosity_catalog} {\it $M/L$ assignment} -- Sketch of the procedures used to test the influence of a choice of a $M/L$ assignment, as explained in the main text. }
\end{figure}

\begin{figure}
  \begin{center}
    \includegraphics*[width=.8\columnwidth]{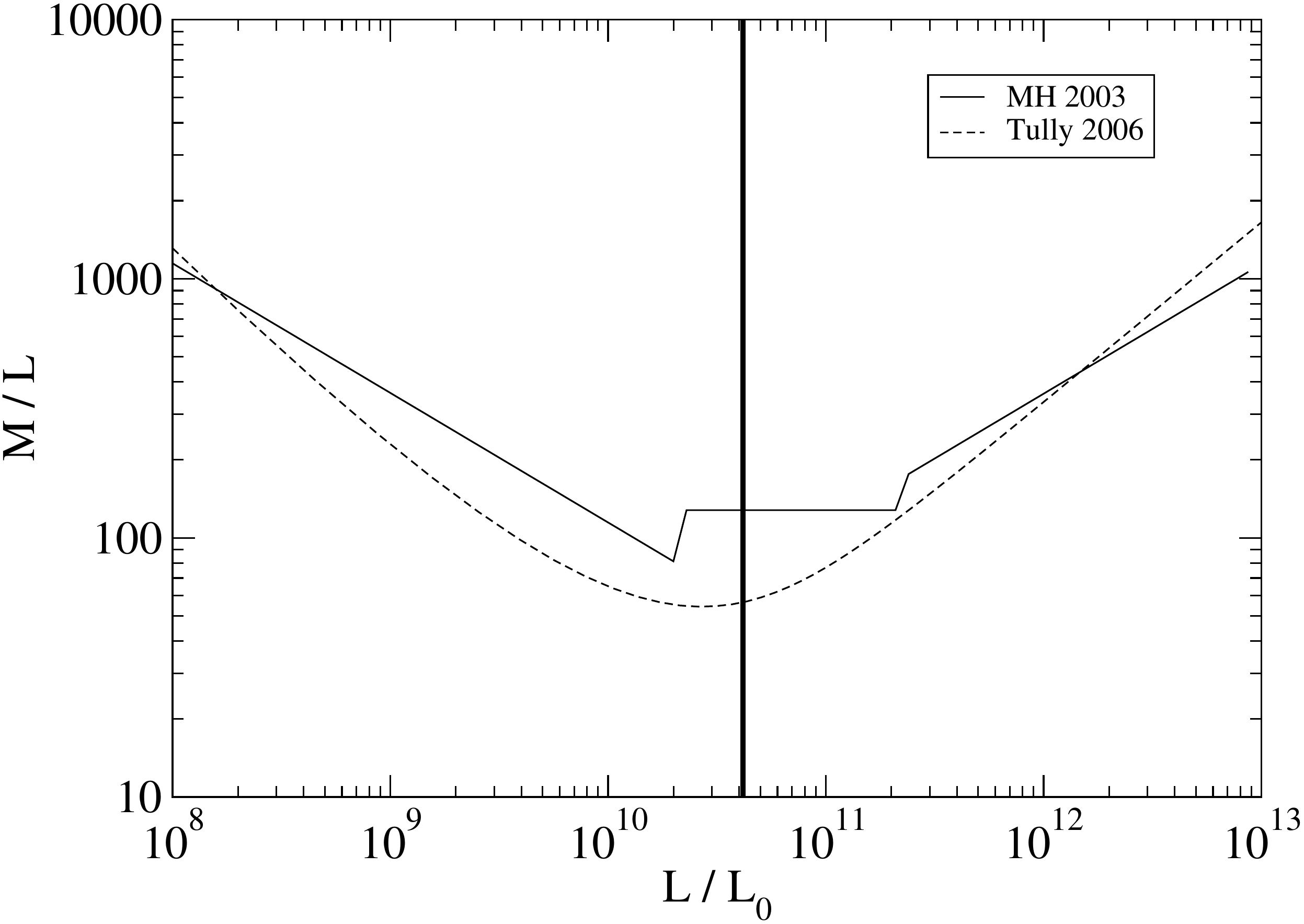}
    \includegraphics*[width=.8\columnwidth]{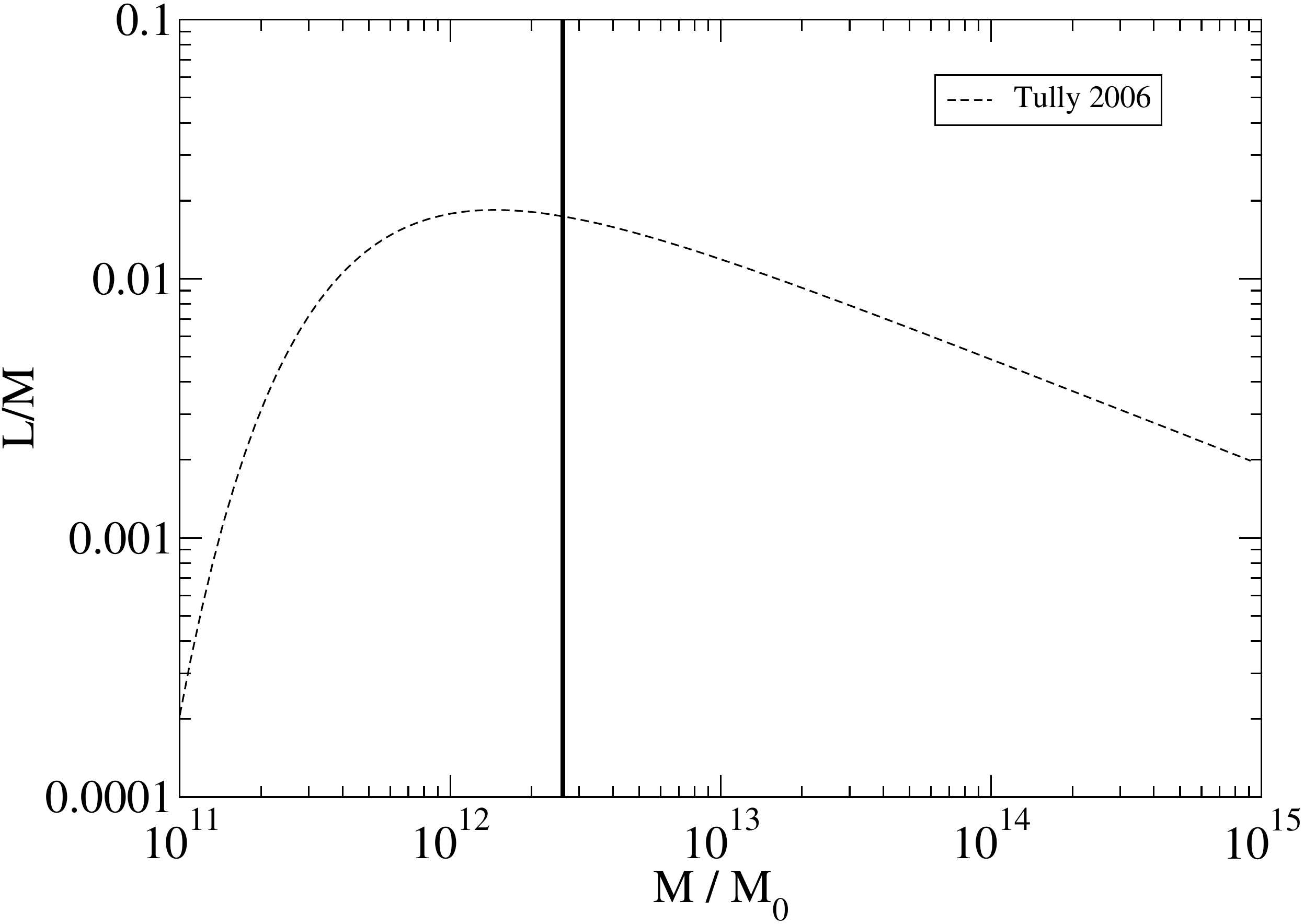}
  \end{center}
  \caption{\label{fig:mass_lights} {\it M/L function} -- The above two
    plots give the forward and inverse mass-to-light functions for
    both \citet{TullyML05} and \citet{Marinoni02} fits. The top panel
    gives the $M/L$ as a function of the luminosity $L$, the bottom
    panel gives $L/M$ as a function of the mass $M$.}
\end{figure}

To test how the choice of mass assignment to galaxies or group of
galaxies affects the results we consider the three following cases, as summarized in Fig.~\ref{fig:flow_luminosity_catalog}:
\begin{enumerate}
  \item {\it T-C case}: a galaxy catalogue is extracted from {\it FullMock}
    by associating a luminosity $L(M)$ to each dark matter
    halo or background particle using Tully's latest best fit of the
    group mass-luminosity relation 
    \citep[][ see Fig.~\ref{fig:mass_lights}]{TullyML05}
    \begin{equation}
      \frac{L_B}{L_\odot} = 2700 \left(\frac{M}{M_\odot}\right)^{0.59}
      \mathrm{e}^{-6\times 10^{11} M_\odot/M},
      \label{eq:tully_ml}
    \end{equation}
    which gives the luminosity in the B band for groups in the mass range $10^{11}\,\,
    \text{M}_\odot \text{ - } 10^{15}\,\,\text{M}_\odot$. Then a new
    mass is given to each tracer assuming 
    \begin{equation}
      M/L = \text{constant,}\label{eq:ml_constant}
    \end{equation}    
    as often used in the litterature, and MAK
    reconstruction is performed on a resampling of this mass
    distribution.
    \medskip
  \item {\it T-MH case}: a less extreme case than assuming
    $M/L=$~constant consists in separating the tracers in three broad
    classes: faint galaxies, luminous galaxies and group/clusters of
    galaxies, as performed by \cite{Marinoni02}, hereafter MH. To do
    this, they used a simple mapping between the Schechter
    luminosity function and the Press-Schechter mass function that
    reads as follows
    \begin{equation}
      \begin{array}{lr}
        M/L = 1.15\;10^7 \left(\frac{L}{L_\odot}\right)^{-0.5} h \frac{M_\odot}{L_\odot} & \frac{L}{L_\odot} < 4\;10^{10} \\
        M/L = 128 h \frac{M_\odot}{L_\odot} & 4\;10^{10} < \frac{L}{L_\odot} < 4\;10^{11} \\
        M/L = 3.6\;10^{-4} \left(\frac{L}{L_\odot}\right)^{0.5} h
        \frac{M_\odot}{L_\odot} & \frac{L}{L_\odot} > 4\;10^{11}
      \end{array}
      \label{eq:marinoni}
    \end{equation}
    as shown in upper panel of Fig.~\ref{fig:mass_lights}. In this
    framework, we generated the same catalog as in {\it T-C} case but
    it was analyzed assuming the $M/L$ function given by
    Eq.~\eqref{eq:marinoni}.
  \item {\it TS-T case}: assuming that we have an unbiased estimator of
    the $M/L$ function, there can still be a scatter around this mean
    value that can increase the errors and also introduce systematic
    bias. We test this by multiplying the mass of each halo of
    {\it FullMock} by a random number $x$ such that $\log_{10} x$ is
    uniformly distributed in 
    $[-1,1]$, prior to MAK reconstruction, which is performed on a
    resampling of the halo catalog following the procedure explained in
    Appendix~\ref{app:subsampling}. Note that the mass of
    background particles remains unchanged during the process, which corresponds to 63\%
    of the matter distribution being unaffected by the
    scattering. However, applying the scatter to small mass haloes only
    introduces a local additional noise which should not have any
    significant consequences on the reconstruction accuracy for which
    deeper potential wells are in fact more critical.
    \medskip
\end{enumerate}

We want to highlight the fact that each of these transformations,
actually corresponding  to transforming the mass of an object of {\it
  FullMock} through a $M\rightarrow L\rightarrow M$ operation, does not
correspond to an identity. One actually gets a new set of masses attached
to each tracer which is different from the original one. Moreover,
the output mass distribution $P_\text{mass,out}(M)$ may be fundamentally different from the
input one $P_\text{mass,in}(M)$. Indeed, computing $P_\text{mass,out}(M)$ is equivalent to
performing a weighted average of $P_\text{mass,in}(M)$. This procedure
induces a global reshaping of the distribution. Consequently, the
statistical properties of the corresponding mass density field
may be affected.

More technically, during the procedure used to construct all the
catalogues above, total mass conservation is enforced.
Note that the total mass depends on $\Omega_\text{m} h^2$, but this
normalization does not affect MAK displacements, which are sensitive
to density contrasts only. Parameters $\Omega_\text{m}$ and $h$ in
fact intervene
while performing velocity-velocity
comparison and while converting distances to velocities
(\S~\ref{sec:malmquist}), respectively.

\begin{figure*}
  \begin{center}
    \begin{tabular}{ccc}
      & \includegraphics[width=.22\linewidth]{refVelSim}  &  \\
      (TS-T) & (T-C) & (T-MH) \\
      \includegraphics[width=.22\linewidth]{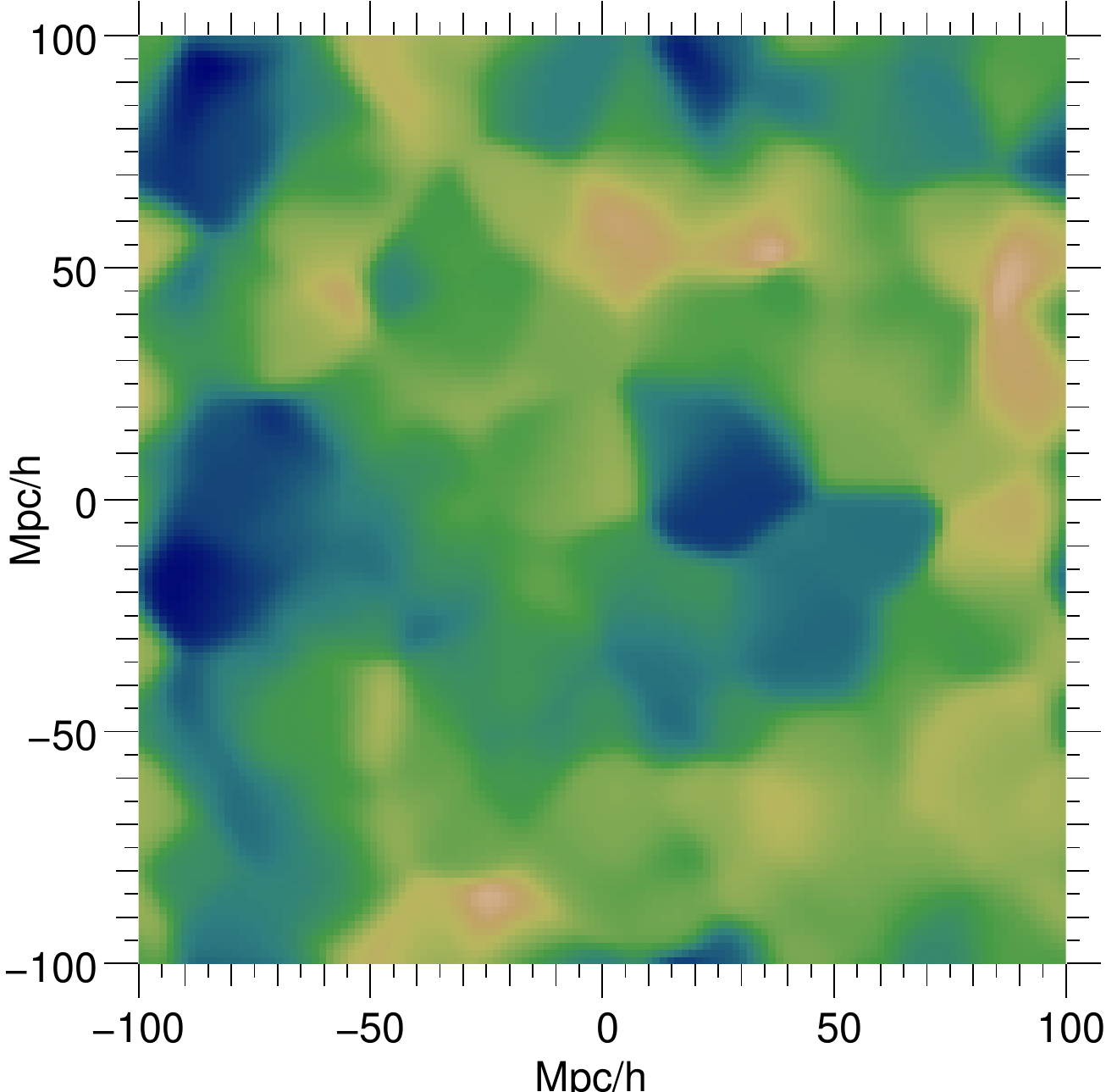} &
      \includegraphics[width=.22\linewidth]{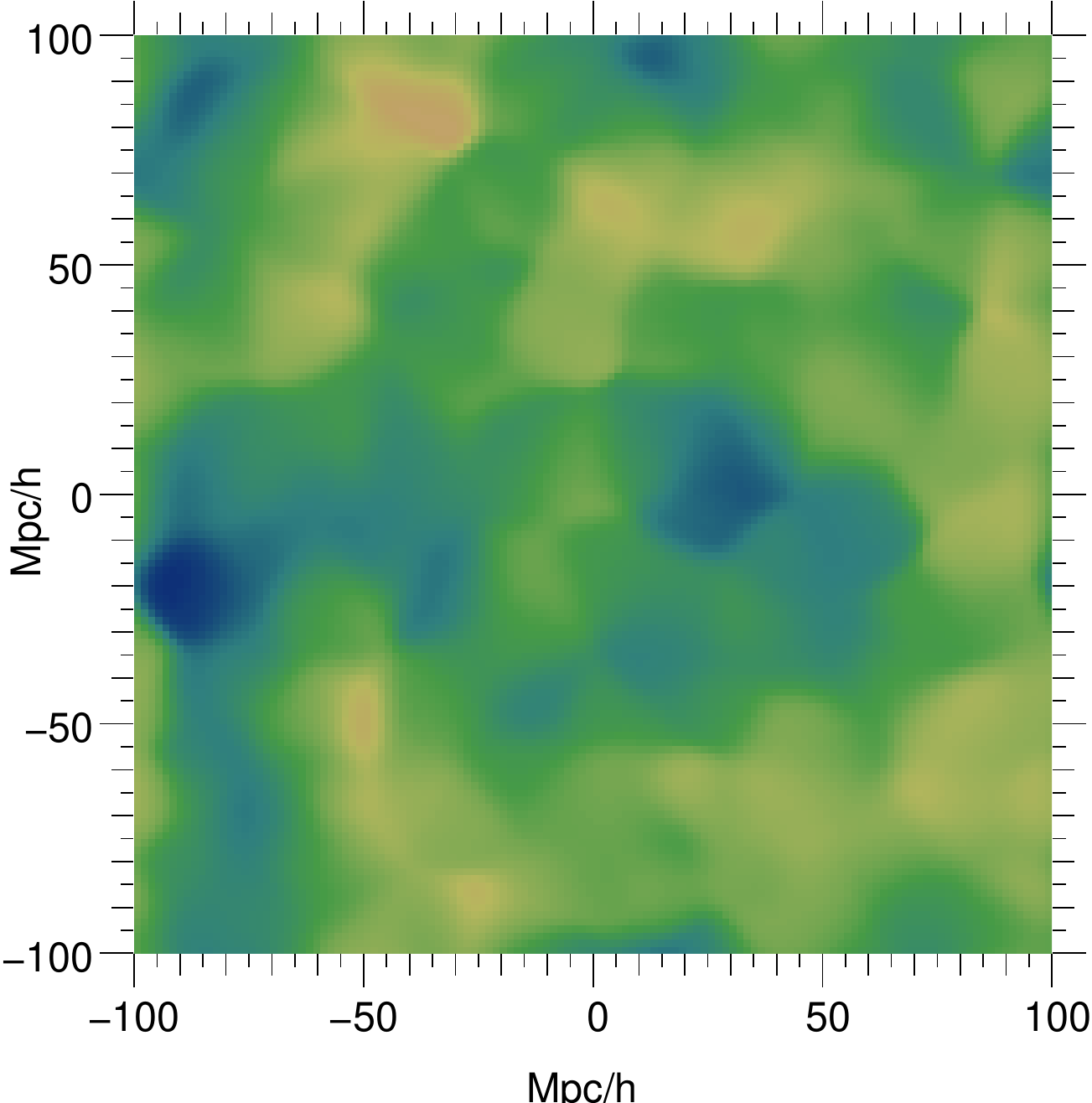} &
      \includegraphics[width=.22\linewidth]{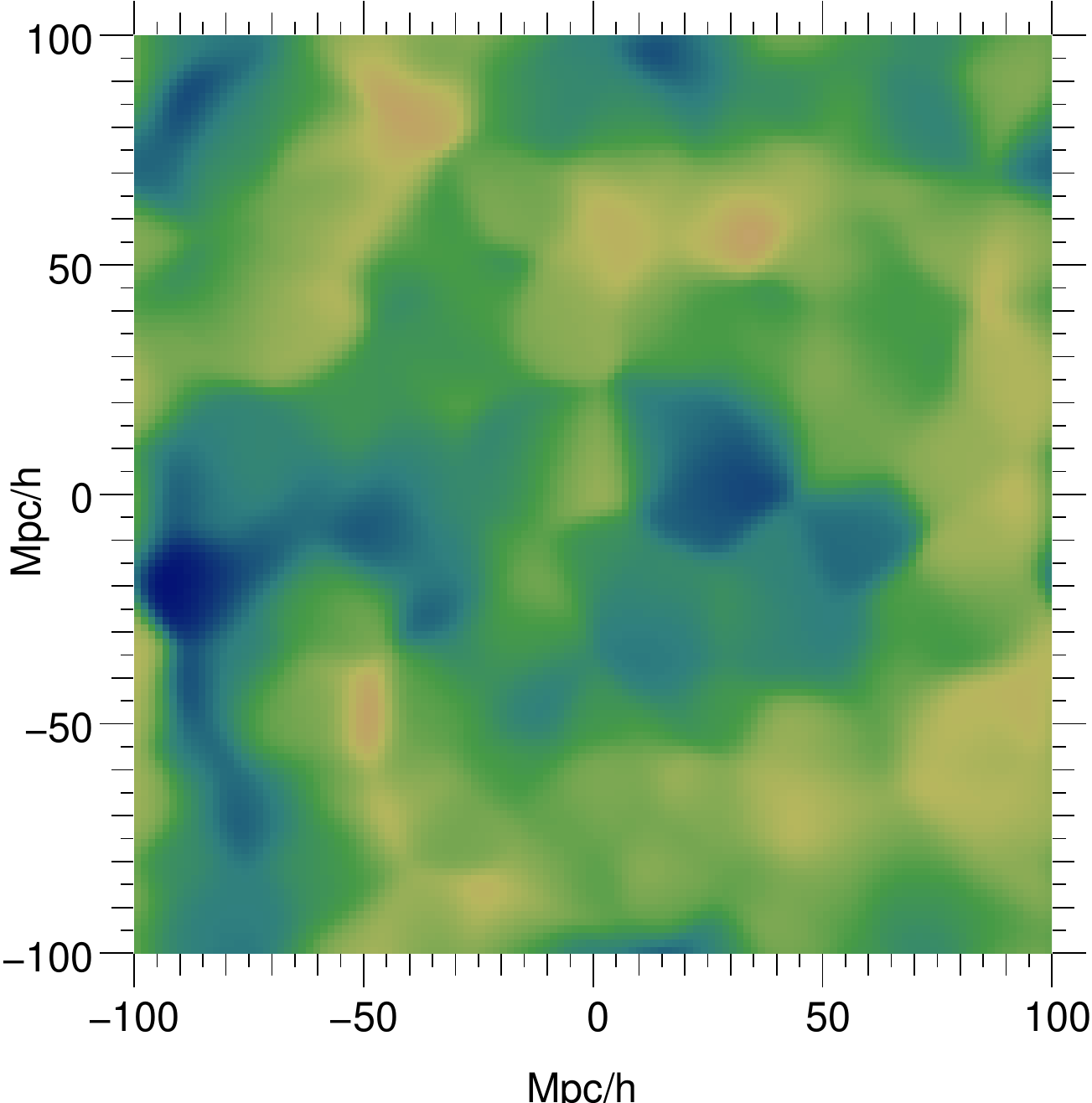} \\
      \includegraphics[width=.22\linewidth]{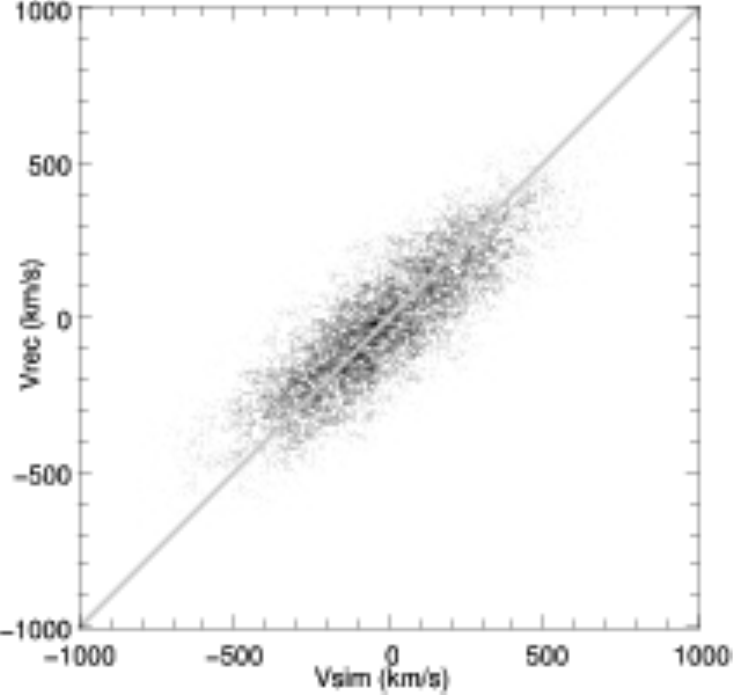} &
      \includegraphics[width=.22\linewidth]{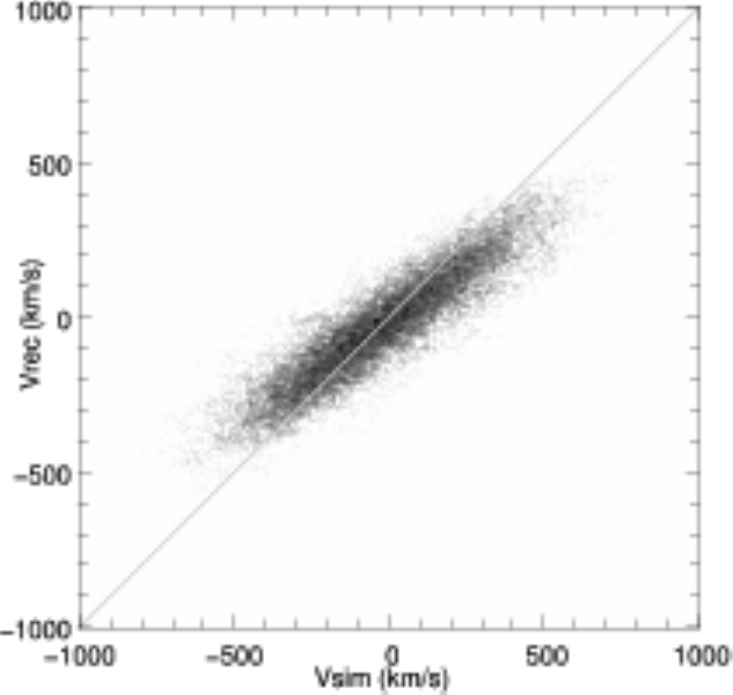} &
      \includegraphics[width=.22\linewidth]{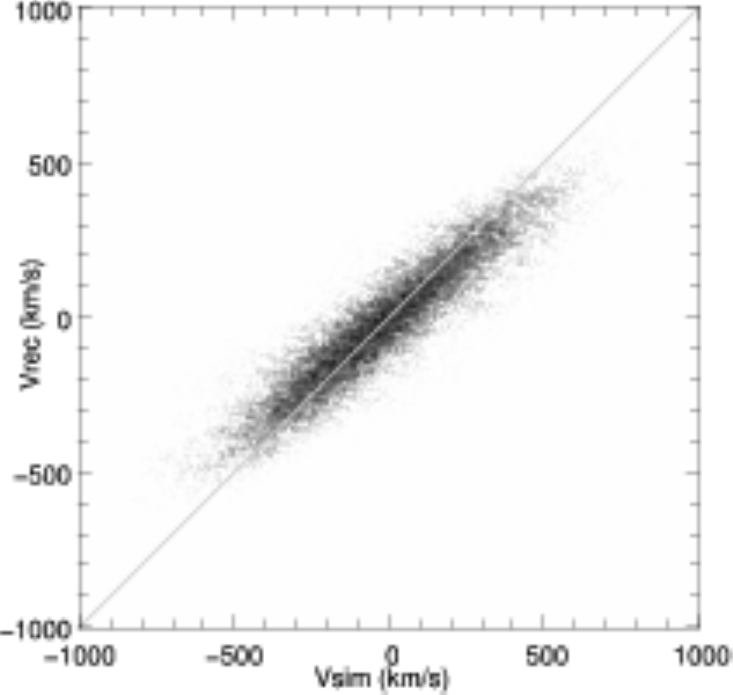}
    \end{tabular}
  \end{center}
  \vspace*{-0.4cm}\caption{\label{fig:ml_diff_assign} {\it $M/L$ bias} -- 
    The top panel gives the expected line-of-sight component $v_r$ of the
    velocity field, smoothed with a 5~$h^{-1}$Mpc Gaussian filter, as
    given by the simulation in a thin slice of the simulation
    containing the observer. The middle panels gives the reconstructed $v_r$
    field, with the same smoothing, after having applied each
    of the transformations specified in
    Fig.~\ref{fig:flow_luminosity_catalog} to {\it FullMock}. 
    The lower panels gives the scatter between the reconstructed and
    simulated peculiar velocities for each of the transformations.    
  }
    
\end{figure*}

\begin{table*}
  \caption{\label{tab:ML_errors} {\it $M/L$ bias effect} -- This table
  gives the results obtained using different statistical tools. We also
  measured $\Omega_\text{m}$ using six different methods: the label $s$ means we
  used the slope estimated by using all objects, the label $\mathfrak{L}$
  is used when $\Omega_\text{m}$ has been determined using the
  likelihood analysis, and the label 1.5$\sigma$ is used when the slope
  is estimated using only the objects within the
  1.5$\sigma$ isocontour of the PDF between reconstructed velocities
  and simulated velocities (method described in \S~\ref{sec:mak_error}). }
  \begin{center}
    \begin{tabular}{|c||c|c|c||c|c|c|c|c|c|}
      \hline
      \multirow{2}{1cm}{Transf.} & \multicolumn{3}{c||}{Velocity} &
      \multirow{2}{1.2cm}{$\Omega_\text{m}$ ($s$)} &
      \multirow{2}{1.2cm}{$\Omega_M$ ($\mathfrak{L}_\text{min}$)} &
      \multirow{2}{1.2cm}{$\Omega_M$ ($\mathfrak{L}_\text{max}$)} 
          & \multirow{2}{1.2cm}{$\Omega_M$ (1.5$\sigma$,$s_\text{med}$)} 
          & \multirow{2}{1.2cm}{$\Omega_M$ (1.5$\sigma$,$s_\text{min}$)} 
          & \multirow{2}{1.2cm}{$\Omega_M$ (1.5$\sigma$,$s_\text{max}$)} \\
      \cline{2-4}
      & $s$ & $r$ & $\sigma$ &  & & \\
      \hline \hline
      None & 0.88 & 0.89 & 0.58 & 0.38  & 0.30 & 0.31 & 0.30 & 0.28 & 0.31 \\
      TS-T & 0.90 & 0.78 & 0.64 & 0.36 & 0.26 & 0.30  & 0.28 & 0.24 & 0.33 \\
      T-MH & 0.80 & 0.80 & 0.60 & 0.45 & 0.33 & 0.38 & 0.36 & 0.32 & 0.40 \\
      T-C & 0.71 & 0.78 & 0.63  & 0.55 & 0.40 & 0.48 & 0.44 & 0.37 & 0.51\\
      \hline
    \end{tabular}
  \end{center}
\end{table*}

As expected, random uncertainty on the mass determination does not
introduce any bias, it only increases the scatter in the measurements
as illustrated by the lower left panel of
Fig.~\ref{fig:ml_diff_assign}. A more important issue is the global
knowledge of the $M/L$ relation. Indeed, it seems that the slope of
this relation influences greatly the results, as illustrated by the middle
and right panels of Fig.~\ref{fig:ml_diff_assign}. Clearly, if the
galaxies follow the Tully formula \eqref{eq:tully_ml}, it is
definitely wrong to assume constant $M/L$ and even the MH fit
introduces a significant bias, although it is well within
the observational errors compared to Eq.~\eqref{eq:tully_ml}.
\correction{
  It must be noted that this bias can be turned into an
  advantage if one does not want to measure $\Omega_{\text{m}}$ but
  the $M/L$ relation. Indeed, WMAP experiment
  \citep{WMAP1map,SpergelCosmo2006} coupled with an analysis of the
  power spectrum of large scale
  density of galaxies \citep{Tegmark2006} gives good constraints on
  the real $\Omega_{\text{m}}$ now. Our method, on the other
  hand, is able to measure the discrepancy between the
  measured $\beta$ and the expected growth factor
  $\beta_{\text{expected}} = \Omega_{\text{m}}^{5/9}$ ({\it i.e.} the
  bias). This measurement may give an idea of
  how wrong is the assumed $M/L$ relation prior to the reconstruction and may push us to
  try different plausible $M/L$ functions. Thus our method is able to
  measure the {\it way} that the matter is distributed in the Universe once
  it is given its average density $\Omega_\text{m}$.
  On the other hand, if the above bias is well understood, this method
  helps at reducing the degeneracy in the determination of
  cosmological parameters. Indeed, our posterior probability on
  $(\Omega_\text{m},h)$ gives a constraint orthogonal \citep[for
    example see the results in][]{MohTu2005} to the one obtained from
  the WMAP experiment and from the galaxy statistics of the SDSS.}

\subsection{Magnitude limitation}
\label{sec:incompleteness}

Magnitude-limited sampling of mass tracers introduces a new type of
problem: flux limitation decreases the mass resolution toward the outer edges of the
catalogue contrary to the homogeneous case studied in \S~\ref{sec:diffuse_mass}. 
Usually, the incompleteness is handled by boosting uniformly the
luminosities of galaxies at a given distance from the
observer \citep{Branchini02}, prior to conversion of luminosities into
masses. This is a fair approach if $M/L=\text{constant}$, modulo the
issues discussed in \S~\ref{sec:diffuse_mass}. However, this method is
in general questionable for non-trivial $M/L$ relations
as in Eq.~\eqref{eq:tully_ml} or if different $M/L$'s are assigned to galaxies
with different types. In these last two cases, the missing mass correction should be applied to
the mass distribution itself instead of the luminosity one, to avoid
systematic errors on mass assignments, hence on reconstructed
velocities. This unfortunately requires a prior assumption on the
value of $\Omega_{\text{m}}$, but only slightly complicates the analyses.

In the observational data, galaxies are separated into two populations: groups\footnote{Groups
are defined here as compact sets of 5 galaxies or more.} of
galaxies \citep{Tully87} and field
galaxies. These two populations should be treated separately, keeping
in mind that the groups are the most critical because their
gravitational influence is much larger than individual field galaxies
and they have better peculiar velocity measurements. 

The full procedure consisting of creating a magnitude-limited mock catalogue and
recovering the mass distribution is detailed in
Appendix~\ref{app:magnitude_limit}. Let us recall that, in our
mock catalogues, groups of galaxies are simulated dark matter haloes
with more than 5 particles while background galaxies are identified
with dark matter particles unassigned to any halo. We list here the
key steps we used to correct for incompleteness: 
\begin{enumerate}
  \item[I.] The total apparent luminosities of groups of galaxies is
    obtained assuming a global or a local Schechter luminosity
    distribution for the considered groups. The intrinsic luminosity is
    computed trivially from the total apparent luminosity and the
    redshift of the group.
  \item[II.] The intrinsic luminosity of the remaining unbound
    galaxies (thus field galaxies) is also determined, straightforwardly.
  \item[III.] Then, masses are estimated by assigning appropriate $M/L$
    to each object of I and II.
  \item[IV.] The local missing mass from undetected background
    galaxies is inferred from the detected mass distribution. This
    requires a prior on $\Omega_{\text{m}}$.
  \item[V.] This missing mass may either be reassigned locally to
    detected field galaxies of II (our choice) or be introduced by the mean of new
    randomly positioned tracers, as discussed in \S~\ref{sec:diffuse_mass}.
\end{enumerate}

\begin{figure}
  \begin{center}
    \includegraphics*[width=\linewidth]{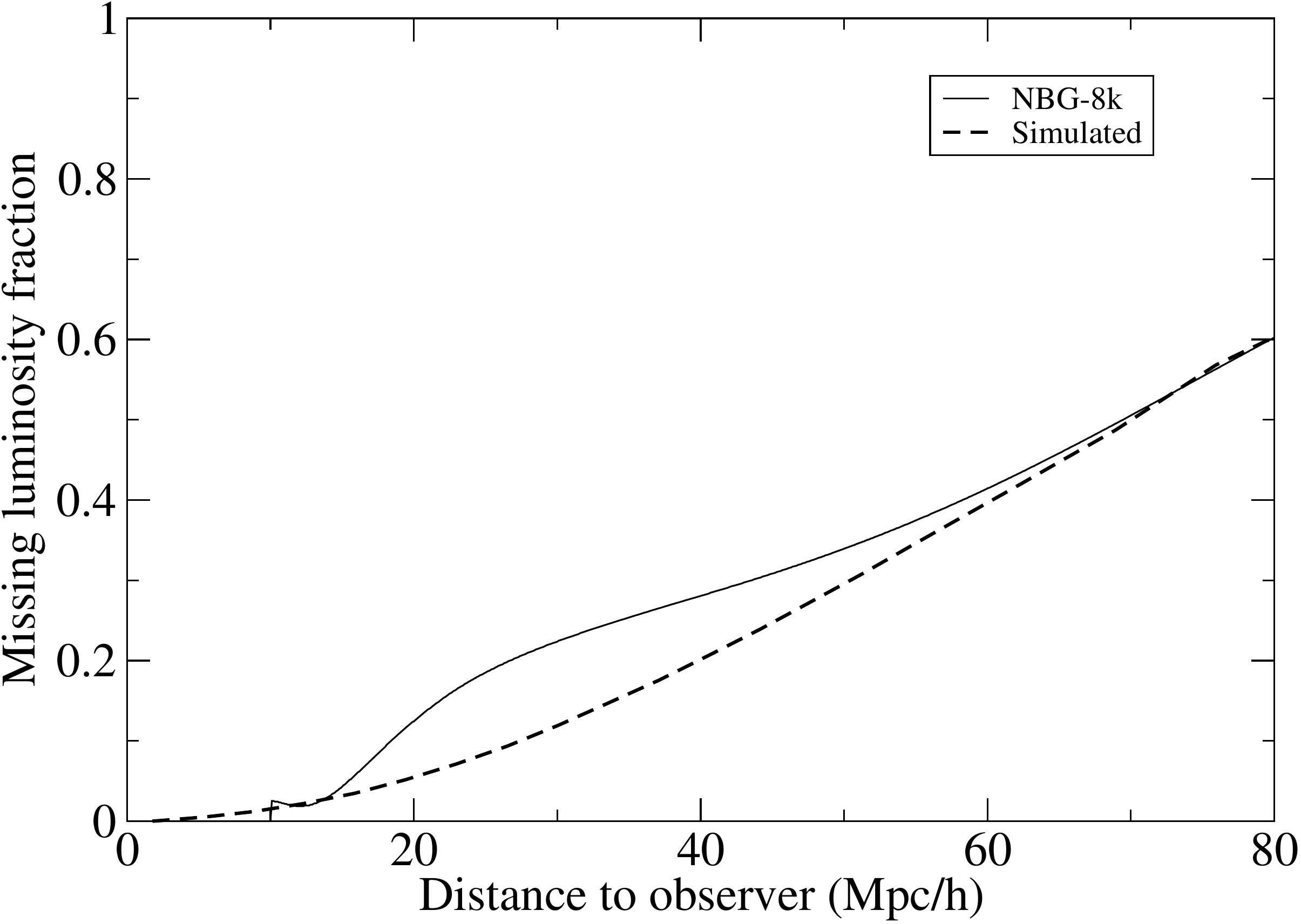}
  \end{center}
  \caption{\label{fig:missing_lum} {\it Magnitude limitation} -- Solid
  line: NBG-8k predicted luminosity incompleteness at the given
  distance from the observer. Dashed line: Simulated luminosity incompleteness in {\it
    8k-mock6}. The incompleteness is expressed in terms of missing
  luminosity fraction at the specified distance.}
\end{figure}

\begin{figure*}
  \begin{center}
    \begin{tabular}{cc}
      {\large Simulated} & {\large Reconstructed} \\
      \includegraphics[width=.4\linewidth]{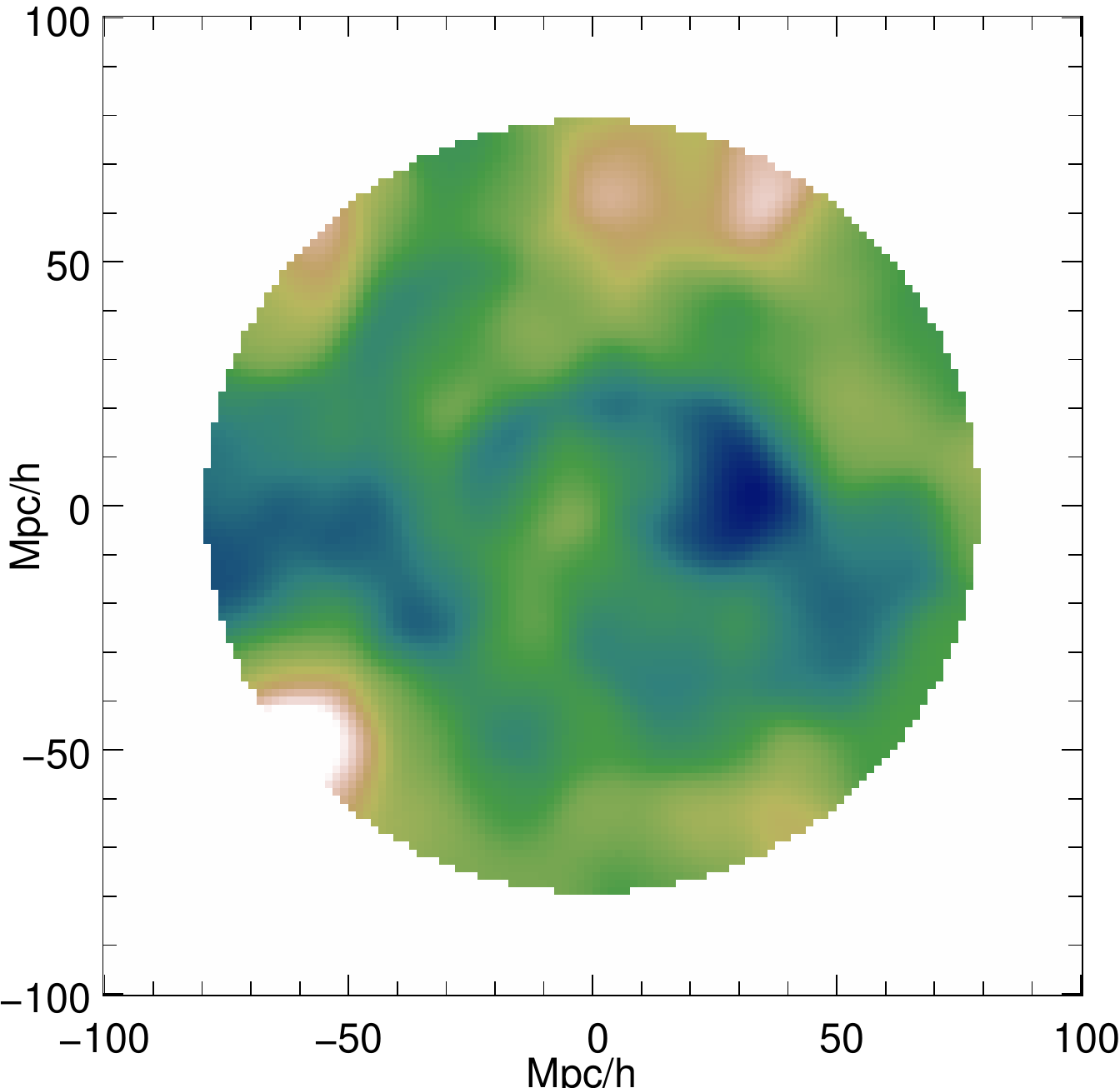} &
      \includegraphics[width=.4\linewidth]{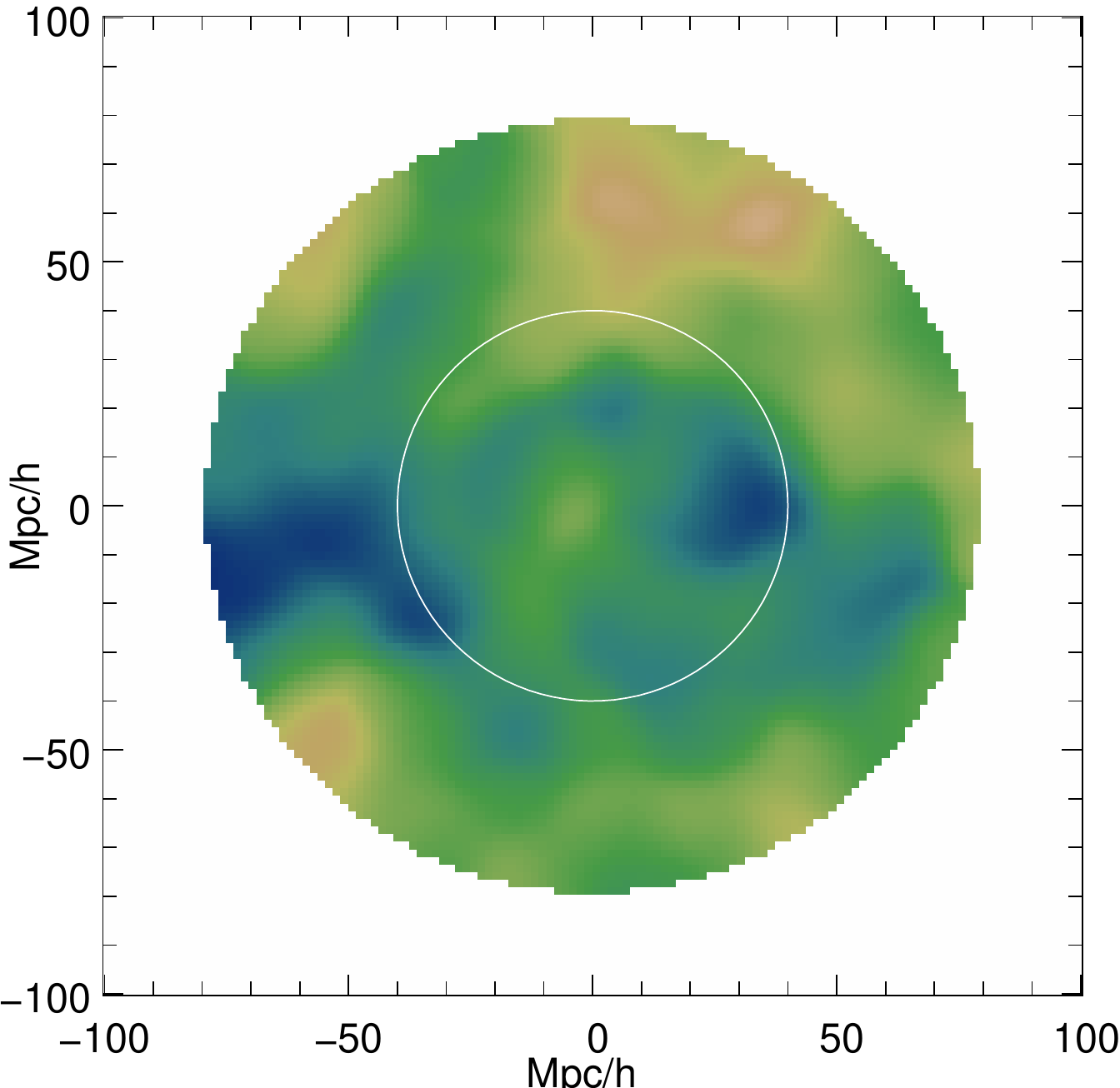} \\
      {\large Comparison in 80~$h^{-1}$Mpc} & {\large
        Comparison in 40~$h^{-1}$Mpc} \\
      \includegraphics[width=.4\linewidth]{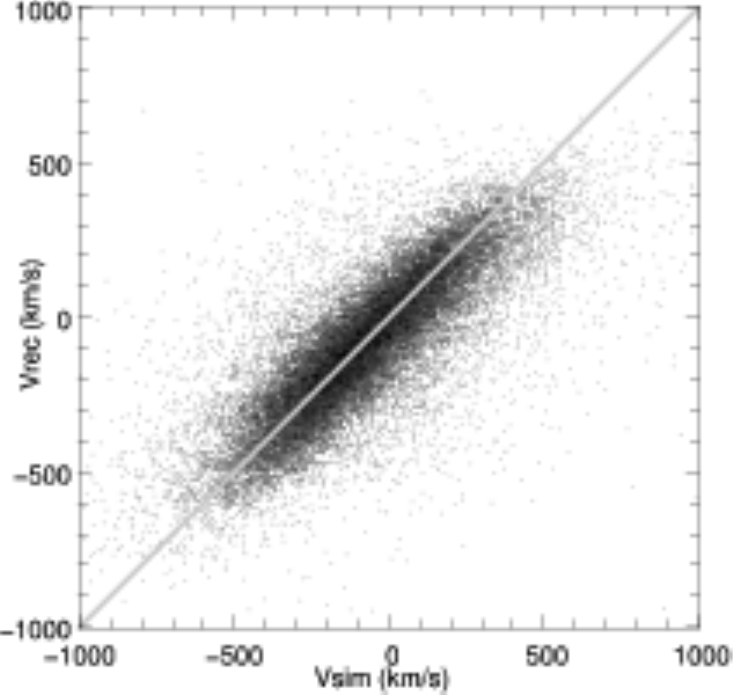} &
      \includegraphics[width=.4\linewidth]{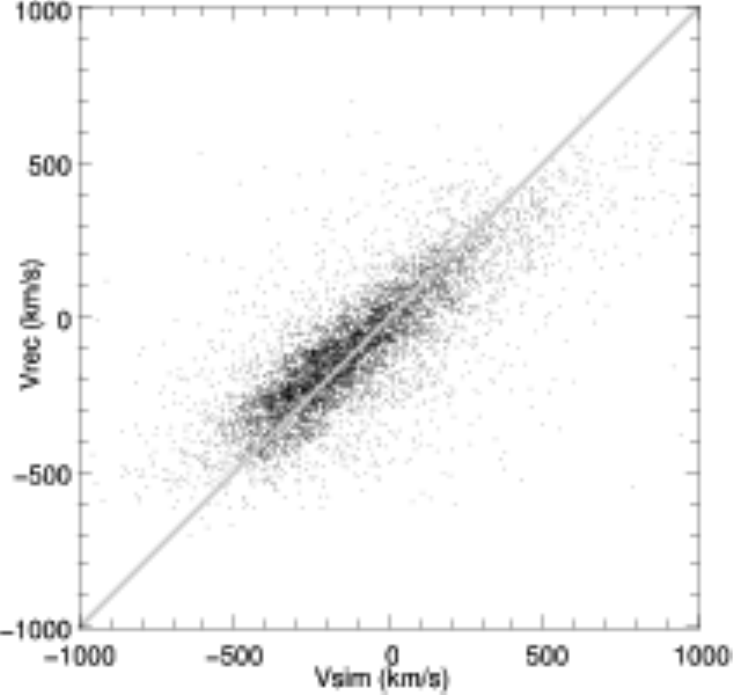}
    \end{tabular}
  \end{center}
  \vspace*{-0.4cm}\caption{\label{fig:incompleteness_scatter} {\it
      Incompleteness: magnitude limitation} -- Top panels: A slice
    of
    the line-of-sight component of the simulated velocity field in 8k-mock6 and
    the reconstructed one, after smoothing with a 5~$h^{-1}$Mpc Gaussian
    window. The displayed slice is chosen to include the observer in
    $(0,0)$.
    The white circle in the right panel gives the size
    of the 40~$h^{-1}$Mpc sphere embedded in the 80~$h^{-1}$Mpc one.
    Bottom panels: The scatter plots compare
    the reconstructed and simulated velocities of objects in the 80~$h^{-1}$Mpc region (left
    panel) and in the 40~$h^{-1}$Mpc volume (right panel).
  }
\end{figure*}

To examine the effects of systematics in the correction for incompleteness,
we use  {\it 8k-mock6} and choose a flux limit such that the resulting
mock catalogue has an incompleteness similar to NBG-8k, as shown in
Fig.~\ref{fig:missing_lum}. 
Results are summarized in Fig.~\ref{fig:incompleteness_scatter} and
in Table~\ref{table:incompleteness}.

The reconstructed radial peculiar velocities $v_\text{r,rec}$ are
behaving extremely well.
On average, the comparison between simulated and reconstructed
velocity fields is surprisingly good in a volume of radius
80~$h^{-1}$Mpc, even though the edge misses locally 98\% of the field
galaxies which represents 60\% of the total mass in our mock
catalogue. It means that, though we keep only 2\% of the field
galaxies, they suffice, in addition to the groups, for a reasonably fair recovery of
 the large-scale peculiar  velocity field.
Note the small bias in the scatter of the lower right panel of
Fig.~\ref{fig:incompleteness_scatter}, resulting in a slightly
larger $\Omega_{\text{m}}=0.38$ than the expected value of 0.30, but in
good agreement with the effective value of 0.35 expected in the
corresponding volume (see \S~\ref{sec:cosmic_var} on cosmic variance effects). This bias might be the consequence of
our treatment of the missing mass coming from undetected tracers as
discussed, in detail in Appendix~\ref{app:magnitude_limit} (point B).

\begin{table*}
  \caption{\label{table:incompleteness} {\it Incompleteness: magnitude
      limitation} -- Column description is given in the caption of Table~\ref{tab:ML_errors}.}
  \begin{center}
    \begin{tabular}{|c|c|c|c|c|c|c|c|c|c|}
      \hline
      \multirow{2}{2cm}{Volume} 
      & \multicolumn{3}{|c|}{Velocity field} &
      \multirow{2}{1.2cm}{$\Omega_\text{m}$ ($s$)} &
      \multirow{2}{1.2cm}{$\Omega_\text{m}$ ($\mathfrak{L}_\text{min}$)}  &
      \multirow{2}{1.2cm}{$\Omega_\text{m}$ ($\mathfrak{L}_\text{max}$)}  
            & \multirow{2}{1.2cm}{$\Omega_\text{m}$ (1.5$\sigma$,$s_\text{med}$)} 
            & \multirow{2}{1.2cm}{$\Omega_\text{m}$ (1.5$\sigma$,$s_\text{min}$)} 
            & \multirow{2}{1.2cm}{$\Omega_\text{m}$ (1.5$\sigma$,$s_\text{max}$)} \\
      \cline{2-4}               
      &     $s$ &   $r$      &   $\sigma$   &  & \\
      \hline
      \hline
      \multirow{1}{2cm}{8k}    
      &  0.86   &  0.77      &   0.64     &  0.39  & 0.26 & 0.31 & 0.29 &
      0.25 & 0.34 \\
      \hline
      \multirow{1}{2cm}{4k}    
      &  0.77   &  0.75       & 0.66       & 0.48  & 0.37 & 0.45 & 0.38 &
      0.30 & 0.47 \\
      \hline
    \end{tabular}
  \end{center}
\end{table*}

\section{Redshift distortion}
\label{sec:redshift_distortion}

The input of MAK reconstruction is the position of objects in real
space as needed by Eq.~\eqref{eq:assign}. However redshift catalogues give us galaxy positions in redshift space,
namely $s_r = H d + v_r$, where $s_r$ is the redshift distance, $d$
is the luminosity distance between the observer and the object and $v_r$
is the line-of-sight peculiar velocity.  
To account for redshift distortions, we must correct for two major
effects:
\begin{itemize}
  \item[-] ``Fingers-of-god'' correspond to an elongation of dense structures along
    the line of sight, such as clusters of galaxies, due to random motions of galaxies
    within these structures. 
  \item[-] Kaiser effect \citep{KaiserEffect87} is a large-scale
    effect coming from the coherent part of the cosmic flows, which,
    for instance, increase the overall density contrast.
\end{itemize}

\begin{table*}
  \caption{\label{tab:redshift_rec} {\it Redshift reconstruction} --
    Column description is given in the caption of Table~\ref{tab:ML_errors}.}
  \begin{center}
    \begin{tabular}{ccccccccc}
      \hline
      $s$ & $r$ & $\sigma$ & 
           \begin{minipage}[c]{1cm} \begin{center} $\Omega_\text{m}$
            { ($s$)} \end{center} \end{minipage}  & 
           \begin{minipage}[c]{1cm} \begin{center} $\Omega_\text{m}$
            { ($\mathfrak{L}_\text{min}$)} \end{center} \end{minipage}  & 
           \begin{minipage}[c]{1cm} \begin{center} $\Omega_\text{m}$
            { ($\mathfrak{L}_\text{max}$)} \end{center} \end{minipage}  & 
          \begin{minipage}[c]{1.2cm} \begin{center} $\Omega_\text{m}$
            {  (1.5$\sigma$,$s_\text{med}$)}\end{center} \end{minipage}  & 
          \begin{minipage}[c]{1.2cm} \begin{center} $\Omega_\text{m}$
            { (1.5$\sigma$,$s_\text{min}$)} \end{center} \end{minipage}  & 
        \begin{minipage}[c]{1.2cm} \begin{center} $\Omega_\text{m}$
            { (1.5$\sigma$,$s_\text{max}$)} \end{center}\end{minipage}  \\
      \hline
      \hline
       0.83 & 0.46 & 0.95 & 0.50 & 0.22 & 0.29 & 0.27 & 0.22 & 0.33 \\
      \hline
    \end{tabular}
  \end{center}
\end{table*}

Fingers-of-god effects can be easily removed by simply
collapsing groups or clusters to a single point, as usually performed
in the literature. However, such a
procedure is generally carried out in a rather ad-hoc way and is
certainly not free of biases. 

The Kaiser effect can be accounted for by modifying the cost
function \eqref{eq:assign} using the Zel'dovich approximation
to infer line-of-sight peculiar velocities as functions of the sought
displacement field \citep{MohTu2005,ValentineTaylor2000}.
 If ${\bf s}({\bf q})$ is the redshift coordinate of a
particle originally at ${\bf q}$ then the total cost \eqref{eq:assign}
of the association $\sigma$ becomes:
\begin{equation}
  I_\sigma = \sum_{i=1}^N \left( \left( {\bf s}_i -
      {\bf q}_{\sigma(i)} \right)^2 - \frac{\beta (2 + \beta)}{\left(1 +
      \beta\right)^2} \frac{\left(\left({\bf s}_i -
          {\bf q}_{\sigma(i)}\right)\cdot{\bf s}\right)^2}{||{\bf
        s}||^2}\right), \label{eq:zassign}
\end{equation}
where $\beta$ is the linear growth factor. Once the redshift
displacement ${\bf \Psi}^\text{s} = {\bf s} - {\bf q}$ has been computed, the reconstructed
radial peculiar velocity of the object $i$ can be obtained by
\begin{equation}
  v^\text{s}_\text{r,rec} = \frac{\beta}{1+\beta} \frac{{\bf
      s}\cdot{\bf \Psi}^\text{s}}{||{\bf s}||}.
\end{equation}
The cost function $I_\sigma$ leads to the exact result in the case of
a Zel'dovich displacement field without shell crossing after redshift
distortion. However, in general, the second term (accounting for redshift distortion)
of Eq.~\eqref{eq:zassign} becomes of the same order as the first term (the
real space cost term) near the origin. In this case, the
reconstruction becomes ill-defined because of the loss of
convexity of functional $I_\sigma$. We expect thus the central part of all
catalogues to be, in general, poorly reconstructed. The size of such a
region is roughly determined by the magnitude $v_\text{obs}$ of the large-scale flow nearby the
observer with respect to the Cosmic Microwave Background. The velocity $v_\text{obs}$ determines the relative
contribution of the first term with respect to the second term of
Eq.~\eqref{eq:zassign}. In practice $v_\text{obs}$ is of the order of
a 
few hundred \kms \citep[for instance the Local Group velocity is
630~\kms,][]{Erdogdu2006} which gives us a region of ``exclusion''
of radius of about a few $h^{-1}$Mpc.\footnote{See e.g. \cite{VelGrav07}
  for a similar discussion.}

Again, MAK reconstruction
fails in regions where shell crossings occur. Projection in redshift
space generates such shell crossings along the line-of-sight. These
shell crossings are dramatic because of their anisotropic
nature. In particular, filaments can cross each other while passing
from real to redshift space, implying the reconstruction will fail
in a large region of the catalogue encompassing the
gravitational influence of these filaments. In this area, most of the
reconstructed radial velocities will have the opposite sign compared to the
true velocity. Of course, shell crossings in redshift
space can have more complex consequences but this simple example
suggests that MAK reconstruction should not work as well in redshift
space as in real space.\footnote{This is also true for the Least-Action
  method for which multiple solutions quickly arises.}

Another problem of this method is that one must assume $\beta$ prior
to the reconstruction. As for \S~\ref{sec:incompleteness}, where we
had to guess the undetected mass, we 
choose a value $\Omega_{\text{m,in}}$, thus an assumed $\beta_\text{in}$, then we make a redshift
reconstruction and measure a $\Omega_{\text{m,out}}$. In practice, the ``true''
$\Omega_{\text{m}}$ of the catalogue was chosen to be the one for which $\Omega_{\text{m,in}} =
\Omega_{\text{m,out}}$, which corresponds to having self-consistent orbits
modeling when doing MAK reconstruction and when one makes a comparison
with measured velocities. 

\begin{figure*}
  \begin{center}
    \begin{tabular}{cc}
      Simulation & Redshift reconstruction \\
      \includegraphics[width=.3\linewidth]{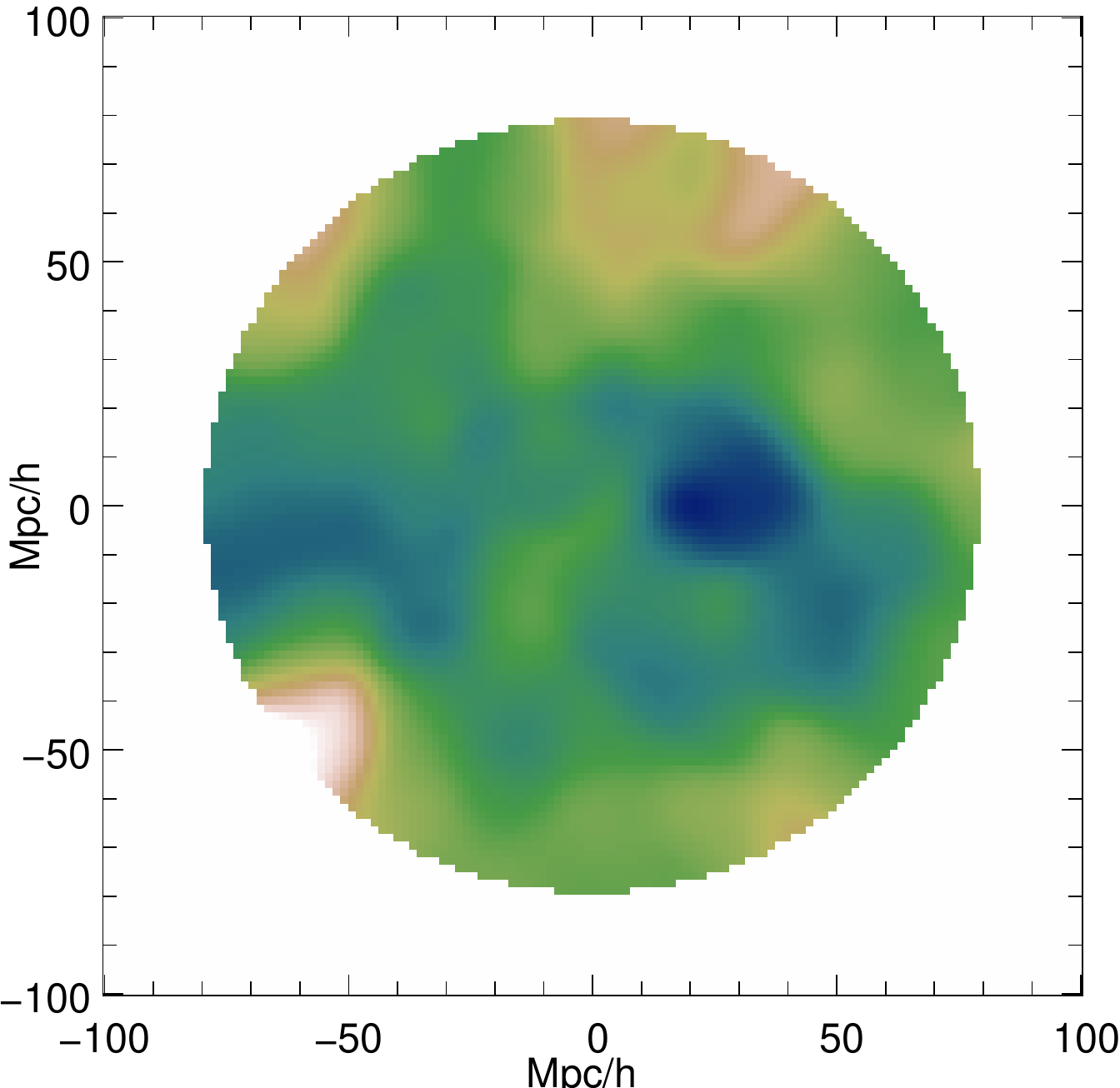} &
      \includegraphics[width=.3\linewidth]{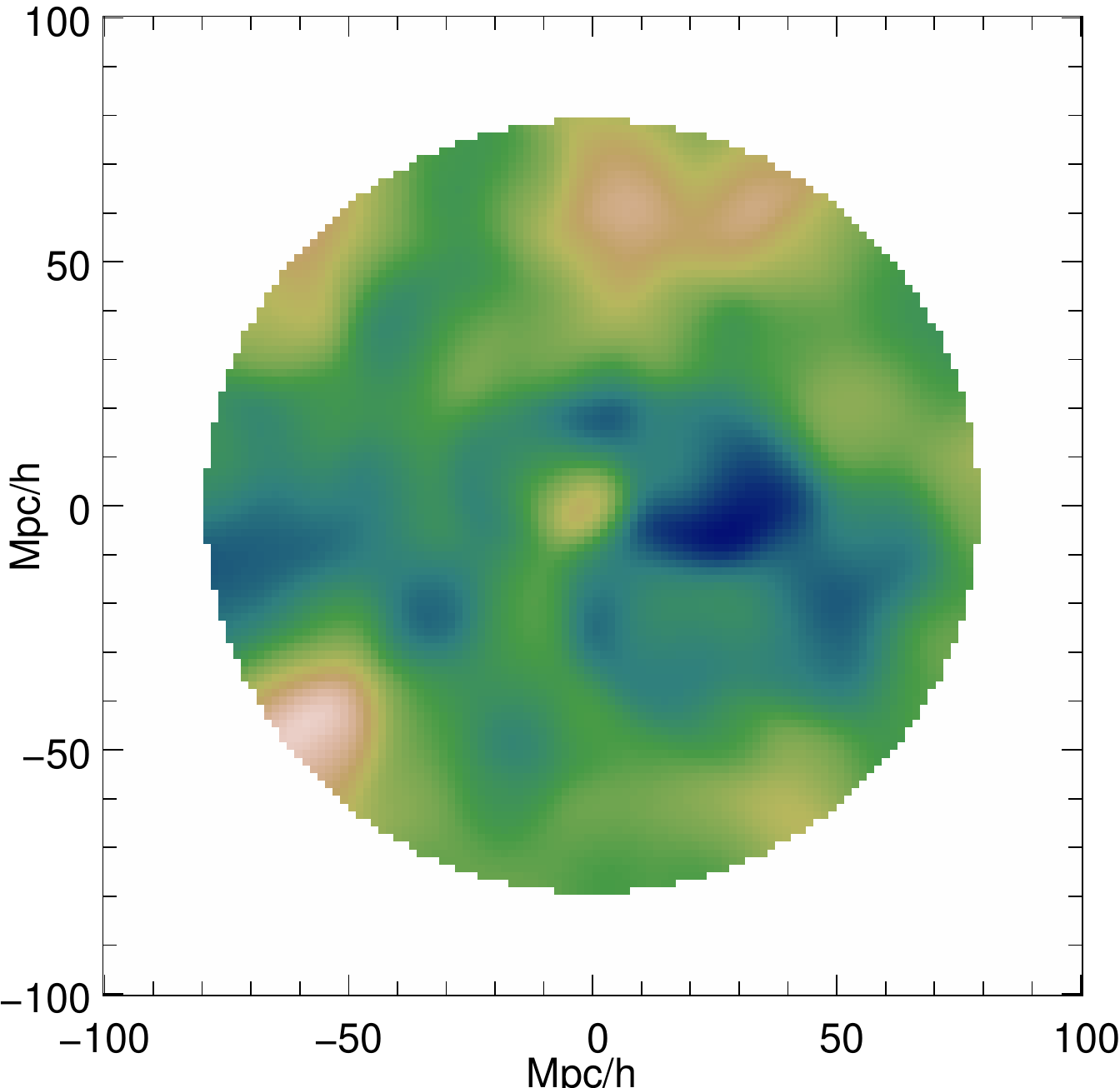} \\[.7cm]
      Real space reconstruction & Redshift reconstruction \\
      \includegraphics[width=.3\linewidth]{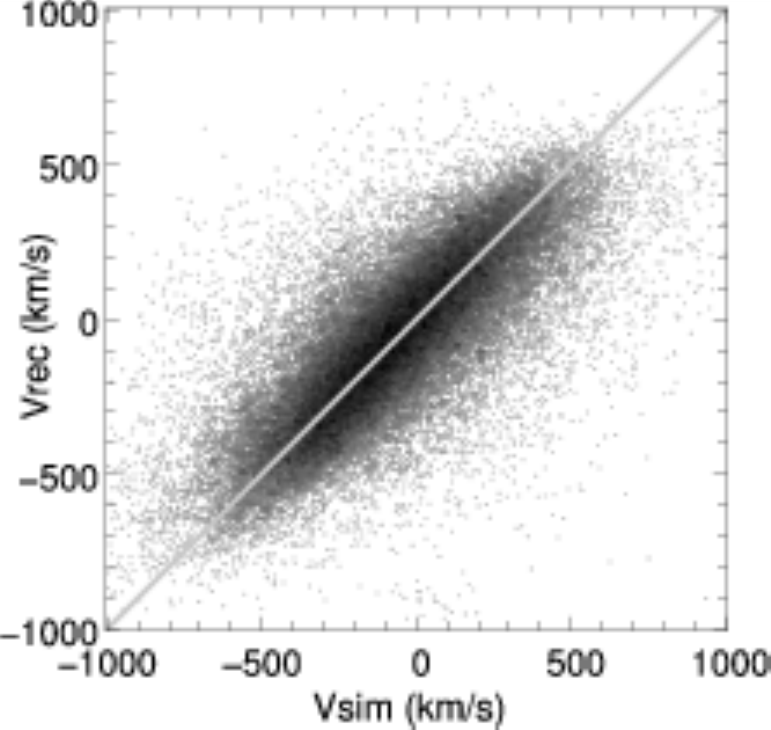} &
      \includegraphics[width=.3\linewidth]{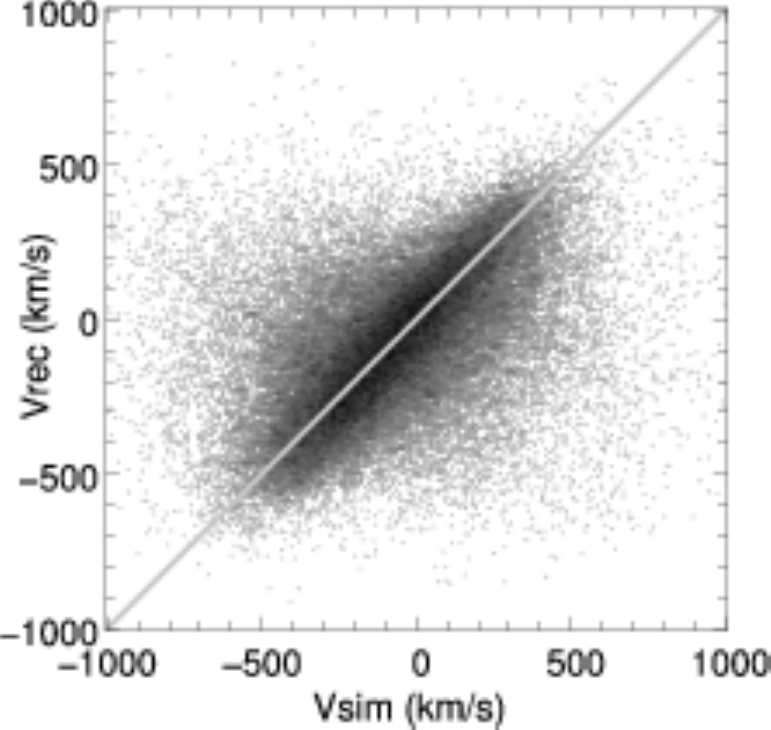} \\
    \end{tabular}
  \end{center}
  \caption{\label{fig:redshift_rec} {\it Redshift distortion
      correction} -- Top panels: A slice of the smoothed velocity
    field $v_\text{r,sim}$ and $v_\text{r,rec}$ is shown in the left
    and right panels, respectively. The two fields have been smoothed
    with a 5~$h^{-1}$Mpc Gaussian window, with objects put at their
    real (simulated and reconstructed) comoving coordinates. Bottom panels: Scatter plots  between
    $v_\text{r}$ and $v_\text{r,rec}$ for individual mass tracers. The
    left panel (right panel) was produced using a real space
    reconstruction (redshift space reconstruction, respectively). In
    both cases, only objects within a sphere of 8000~\kms are shown. }
\end{figure*}

Fig.~\ref{fig:redshift_rec} shows both reconstructed and simulated
velocity fields and the scatter between $v^\text{s}_\text{r,rec}$ and
$v_\text{r,sim}$. The first impression when comparing the two top
panels of Fig.~\ref{fig:redshift_rec} is that the redshift
reconstruction behaves really well. However, some potentially worrying
localized features are present: 
\begin{itemize}
  \item[-] Some important structures have their velocities badly
    reconstructed. Two important examples are the green-yellowish finger just
    above the center of the upper right panel of
    Fig.~\ref{fig:redshift_rec} and the big velocity peak at the
    top of this same panel. In the left panel, these two structures
    are not so prominent. The difference can be understood by studying the
    impact of the Kaiser effect on the reconstructed velocity
    field. Basically, two nearby filaments can merge in redshift space
    and give birth to a filament with a higher apparent density. The
    reconstruction is not able to separate these two filaments, which
    leads to an area with higher reconstructed velocities than the
    true ones. Thus, we expect in observational data to meet problems in the neighbourhood of the Great Wall, which is
    a supercluster of filaments compressed by redshift distortion.
  \item[-] The velocity field in the immediate (5-10~$h^{-1}$Mpc) neighbourhood of the mock
    observer has lost its spatial structure and even presents a
    spurious peak. This is, most
    unfortunately, an expected problem that is linked to the above
    discussion on the problems of $I_\sigma$ near the
    observer. Indeed, in the neighbourhood of the observer, $I_\sigma$
    becomes singular and the reconstruction misses, most likely,
    the right orbits. Analysing  the smoothed velocity
    field seems to show that this effect looks in practice much like the one
    just above: the reconstructed velocity field
    may be boosted by the merging of different structures in the
    neighbourhood of the observer. 
  \item[-] The lower right panel presents two additional off-diagonal
    tails compared to lower left panel. As discussed earlier, these
    tails are due to shell-crossings occuring along the lines-of-sight
    when passing from real to redshift space. These extra
    shell-crossings result in  some reconstructed velocities acquiring
    a sign opposite to the true velocities.
\end{itemize}

\begin{figure}
  \begin{center}
    \includegraphics*[width=.9\linewidth]{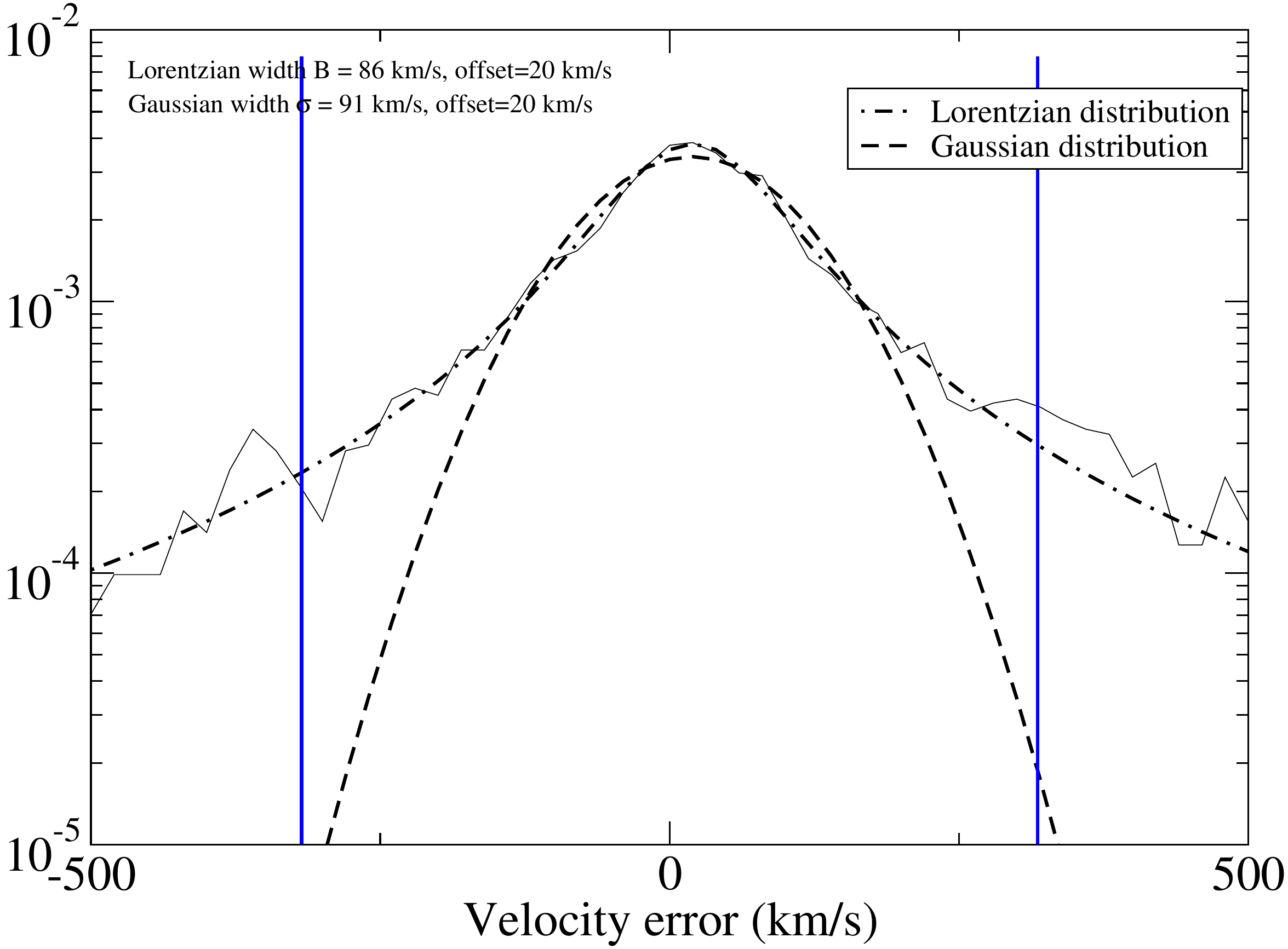}
  \end{center}
  \caption{\label{fig:red_reconstructed_velocity_stat} {\it Error
      distribution of the reconstructed velocity field, redshift
      space} -- 
    Same as in
    Fig.~\ref{fig:reconstructed_velocity_stat} but the solid curve
    corresponds to the
    probability distribution of the quantity $v^\text{s}_\text{r,rec} -
    v_\text{r,sim}$, where $v^\text{s}_\text{r,rec}$ and $v_\text{r,sim}$ are the line-of-sight
    reconstructed and simulated velocities, respectively.}
\end{figure}

Similarly as in \S~\ref{sec:mak_error}, we have computed in
Fig.~\ref{fig:red_reconstructed_velocity_stat} the distribution of
differences $P^\text{s}_\text{VE}$ between $v^\text{s}_\text{r,rec}$
and $v^\text{s}_\text{r,sim}$, for a redshift reconstruction applied
on 8k-mock6 based on a $64^3$ mesh.\footnote{The handling of
  the finiteness of the catalogue volume is handled in \S~\ref{sec:lag_volume} } Though the
distribution is of course wider than in
Fig.~\ref{fig:reconstructed_velocity_stat}, the previously drawn
conclusions are still valid. $P^\text{s}_\text{VE}$ is better fitted by
a Lorentzian distribution with $B = 86$~\kms than by a Gaussian of
width $\sigma=91$~\kms, particularly in the tails.

To check the effects of redshift distortion on the quality of the
reconstruction, one can compare Table~\ref{tab:redshift_rec} to the first
row of Table~\ref{tab:ML_errors}. As usual, the $s$ parameter is slightly biased
below unity due to nonlinear effects discussed in
\S~\ref{sec:mak_error}, which seem, not surprisingly, to be slightly
enhanced by  redshift distortions. The appeareance of the off-diagonal
tails in the lower right panel of Fig.~\ref{fig:redshift_rec} increases
the level of scattering, hence the correlation coefficient $r$
decreases and the signal-to-noise ratio increases.
Reducing the analysis to the region inside 1.5$\sigma$ isocontour
greatly improves the results, as expected, but still leads to a value
of $\Omega_\text{m}$ slightly biased to lower values,
$\Omega_\text{m}=0.27$. 

\section{Effects of catalogue geometry}
\label{sec:edge_effects}

In practice, real galaxy catalogues are not spatially periodic as is our
simulation. They represent a region of finite volume with non-trivial
geometry. In particular, two kinds of problems arise:
\begin{itemize}
\item[-] {\it Edge effects} -- 
  Reconstruction of the galaxy trajectories without any piece of
  information on what may affect them dynamically from the outer
  parts of the catalogue is likely to introduce significant
  sources of errors, possibly systematic. We separate here edge
  effects into two subclasses: the effects of the obscuration by our
  galaxy, which defines a Zone of Avoidance (hereafter ZOA) and the
  effects of finite depth of the catalogue. 
  These two effects need a separate treatment detailed in
  \S~\ref{sec:zoa} and \S~\ref{sec:lag_volume}.
\item[-] {\it Cosmic variance} -- The finite volume of the accessible
  part of the Universe might be a potentially unfair realization
  of the random process underlying  the properties of the large
  scale matter distribution. We must investigate whether our
  method, including handling of edge effects, is robust to the
  recovering of the
  statistical properties of the whole Universe from observations of only a fraction of it.
\end{itemize}

\subsection{Zone of avoidance}
\label{sec:zoa}

Dust present in the Milky Way's galactic plane highly attenuates the
light, thus galaxy catalogues generally do not provide any data in
this direction (approximately the region within $|b| < 5\;\deg$, where
$b$ is the galactic latitude) of the ZOA.
This strong attenuation introduces a boundary effect, which has the
unpleasant feature of being present at any distance from the observer
and may thus severely affect  the measurements.
As this area is nonetheless relatively small, particularly at low
redshift, a simple correction should be able to greatly remove the
boundary effect in the inner region of the catalogue.

Simulating the effect is made easy by putting an observer at the
center of the simulation volume and by removing all mass
tracers in the neighbourhood of the galactic plane $z=0$, {\it i.e.}
which have $|b| < \alpha$. This gives us {\it FullMockZOA}.\footnote{$\alpha=5\;\deg$ in our case.}

\begin{figure*}
  \begin{center}
    \begin{tabular}{cc}
      \includegraphics[width=.3\linewidth]{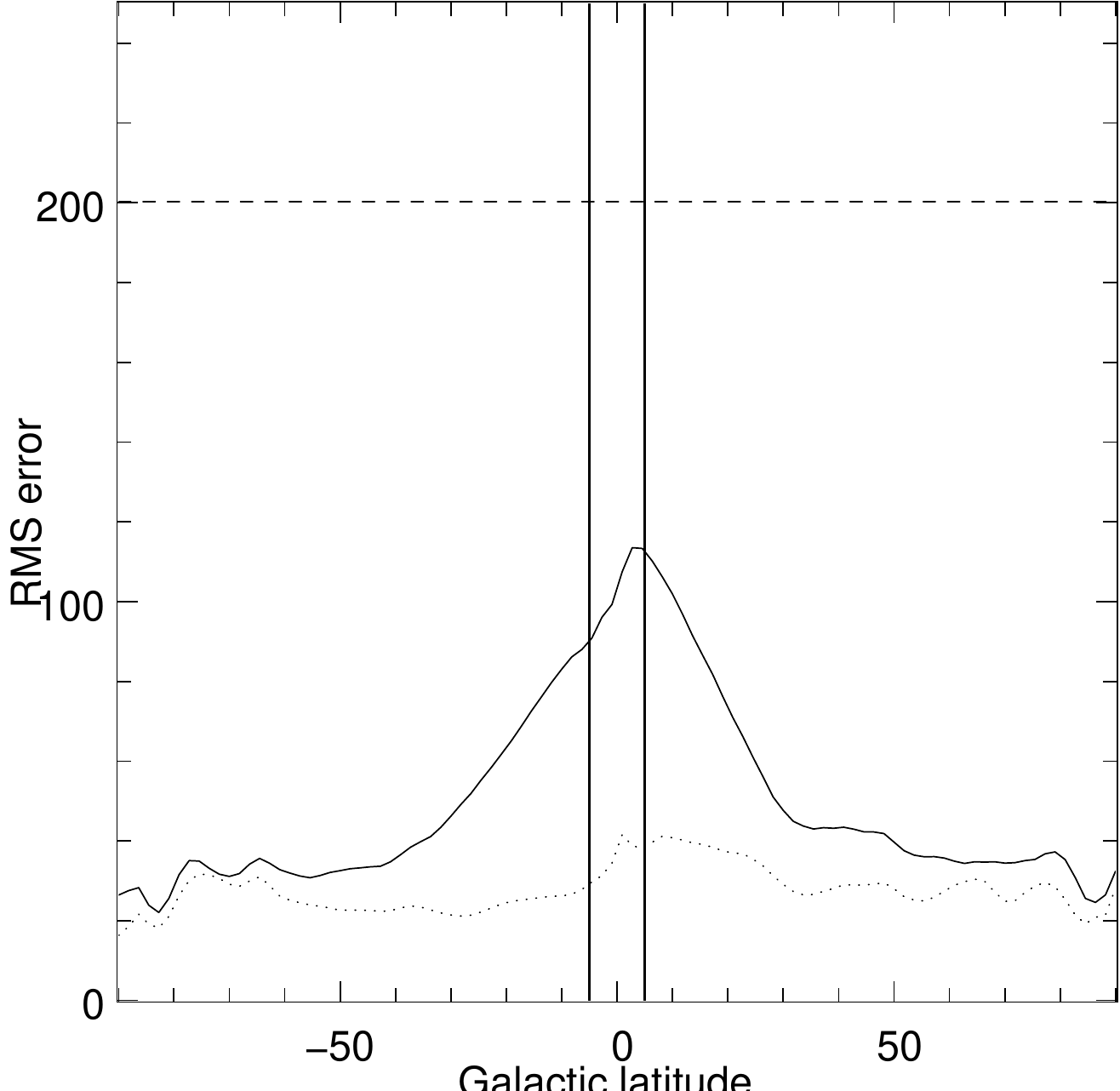} &
      \includegraphics[width=.31\linewidth]{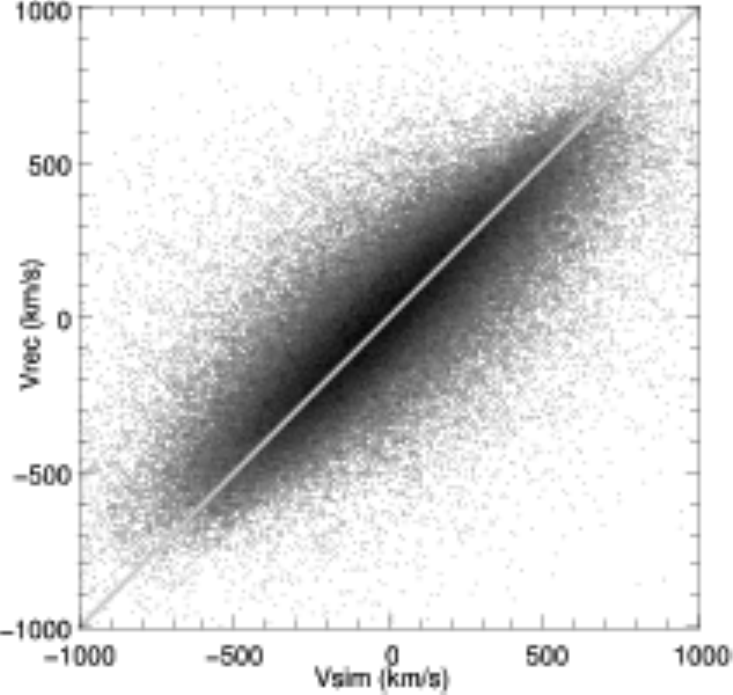}
    \end{tabular}
  \end{center}
  \caption{\label{fig:zoa_errors} {\it Zone of Avoidance / velocity
      field} -- Left panel: Binned average RMS (Root Mean Square) error on the
    smoothed velocity field. As usual, the velocity field window has
    been smoothed with a 5~Mpc/h Gaussian window. Each point is
    computed by averaging the square of the deviation of the velocity field along the
    line-of-sight and for all line-of-sights having belonging to the
    same $\sin(b)$ bin, where $b$ is the ``galactic'' latitude. The
    solid line gives the RMS error in the presence of a 
    zone-of-avoidance at $b=0$. The dotted line gives the RMS error
    for a reconstruction on a catalogue without ZOA. The dashed line
    gives the RMS of the smoothed velocity field itself. Right panel:
    Scatter plots between $v_\text{r,rec}$ and $v_\text{r,sim}$ for
    individual mass tracers.} 
\end{figure*}

Though more advanced ways of filling the ZOA exists \citep[e.g.,
][]{Lahav94,Fontanot03}, this latter is here sufficiently small to be dealt with by
the following simple algorithm. Since the statistical properties of the
galaxies should not change across the boundaries of the ZOA, the
objects in its neighbourhood can be used to fill the zone. We build new mass
tracers to fill the obscured area by applying a locally planar symmetry 
transformation to the galaxies and groups with $-3\alpha < b <
-\alpha$ according to 
the ``plane'' $-\alpha$. We execute the same operation on objects with
$+\alpha < b < +3 \alpha$ but according to the ``plane'' $+\alpha$. In the
end, the masses of the copied haloes in the ZOA are divided by two and we only
take half of the field galaxies. This method has been used previously
to fill the zone of avoidance in NBG-3k \citep{Shay95} and NBG-8k.
This folding procedure has been applied to {\it
  FullMockZOA}, slightly moving some of the newly created objects to enforce
the periodicity of the simulation box to avoid mixing the effect of the
ZOA with other boundary effects. The results are presented in
Fig.~\ref{fig:zoa_errors}. As expected, the ZOA has a clear impact on
errors of the reconstructed velocities.

The typical errors on the reconstructed velocities, represented in the
left panel of this figure,  rise substantially in the vicinity of
the obscured area.
 Fortunately, they remain well below the natural
velocity dispersion of the simulation (dashed line). As we are comparing velocity fields filtered
with a 5~$h^{-1}$Mpc Gaussian window, we expect the reconstructed
velocity {\it field} to be nearly error free for all points nearer than about
60~$h^{-1}$Mpc.\footnote{This corresponds to taking a $5\deg$ wide ZOA
  and computing at what distance the window is smaller than the ZOA.}
It is also fortunate we have not introduced an extra
bias using the filling algorithm, as shown both by
comparing Table~\ref{fig:table_zoa} to the first row of
Table~\ref{tab:ML_errors} and looking at the scatter plot in the right panel of the
Fig.~\ref{fig:zoa_errors}. We nonetheless highlight that the edge effect is
not at all localized  near the ZOA but extends quite far away and
becomes negligible only for $|b| >
20\;\deg$. Table~\ref{fig:table_zoa} shows that the above extra noise
does not have any impact on the measured $\Omega_\text{m}$. 

\begin{table*}
   \caption{\label{fig:table_zoa} {\it Zone of avoidance} -- Noise and
    biasing summary. Column description is given in the caption of
    Table~\ref{tab:ML_errors}. }
 \begin{center}
    \begin{tabular}{ccccccccc}
      \hline
      $s$ & $r$ & $\sigma$ &
      \begin{minipage}[c]{1.4cm} \begin{center}
          $\Omega_\text{m}$ ($s$) \end{center} \end{minipage} &
      \begin{minipage}[c]{1.4cm} \begin{center}
          $\Omega_\text{m}$ ($\mathfrak{L}_\text{min}$) \end{center} \end{minipage} &
      \begin{minipage}[c]{1.4cm} \begin{center}
          $\Omega_\text{m}$ ($\mathfrak{L}_\text{max}$) \end{center} \end{minipage} &
      \begin{minipage}[c]{1.4cm} \begin{center} $\Omega_\text{m}$
          (1.5$\sigma$,$s_\text{med}$) \end{center} \end{minipage} &
      \begin{minipage}[c]{1.4cm} \begin{center} $\Omega_\text{m}$
          (1.5$\sigma$,$s_\text{min}$) \end{center} \end{minipage} &
      \begin{minipage}[c]{1.4cm} \begin{center} $\Omega_\text{m}$
          (1.5$\sigma$,$s_\text{max}$) \end{center} \end{minipage} \\
      \hline \hline
      0.89 & 0.79 & 0.61 & 0.37 & 0.30 & 0.35 & 0.32 & 0.285 & 0.36 \\
      \hline
    \end{tabular}
  \end{center}
\end{table*}

\subsection{Lagrangian domain}
\label{sec:lag_volume}

\begin{figure*}
  \begin{center}
    \begin{tabular}{ccc}
      \hline
      \multicolumn{3}{c}{Current density field} \\
      { (a) TrueDom} & {(b) NaiveDom} & {(c) PaddedDom}      \\[-.1cm]
      \hline
      \includegraphics[width=.25\linewidth]{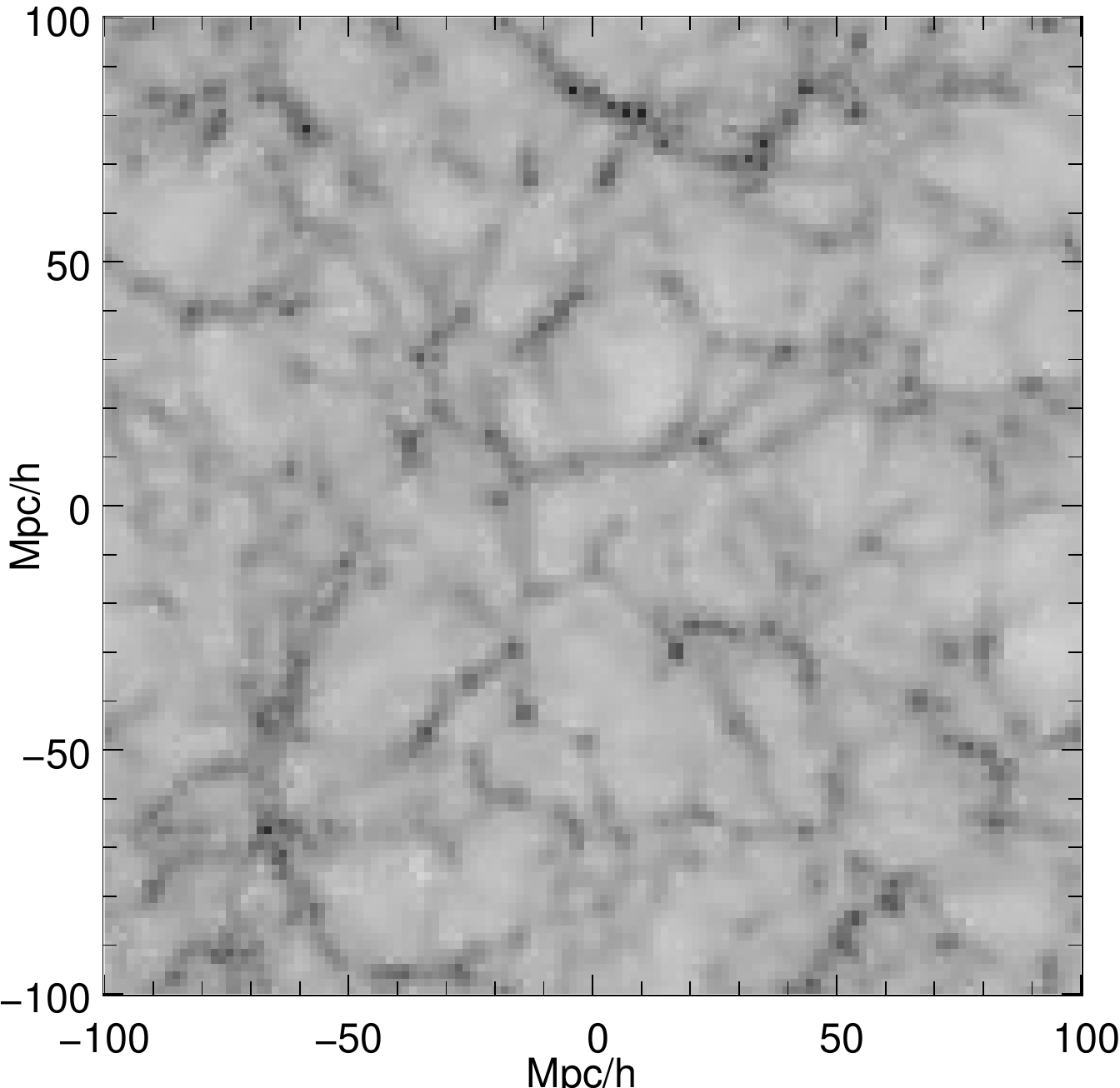} &
      \includegraphics[width=.25\linewidth]{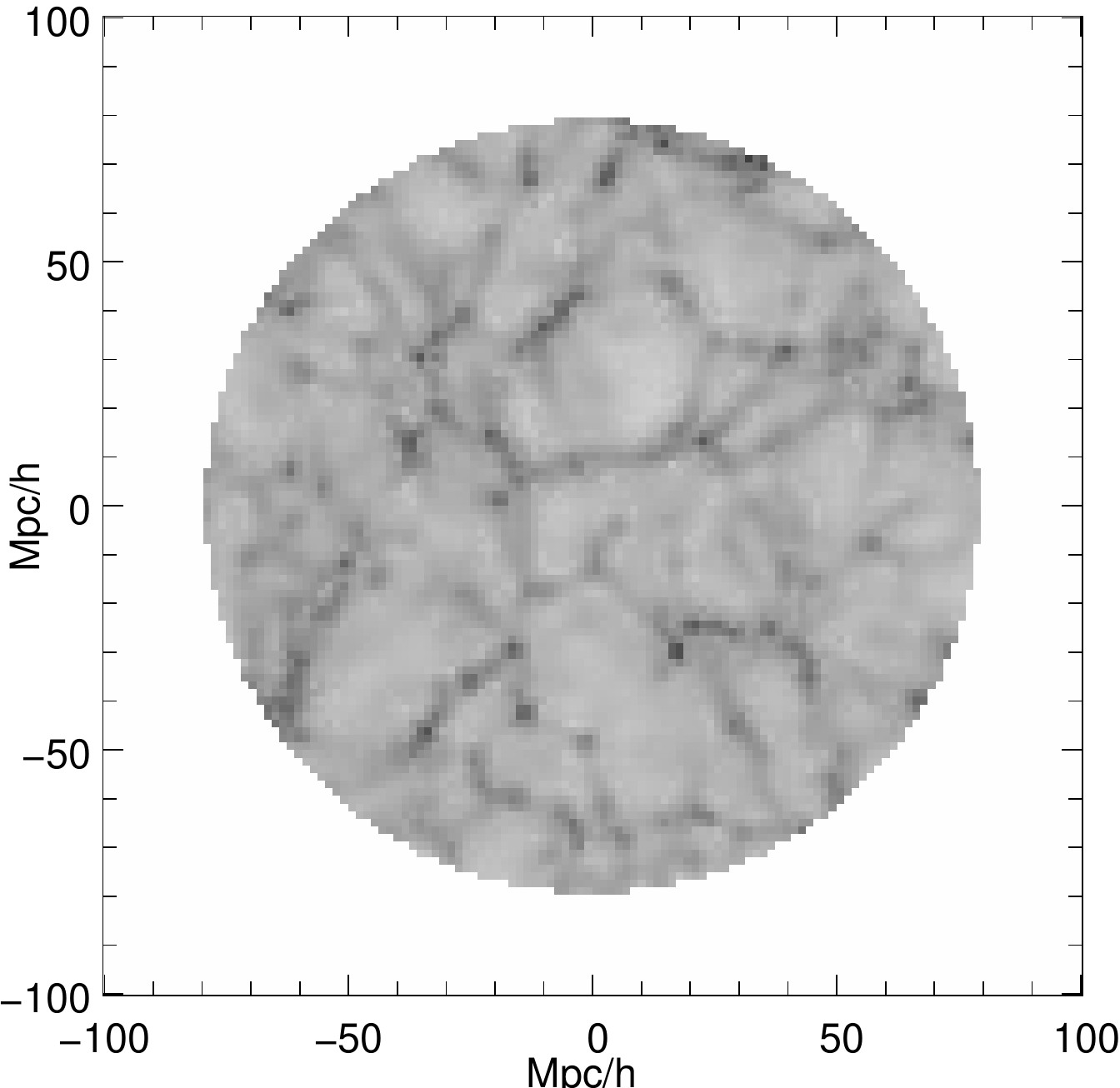} &
      \includegraphics[width=.25\linewidth]{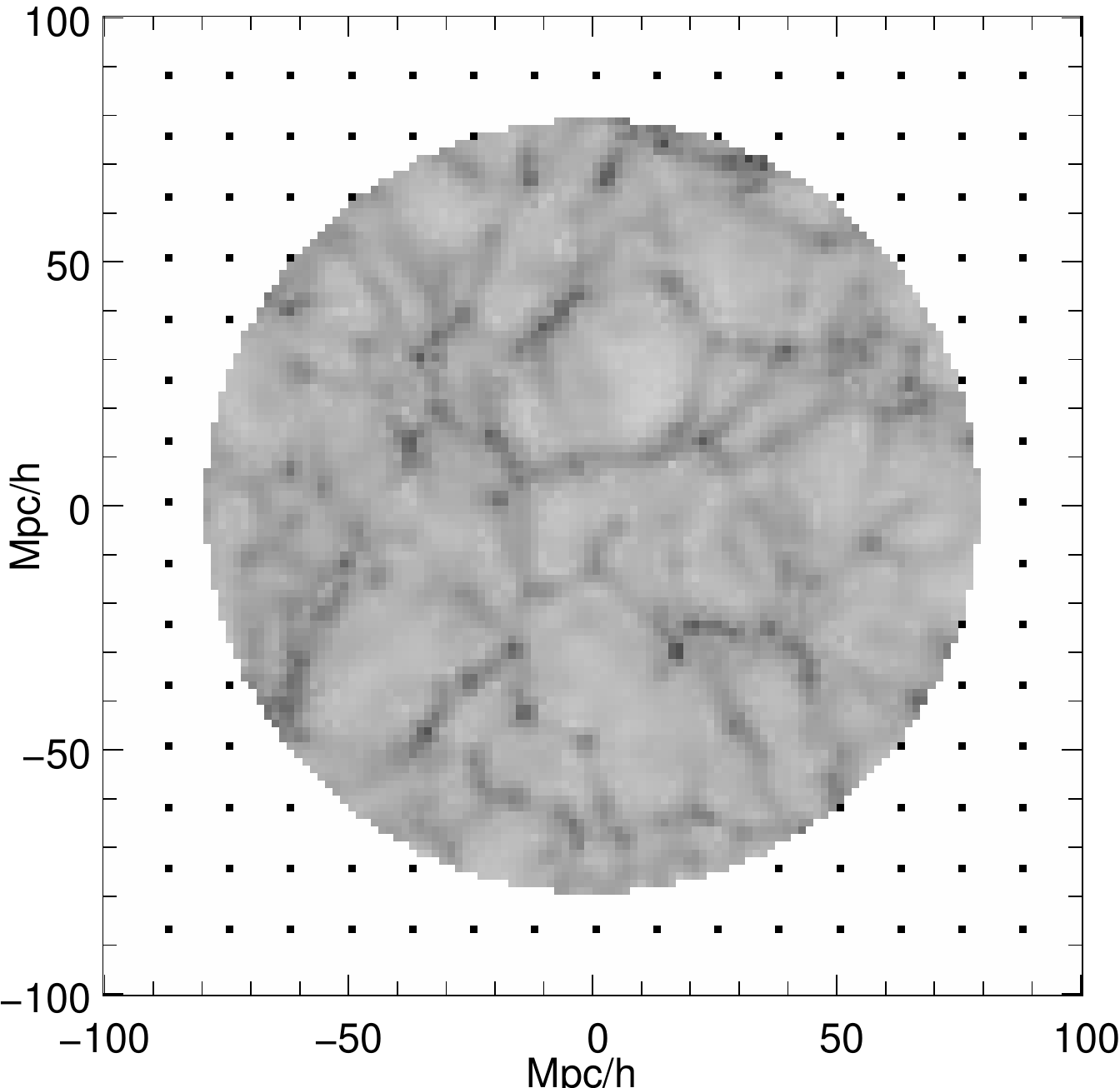} \\
      \hline
      \multicolumn{3}{c}{Reconstructed velocity fields} \\[-.1cm]
      \hline
      \includegraphics[width=.25\linewidth]{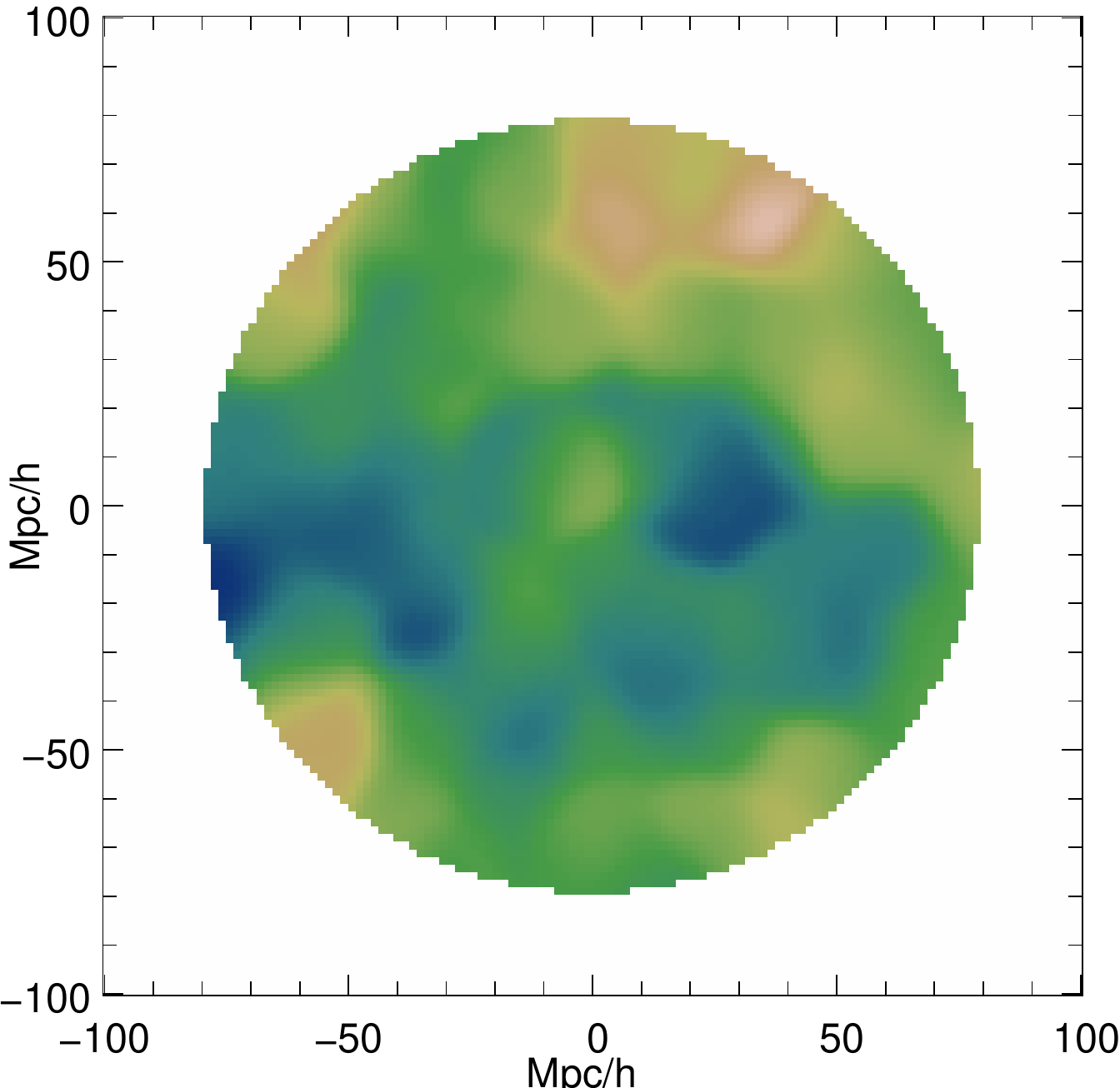} &
      \includegraphics[width=.25\linewidth]{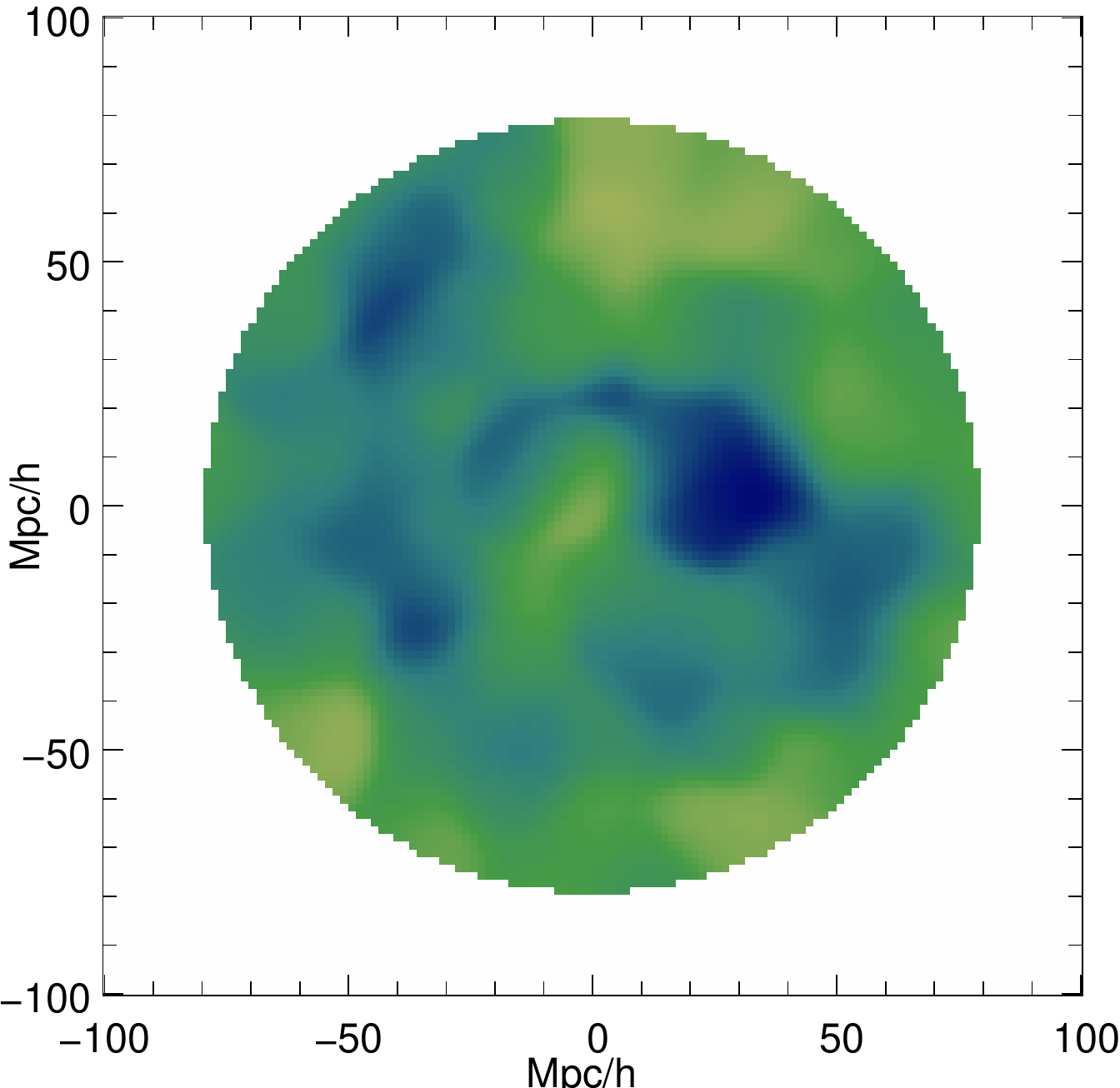} &
      \includegraphics[width=.25\linewidth]{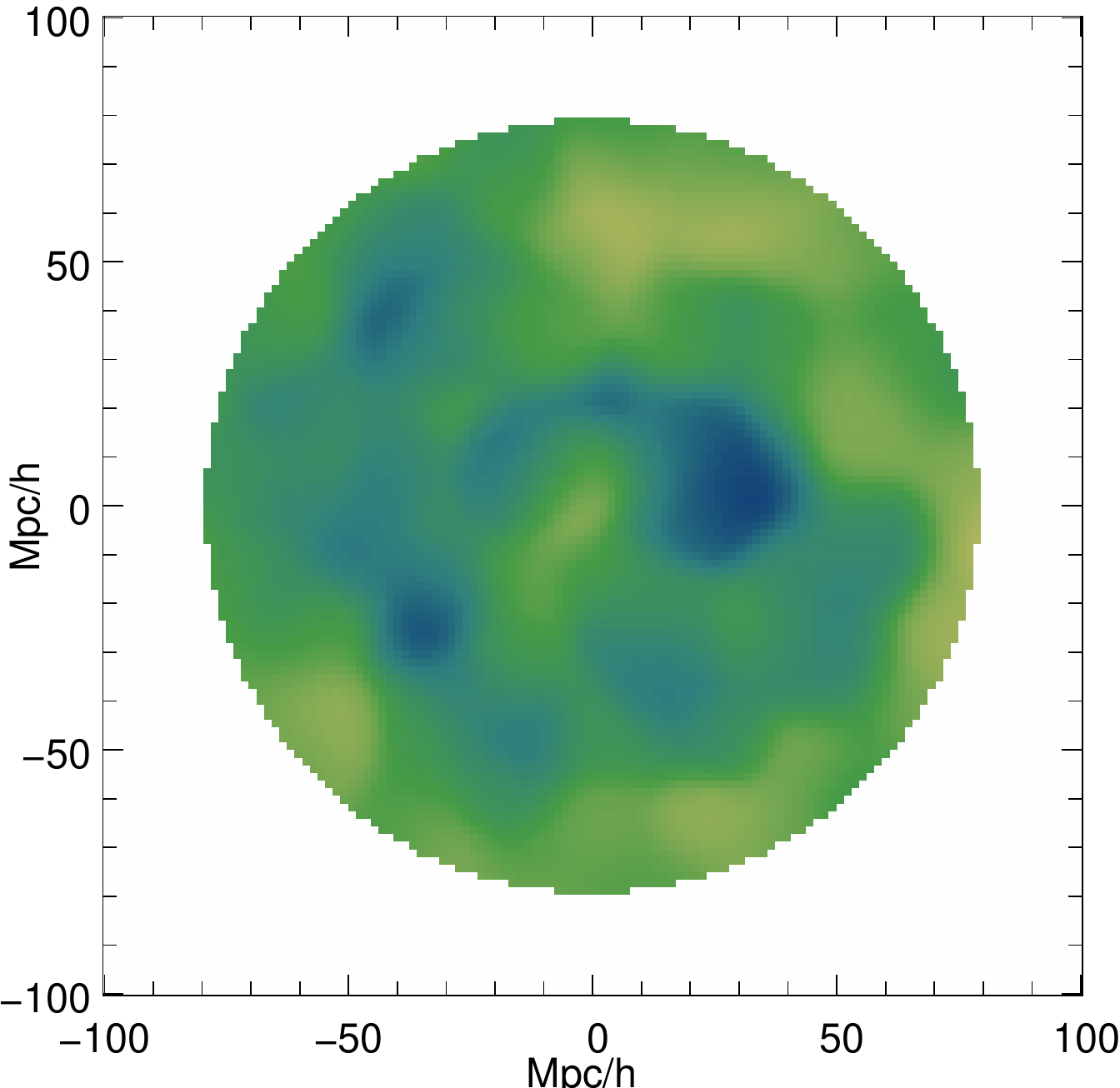}
      \\
      \hline
      \multicolumn{3}{c}{In 8000 \kms} \\[-.1cm]
      \hline
      \includegraphics[width=.25\linewidth]{fig12c15g} &
      \includegraphics[width=.25\linewidth]{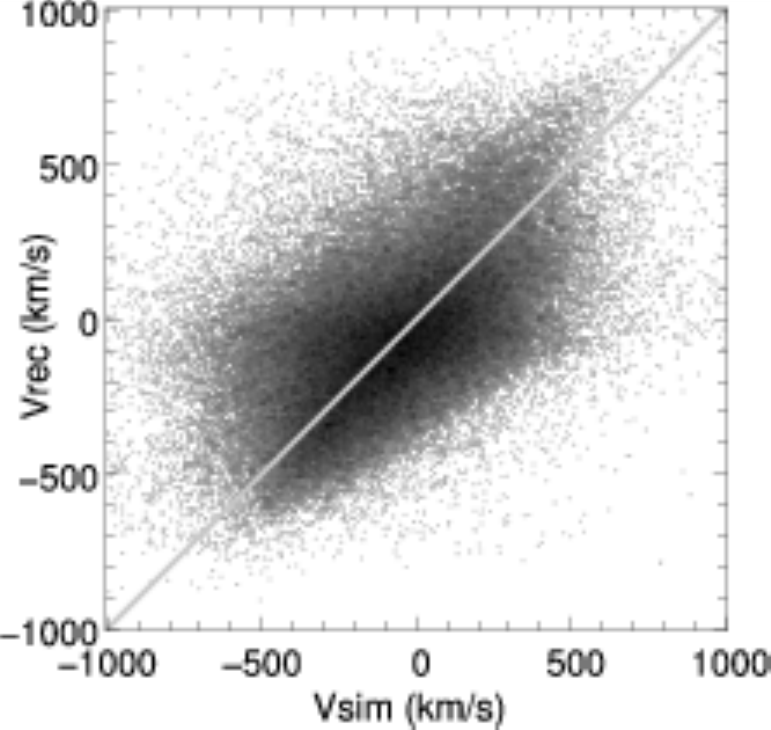} &
      \includegraphics[width=.25\linewidth]{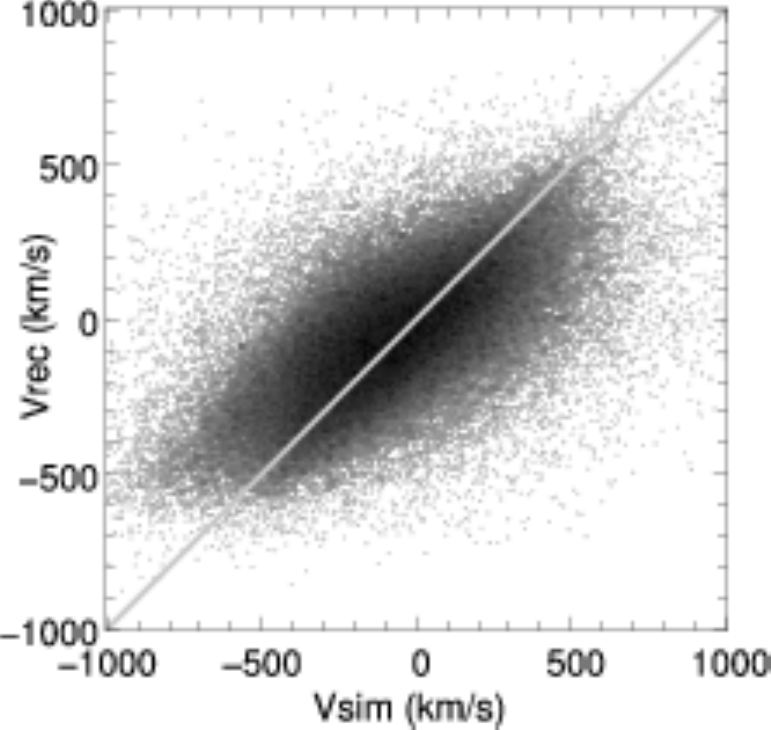} \\
      \hline
      \multicolumn{3}{c}{In 4000 \kms} \\[-.1cm]
      \hline
      \includegraphics[width=.25\linewidth]{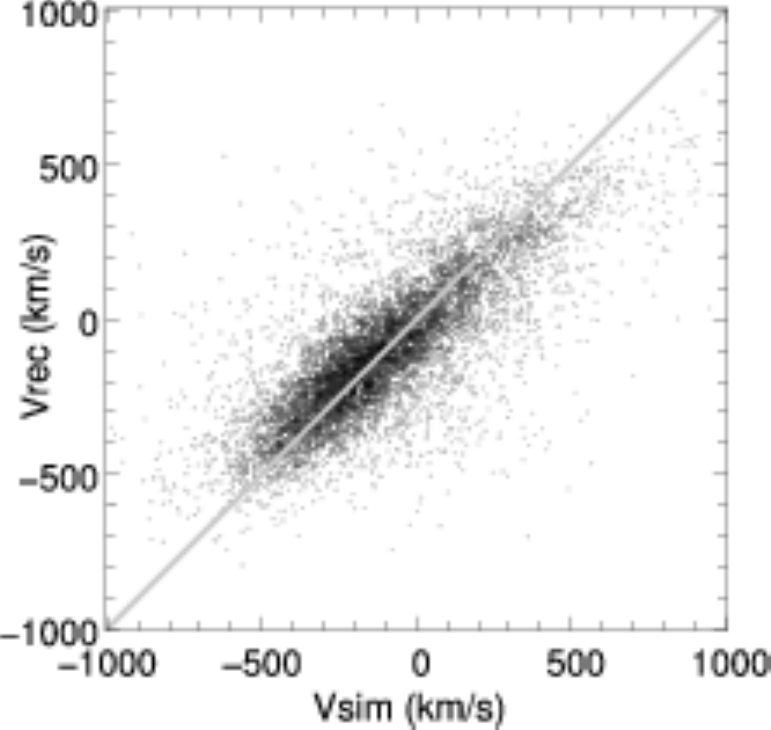} &
      \includegraphics[width=.25\linewidth]{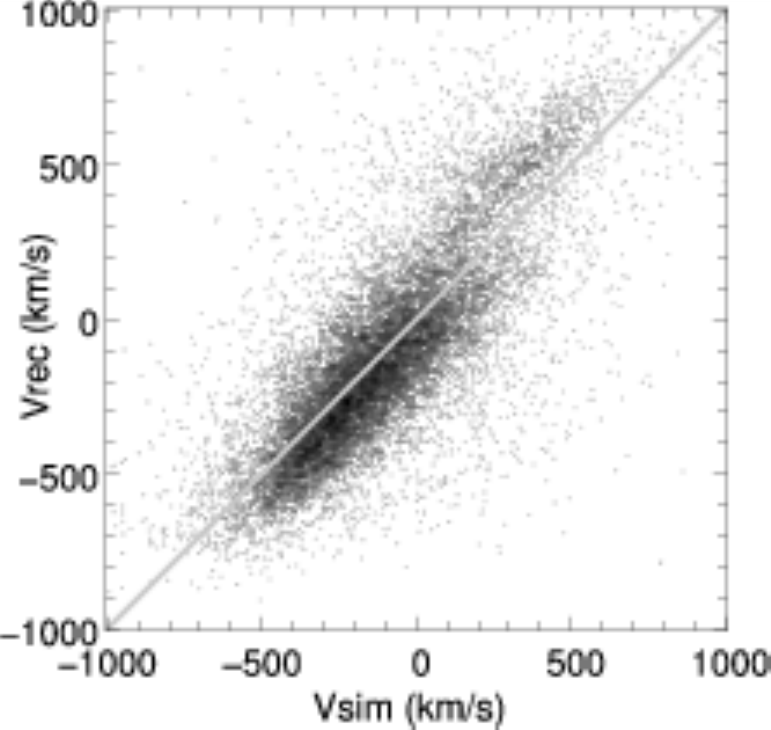} &
      \includegraphics[width=.25\linewidth]{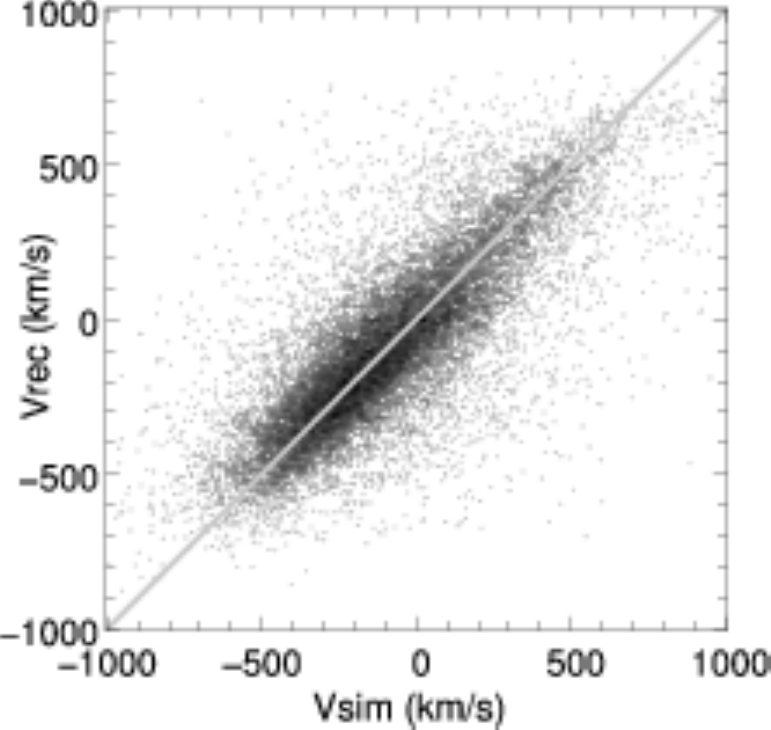} 
    \end{tabular}
  \end{center}
  \vspace*{-0.4cm}
  \caption{ \label{fig:reconstruct_mock6_induced_ic} {\it Lagrangian
      domain / without redshift distortions} -- This figure summarises
      the results obtained on reconstructions that have limited
      information on the Lagrangian domain. The left column
      illustrates the {\it TrueDom} reconstruction, The middle column the
      {\it NaiveDom} one, and the right column the {\it PaddedDom} one. For space
      occupation reasons, the original velocity field given by the simulation is not remembered
      but can be found in Fig.~\ref{fig:incompleteness_scatter}. The {\it
      top row} illustrates the three schemes for handling boundary
      effects on the density field: in the left column one retains
      information of large scale tidal fields, in the middle column
      one cuts the catalogue spherically and does a reconstruction
      on it, in the right column one pads the spherically-cut
      catalogue with particles homogeneously distributed on a grid.
      The {\it second row} gives the reconstructed velocity field in
      each case, smoothed with a 5~$h^{-1}$Mpc Gaussian window as
      usual. The color coding is the same as for the other figures,
      {\it i.e.} dark blue is -1000~\kms and white is +1000~\kms.
      The {\it third row} compares the individual (not smoothed)
      reconstructed and simulated velocities of objects in the
      8k-mock6 catalogue. The {\it fourth row} does the same
      comparison but objects lying only in the 4000~\kms region of the
      8k-mock6 catalogue.
  }
\end{figure*}
 
\begin{figure*}
  \begin{center}
    \begin{tabular}{ccc}
      \hline
      \multicolumn{3}{c}{Current density field} \\
      { (a) TrueDom} & { (b) NaiveDom} & { (c) PaddedDom}   \\
      \hline
      \includegraphics[width=.25\linewidth]{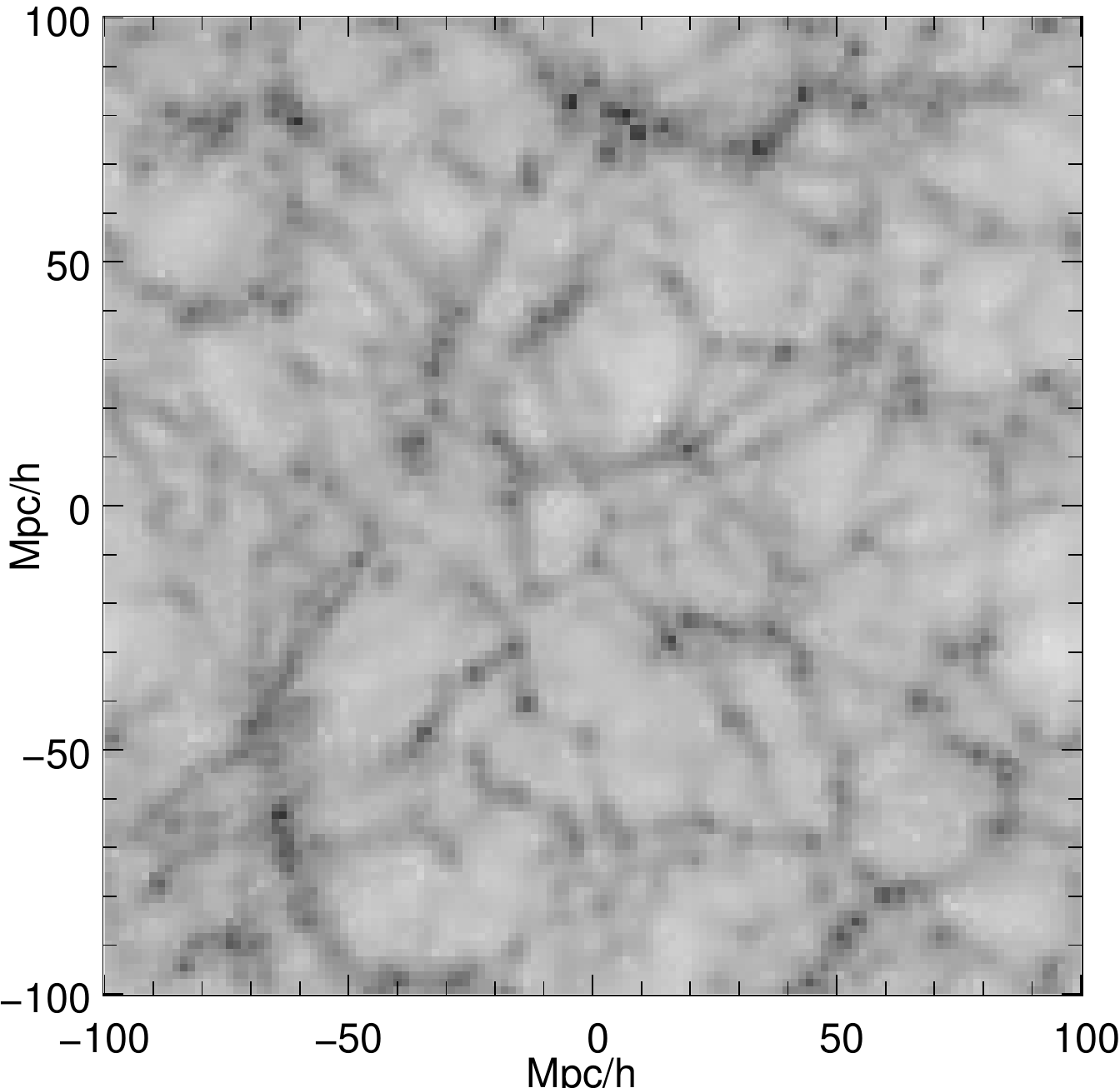} &
      \includegraphics[width=.25\linewidth]{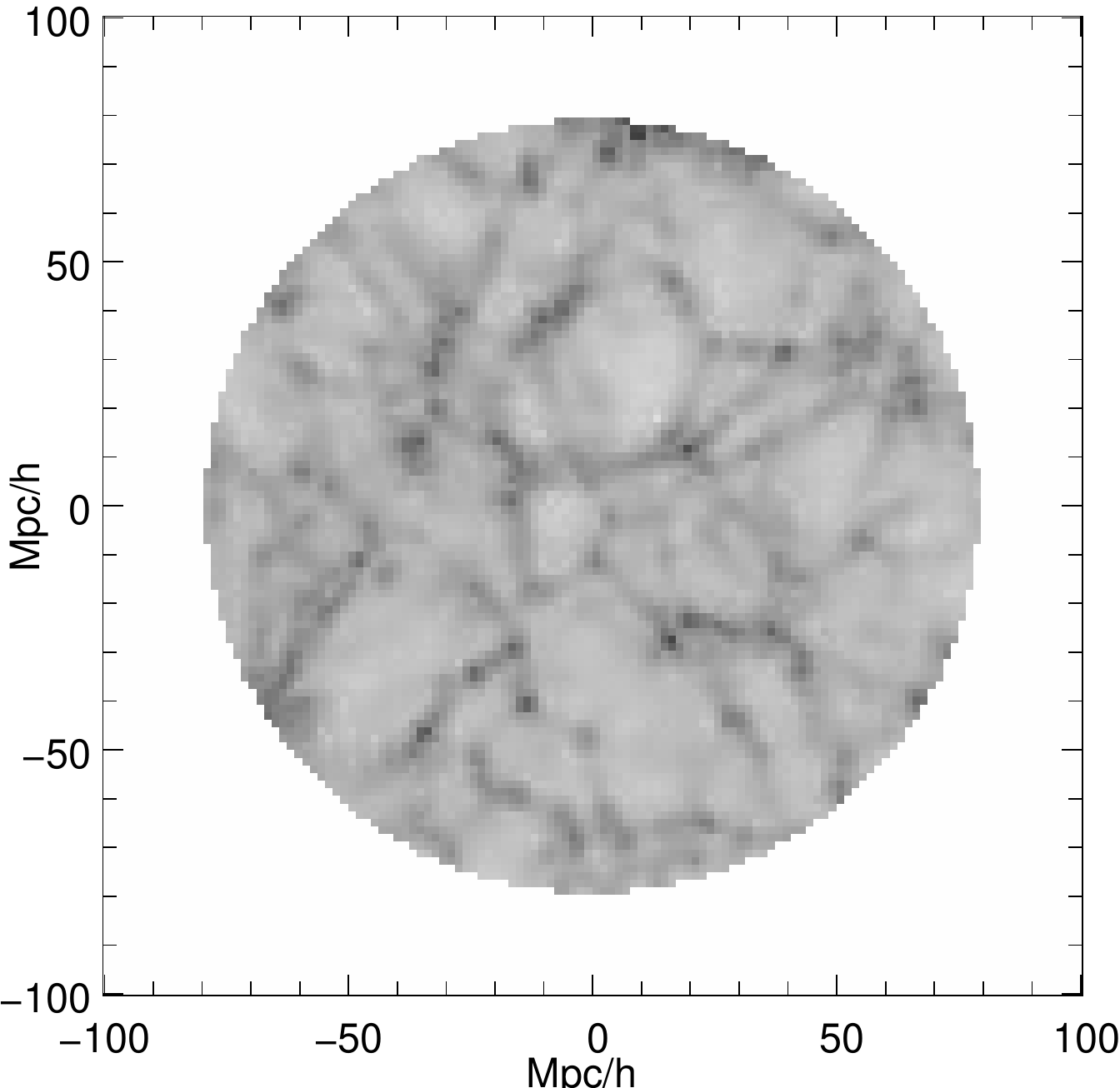} &
      \includegraphics[width=.25\linewidth]{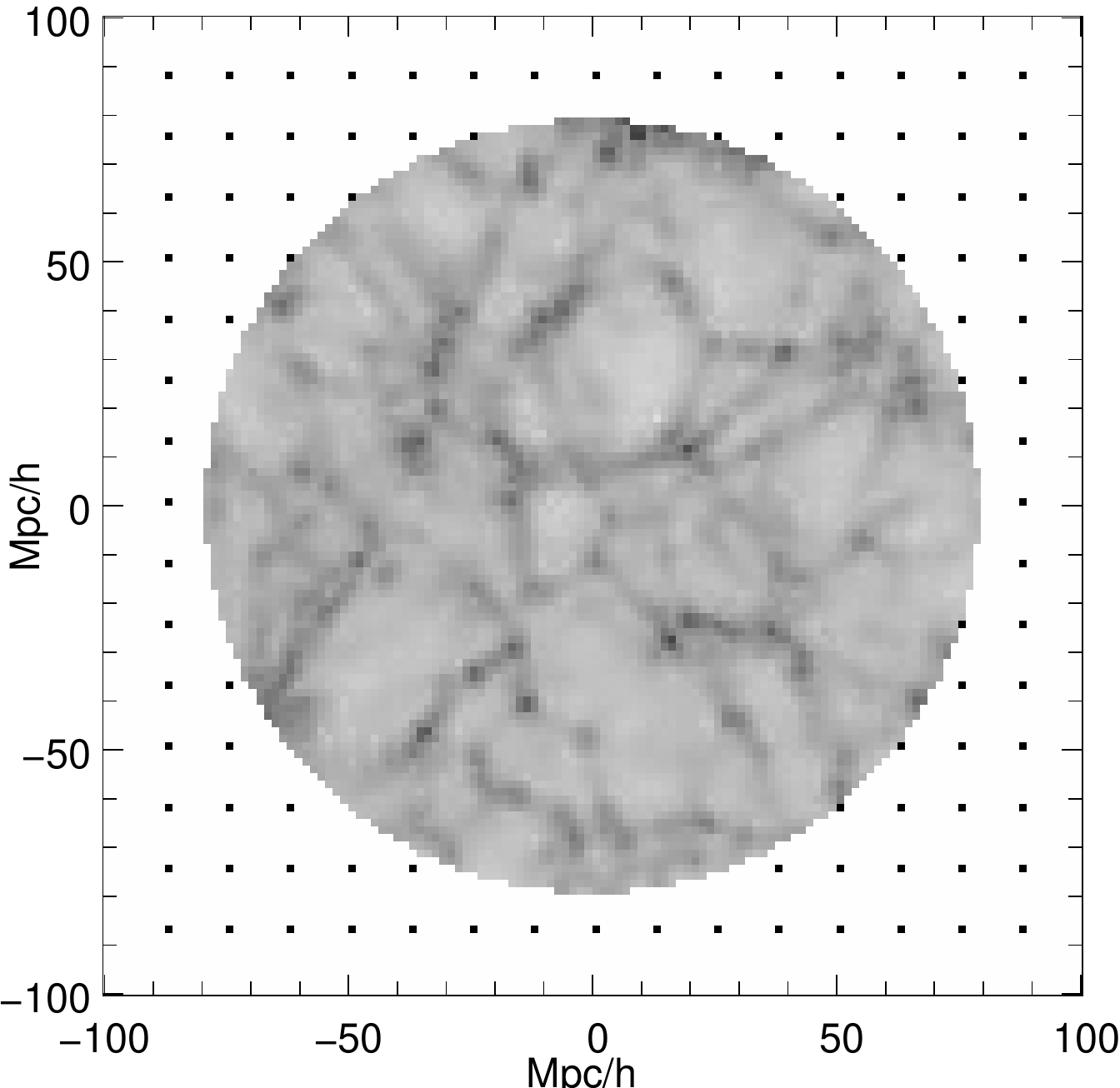} \\
      \hline
      \multicolumn{3}{c}{Reconstructed velocity fields} \\[-.1cm]
      \hline
      \includegraphics[width=.25\linewidth]{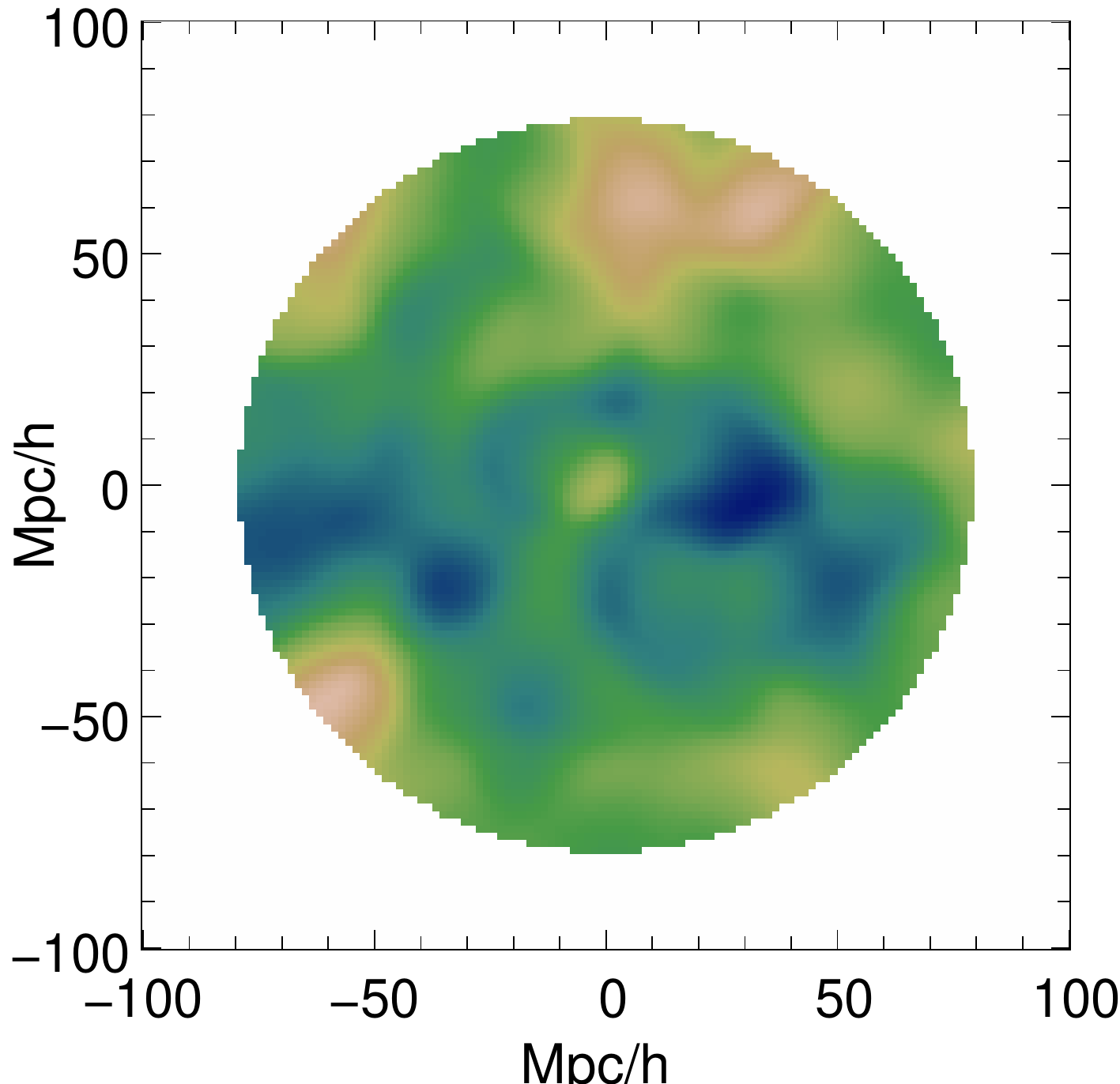} &
      \includegraphics[width=.25\linewidth]{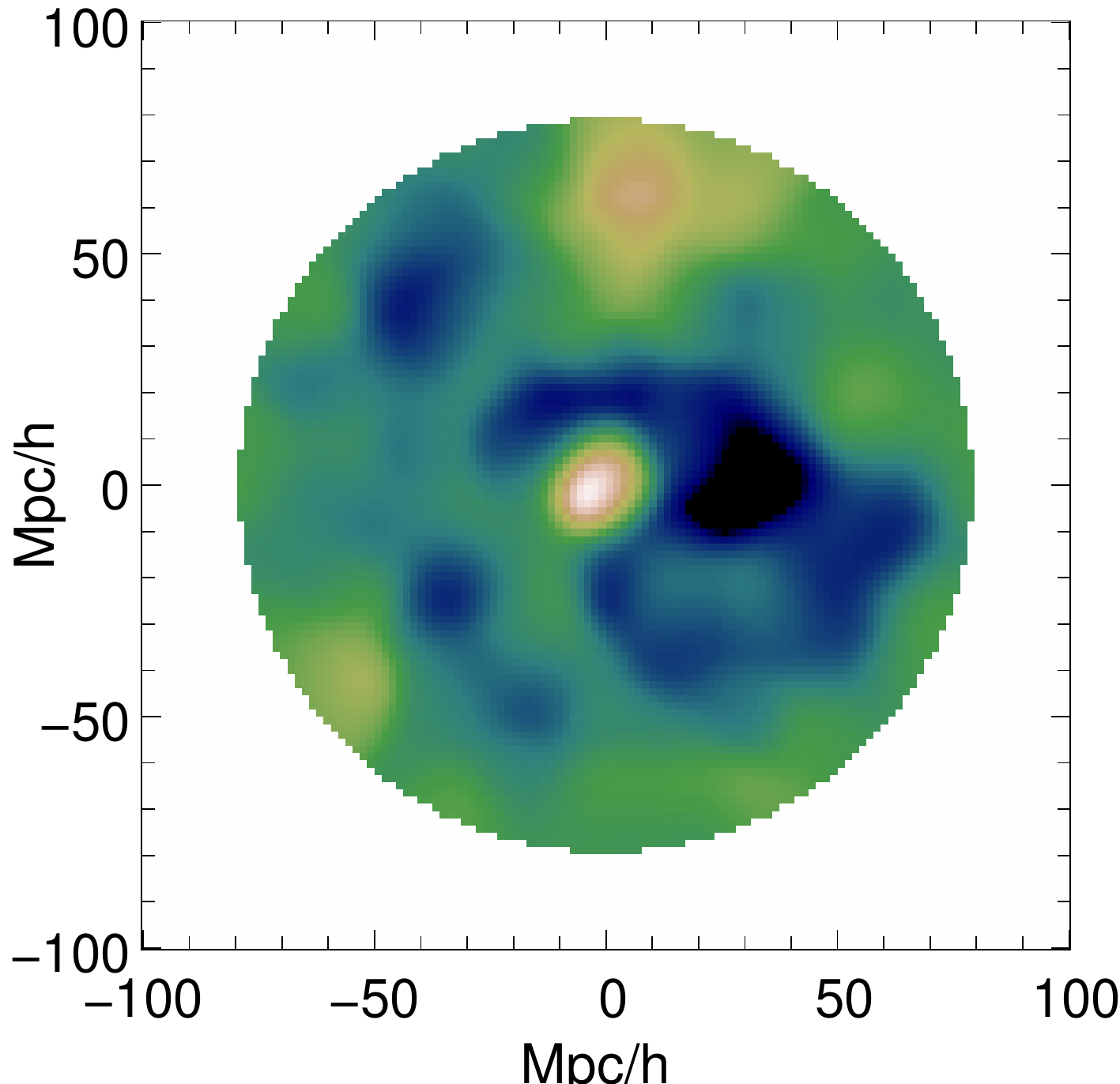} &
      \includegraphics[width=.25\linewidth]{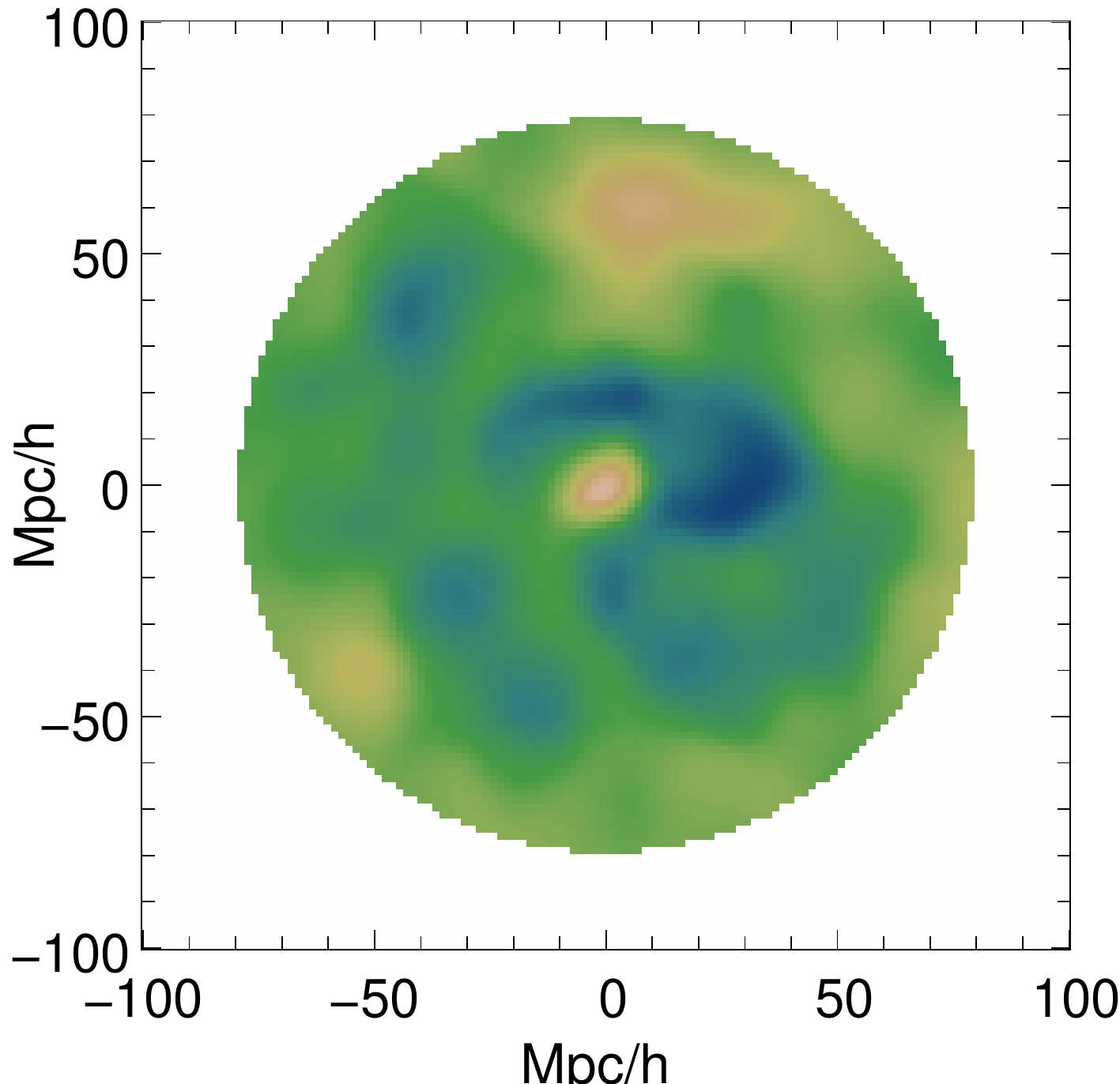}
      \\
      \hline
      \multicolumn{3}{c}{In 8000 \kms} \\[-.1cm]
      \hline
      \includegraphics[width=.25\linewidth]{fig12d} &
      \includegraphics[width=.25\linewidth]{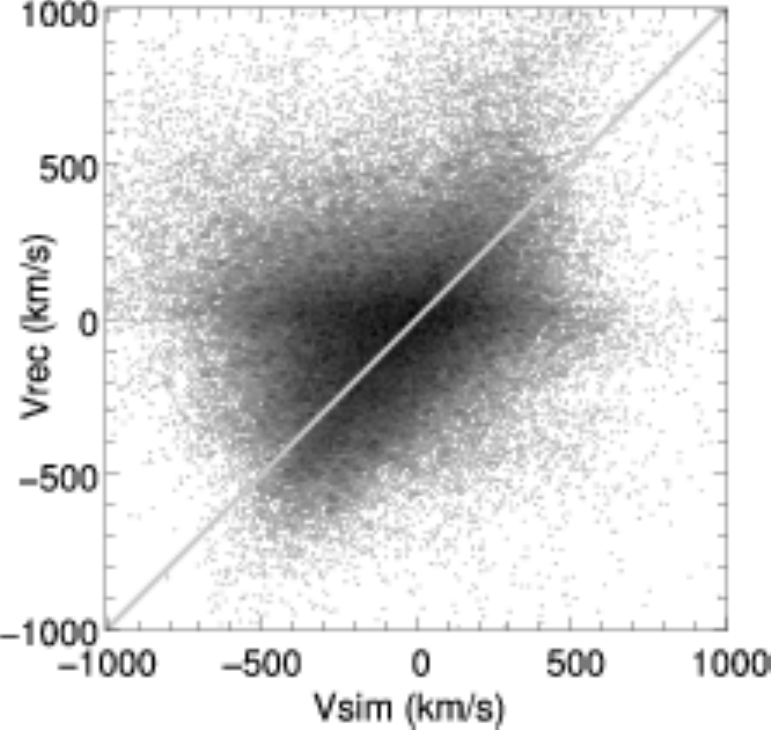} &
      \includegraphics[width=.25\linewidth]{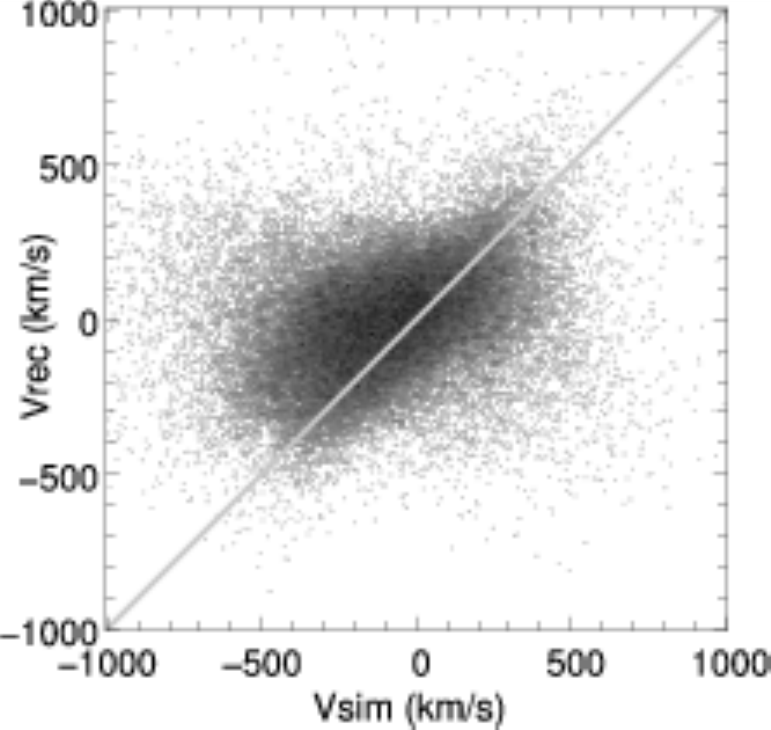} \\
      \hline
      \multicolumn{3}{c}{In 4000 \kms} \\[-.1cm]
      \hline
      \includegraphics[width=.25\linewidth]{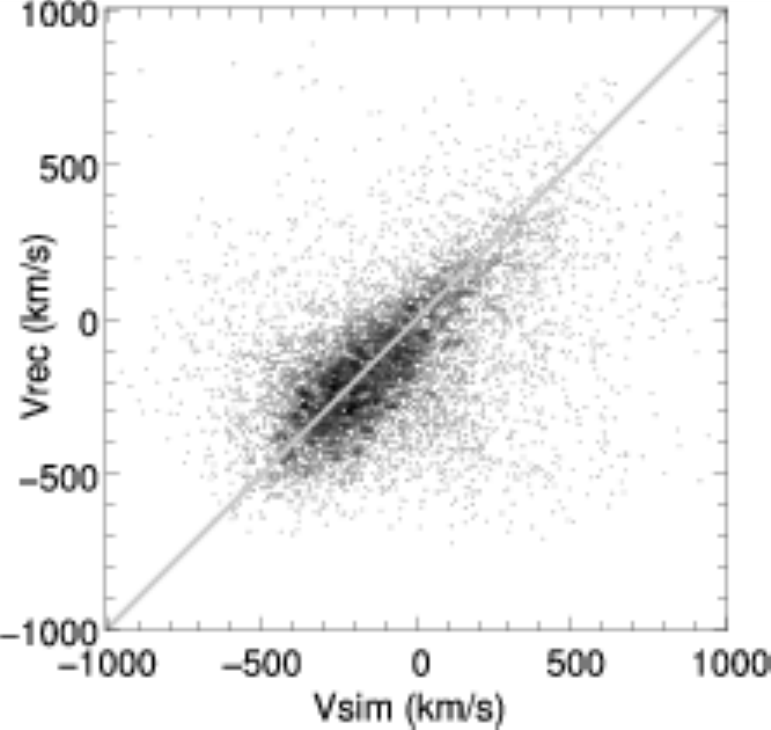} &
      \includegraphics[width=.25\linewidth]{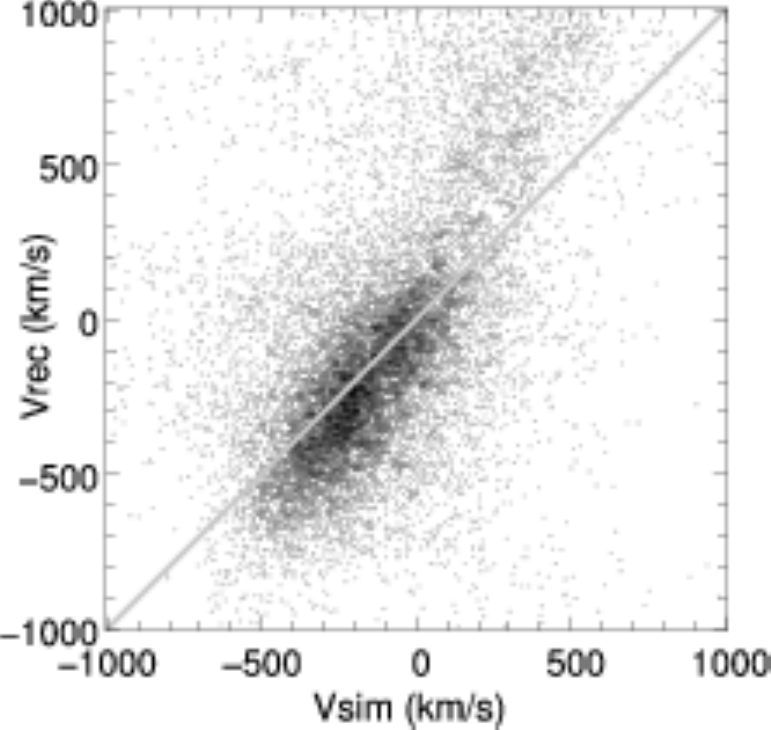} &
      \includegraphics[width=.25\linewidth]{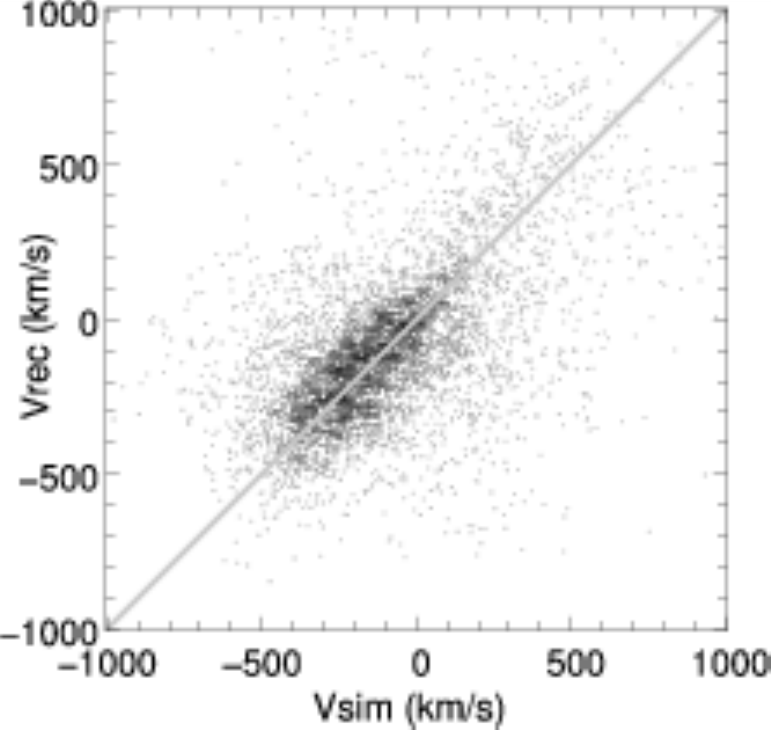} 
    \end{tabular}
  \end{center}
  \vspace*{-0.4cm}
  \caption{ \label{fig:reconstruct_mock6_red_induced_ic}{\it Lagrangian
      domain / with redshift distortions} -- Same as
    Fig.~\ref{fig:reconstruct_mock6_induced_ic}, but for mock
      catalogues including redshift distortion. }
\end{figure*}

The inputs to MAK reconstruction are the present coordinates of the
objects, {\it i.e.} ${\bf x}$ in Eq.~\eqref{eq:assign} or ${\bf s}$ in
Eq.~\eqref{eq:zassign}, and the knowledge of the Lagrangian domain, {\it
i.e.} ${\bf q}$ in Eq.~\eqref{eq:assign} or \eqref{eq:zassign}. Redshift
catalogues give the present ``positions'' of the objects, {\it i.e.}
${\bf s}$ in Eq.~\eqref{eq:zassign}, however we have no observations that
would give us the corresponding Lagrangian domain ${\bf q}$.
We are thus limited to make guesses, though in the end, for huge
catalogues, the details of the guess does not matter as gravitational
forces are screened on large scales by the nearly homogeneous distribution of matter in the
universe. Consequently, what happens at the boundaries should not strongly affect the
central part of the catalogue though some guesses may be better at
confining the edge effects on the boundaries. 
The naive solution is to assume that the Lagrangian domain is not
so different from the volume of the catalogue itself. This assumption
only begins to be a good approximation for volume enclosed in a sphere for which radius is
big enough. For our 80~$h^{-1}$Mpc sample, the mass going in and
out of the volume (from initial to present time) already represents about 16\% of the total mass.
For a 40~$h^{-1}$Mpc sphere, the mass flow is even greater:
it may vary between 30\% and 63~\% of the total mass depending on
the 8k-mock catalogue considered. 
Though tidal field and cosmic variance effects becomes negligible on a 80~$h^{-1}$Mpc
scale, they still affect the boundaries of the Lagrangian domain of a
given catalogue in a non-trivial way.
As we shall show
these problems are further enhanced by redshift distortion.

To achieve a meaningful comparison, we have run a reconstruction on
8k-mock6 using the Lagrangian domain given by the simulation; this
reconstruction is called {\it TrueDom}. Now, we confront
the results of {\it TrueDom} for two different reconstruction setups that
try to recover the Lagrangian domain:
\begin{itemize}
  \item {\it NaiveDom} reconstruction is obtained by assuming a naive spherical
    Lagrangian domain for 8k-mock6. In that case, all the mass that is
    presently in the 8k-mock6 catalogue was uniformly in a sphere of radius
    80~$h^{-1}$Mpc. Equivalently, it means no significant mass
    flow must have gone through the comoving boundaries in the past.

  \item {\it PaddedDom} reconstruction is obtained by padding
    homogeneously the 8k-mock6 catalogue. The padding is chosen such
    that the final MAK mesh that will be reconstructed is an inhomogeneous
    cube (as in right panel of the second row of
    Fig.~\ref{fig:reconstruct_mock6_induced_ic} and
    \ref{fig:reconstruct_mock6_red_induced_ic}). The cube must be
    sufficiently big to absorb density fluctuations present at the boundary
    of the catalogue (typically a 20~$h^{-1}$Mpc buffer zone is
    needed). With real data, 
    we are bound to assume that the catalogue is totally representative of
    the whole universe, {\it i.e.} its effective mean matter density is
    equal to $\Omega_\text{m}$. 
    
\end{itemize}

Fig.~\ref{fig:reconstruct_mock6_induced_ic} shows the result of a {\it
TrueDom}, {\it NaiveDom} and {\it PaddedDom} reconstruction applied to
8k-mock6 in the absence of redshift distortion. Fig.~\ref{fig:reconstruct_mock6_red_induced_ic} gives the
same reconstructions when applied to a redshift
catalogue. Table~\ref{table:lagvol} summarises the value of the
moments of $P(v_\text{r,sim},v_\text{r,rec})$ for different cases.
We will now first confront the results of real space reconstructions,
and second redshift space reconstructions.

 {\it TrueDom} reconstruction does not yield any significant bias at
80~$h^{-1}$Mpc. However, at 40~$h^{-1}$Mpc, cosmic variance effects introduce a
noticeable systematic error in the direction of higher
$\Omega_\text{m}$ that will be discussed in
\S~\ref{sec:cosmic_var}. Compared to {\it TrueDom}, {\it NaiveDom}
gives good overall results though the central blue region of {\it
TrueDom} turns to dark blue in {\it NaiveDom}, which would suggest the
velocity field is biased. This analysis is confirmed by looking at the
bottom scatter plot. The $\Omega_\text{m}$ measurement
(Table~\ref{table:lagvol}) is underestimated by about 26\%
even in the central region of the catalogue which is normally less affected by
boundary effects. {\it PaddedDom}, on the other hand, does not yield
such a sharp discrepancy in the middle of 8k-mock6, namely in the
4k-mock6 region.  Both the bottom scatter plot and the $\Omega_\text{m}$
measurement confirm that the 
reconstructed velocities are nearly bias-free in the central
region. As expected, the velocities in the neighbourhood of the 
boundaries are completely wrong for the two methods.

Now, the catalogues are cut in redshift space. Redshift distortion biases
the velocity distribution of objects on the catalogue boundary: the
catalogue receive more infalling objects than outfalling ones. In some
cases, one may even find objects seemingly artificially separated from
the main volume of the catalogue (they look ``disconnected''). In
those cases, the hypothesis of convexity is definitely lost for those
objects. This problem will enhance boundary problems.
 The case of {\it TrueDom}
reconstruction has been discussed in \S~\ref{sec:redshift_distortion}.
As previously, the peculiar velocities in {\it NaiveDom} and in {\it
  PaddedDom} are largely uncorrelated in the full 8k-mock6 volume
(Fig.~\ref{fig:reconstruct_mock6_red_induced_ic}). 
However, peculiar velocities reconstructed by {\it NaiveDom}
are more strongly overestimated than
by using {\it PaddedDom}'s, as shown in Table~\ref{table:lagvol}. For {\it
  NaiveDom}, the scatter is plagued by a horizontal alignment in
Fig.~\ref{fig:reconstruct_mock6_red_induced_ic}, mid-lower panels,
which is a signature of a strong edge effect. This spurious alignment was already
present, though much less apparent, in the real space case.
 On the other hand, {\it PaddedDom} does
not present this feature but only a large scatter.
 We have verified that objects belonging  the horizontal alignment
 are essentially near the 80~$h^{-1}$Mpc boundary, contrarily to velocities
 reconstructed using {\it PaddedDom} which are more or less uniformly
 distributed and essentially uncorrelated to simulated velocities.\footnote{This
   behaviour is expected from an algorithmic point of view. The
   objects nearby the boundary cannot acquire any displacement using
   MAK because of the ``pressure''/competition of objects inside the
   sphere. This problem is further enhanced in redshift space because
   generally these objects come from outside the sphere and are selected
   because their infall velocity is high. In {\it NaiveDom}, they cannot
   escape from the assumed spherical Lagrangian domain which thus
   leads to zeroing their velocity. On the other hand, {\it PaddedDom} is
   much less strict on the boundary, which leaves the freedom for MAK
   reconstruction to have a non-zero velocity even for objects on the
   boundary of the catalogue.} 
 This means that {\it PaddedDom} is at least better at screening edge
 effects than {\it NaiveDom} in the sense the errors are more evenly
 distributed and less systematic. Though impressively low
 in the last two rows of Table~\ref{table:lagvol}, the correlation 
 coefficient $r$ is actually spoiled by the long tails of the PDF
 shown in the scatter plots in
 Fig.~\ref{fig:reconstruct_mock6_red_induced_ic}.  Concerning
 $\Omega_\text{m}$, {\it NaiveDom} seems less robust to produce an
 unbiased estimation than {\it PaddedDom}. Indeed, looking at
 Table~\ref{table:lagvol}, one may note that the interval delimited by
 $s_\text{med}$, $s_\text{min}$ and $s_\text{max}$ nearly does not contain $\Omega_\text{m}=0.30$
 for {\it NaiveDom}/Real space/40~$h^{-1}$Mpc, and does not contain it at all for
 {\it NaiveDom}/Redshift space. On the contrary, $\Omega_\text{m}=0.30$ is
 always selected by the three $s$ parameters using {\it PaddedDom} reconstruction.
In the rest of this paper, whenever it is needed, we will thus use the {\it PaddedDom} reconstruction.

\begin{table*} 
  \caption{\label{table:lagvol} {\it Lagrangian volume} -- Residual error
    after the correction. Description for some columns is given in the
    caption of Table~\ref{tab:ML_errors}. ``Radius'' gives the spatial
    size of the sphere on which the velocity-velocity comparison is
    conducted. ``Reconstruction type'' indicates the type of
    Lagrangian domain reconstruction and whether it is mixed with
    redshift distortion effect. Details on the meaning of each name are
    given in \S~\ref{sec:lag_volume}. } 
  \begin{center}
    \begin{tabular}{|c|c|c|c|c|c|c|c|c|c|c|}
      \hline
      \multirow{2}{3.8cm}{ Reconstruction type} & \multirow{2}{1.2cm}{Radius ($h^{-1}$Mpc)} &
      \multicolumn{3}{c||}{Velocities} &
      \multirow{2}{1cm}{$\Omega_{\text{m}}$ ($s$)} &
      \multirow{2}{1cm}{$\Omega_{\text{m}}$ ($\mathfrak{L}_\text{min}$)} &
      \multirow{2}{1cm}{$\Omega_{\text{m}}$ ($\mathfrak{L}_\text{max}$)} &
      \multirow{2}{1.1cm}{$\Omega_{\text{m}}$ (1.5$\sigma$,$s_\text{med}$)} &
      \multirow{2}{1.1cm}{$\Omega_{\text{m}}$ (1.5$\sigma$,$s_\text{min}$)} &     
      \multirow{2}{1.1cm}{$\Omega_{\text{m}}$ (1.5$\sigma$,$s_\text{max}$)} \\     
      \cline{3-5} & & $s$ & $r$ & $\sigma$ &  \\ 
      \hline \hline

      \multirow{2}{4cm}{TrueDom / Real space} & 80 & 0.91 & 0.77
      & 0.66 & 0.35 &  0.28 & 0.31 & 0.27 & 0.233 & 0.32  \\
      & 40 & 0.80 & 0.76 & 0.65 & 0.45 & 0.28 & 0.38 & 0.35 & 0.28 & 0.43 \\
      
      \hline

      \multirow{2}{4cm}{NaiveDom / Real space} & 80 & 0.87 & 0.52 &
      0.92 & 0.38 
      &  0.20 &0.28 & 0.42 & 0.20 & 0.87 \\
      & 40 & 1.11 & 0.77 & 0.73 & 0.25 & 0.20 &0.24 &  0.244 & 0.19 & 0.31 \\

      \hline

      \multirow{2}{4cm}{PaddedDom / Real space} & 80 & 0.73 &
      0.65 & 0.77 & 0.53 &  0.36 & 0.48 & 0.45 & 0.27 & 0.75 \\
      & 40 & 0.91 & 0.77 & 0.64 & 0.35 & 0.28 & 0.34 & 0.32 & 0.26 & 0.38  \\
      \hline
      NaiveDom / Redshift space & 40 & 1.49 & 0.51 & 1.31 & 0.11 & 0.15 & 0.26
      & 0.20 & 0.12 & 0.37 \\
      PaddedDom / Redshift space & 40 & 0.93 & 0.53 & 0.94 & 0.36 & 0.18 &
      0.34 & 0.38 & 0.20 & 0.79 \\
      \hline
    \end{tabular}
  \end{center}
\end{table*}

\subsection{Cosmic variance}
\label{sec:cosmic_var}

\begin{figure*}
  \begin{center}
    \begin{tabular}{ccc}
      {\Large 8k-Mock6} & {\Large 8k-Mock7} & {\Large 8k-Mock12} \\
      \includegraphics[width=.3\linewidth]{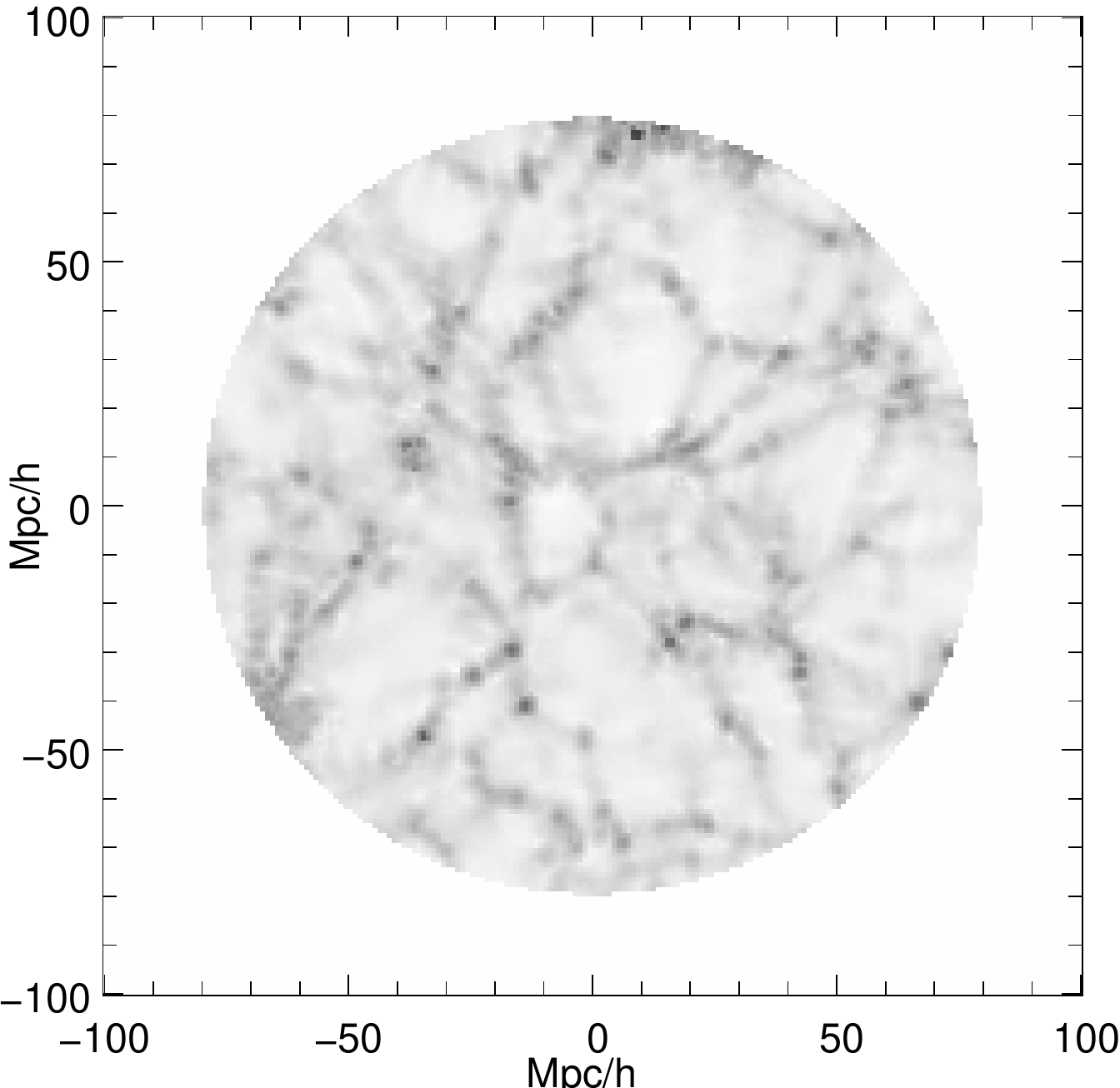} &
      \includegraphics[width=.3\linewidth]{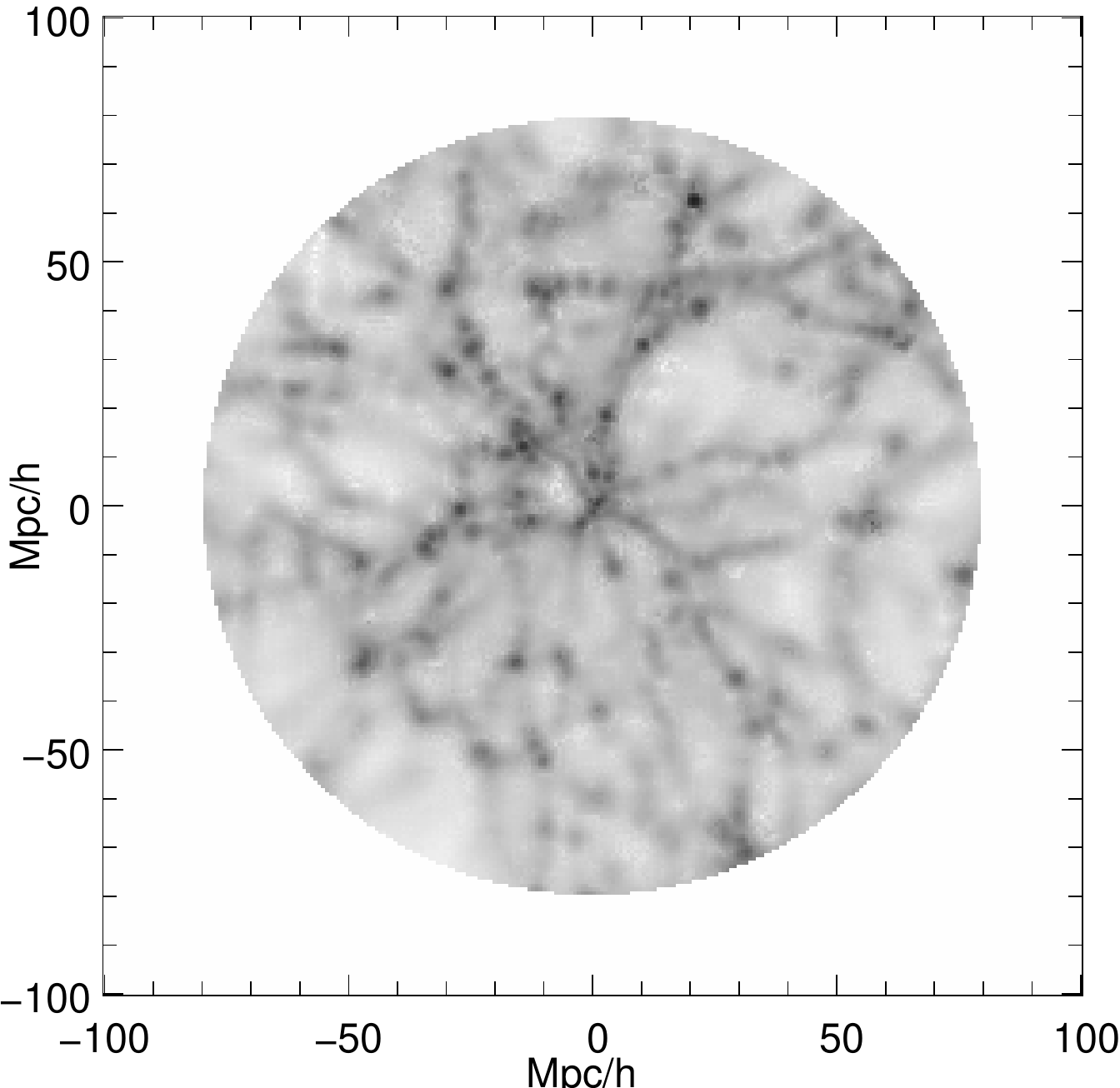} &
      \includegraphics[width=.3\linewidth]{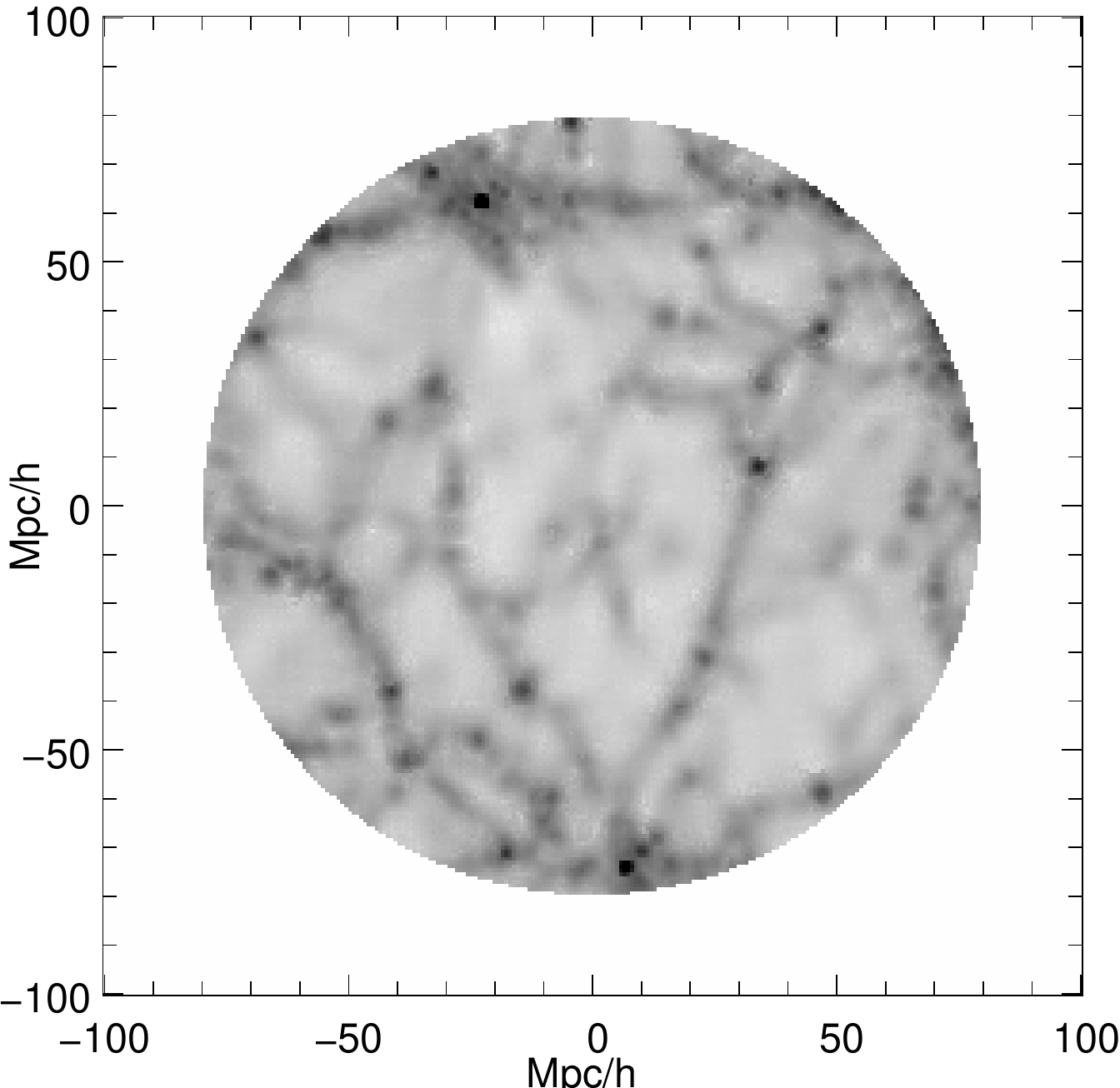} \\
      \includegraphics[width=.3\linewidth]{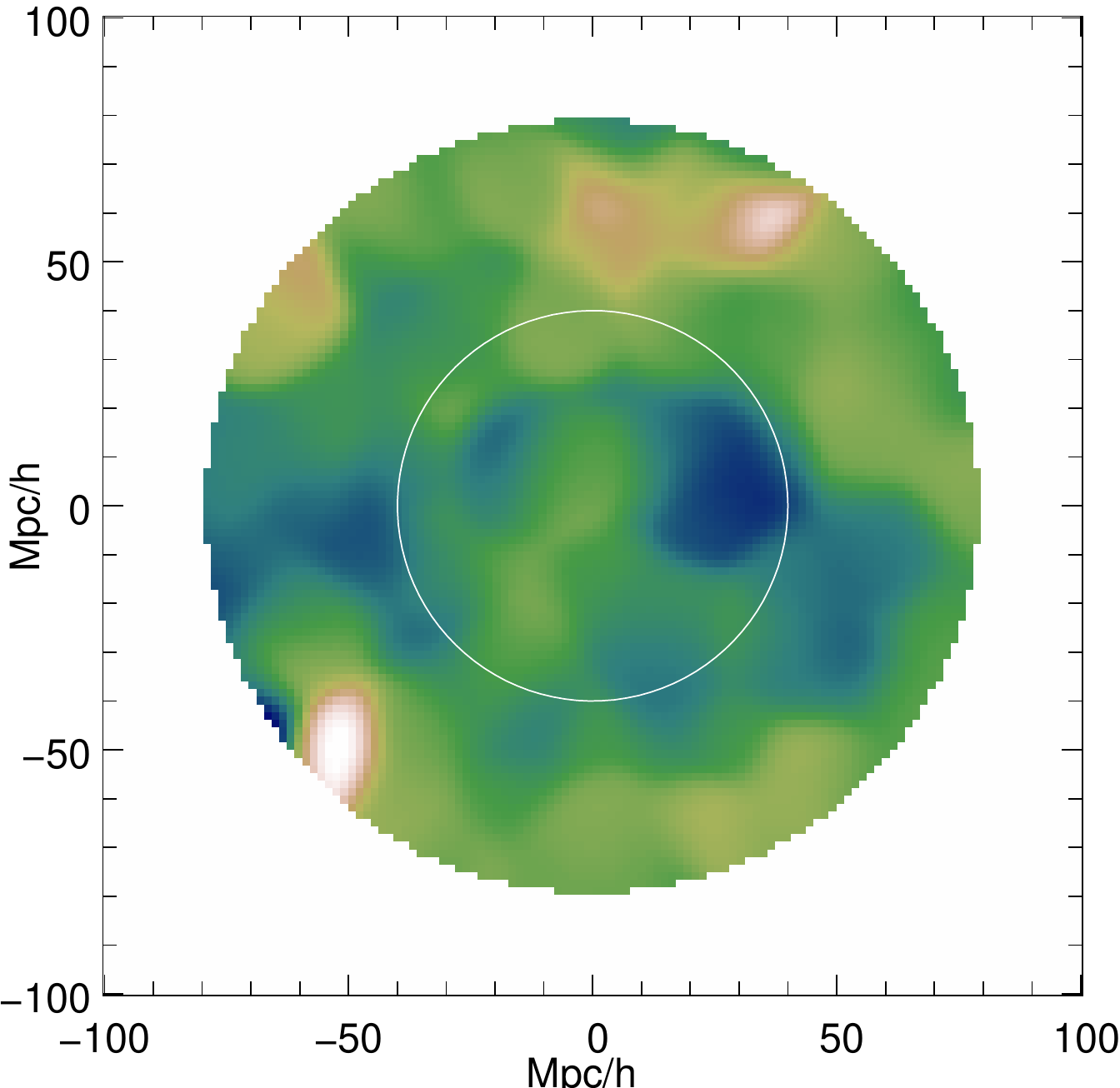} &
      \includegraphics[width=.3\linewidth]{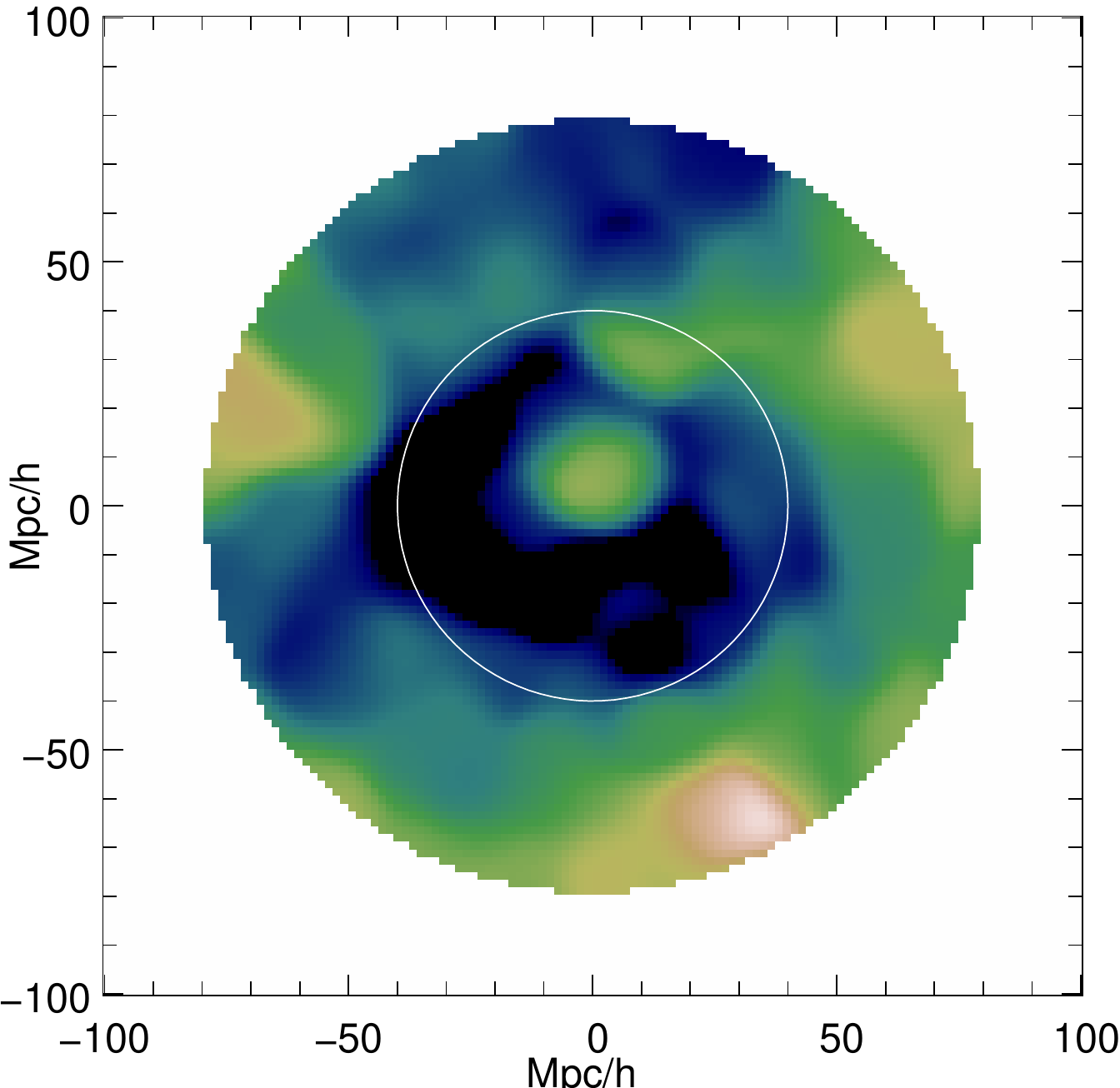} &
      \includegraphics[width=.3\linewidth]{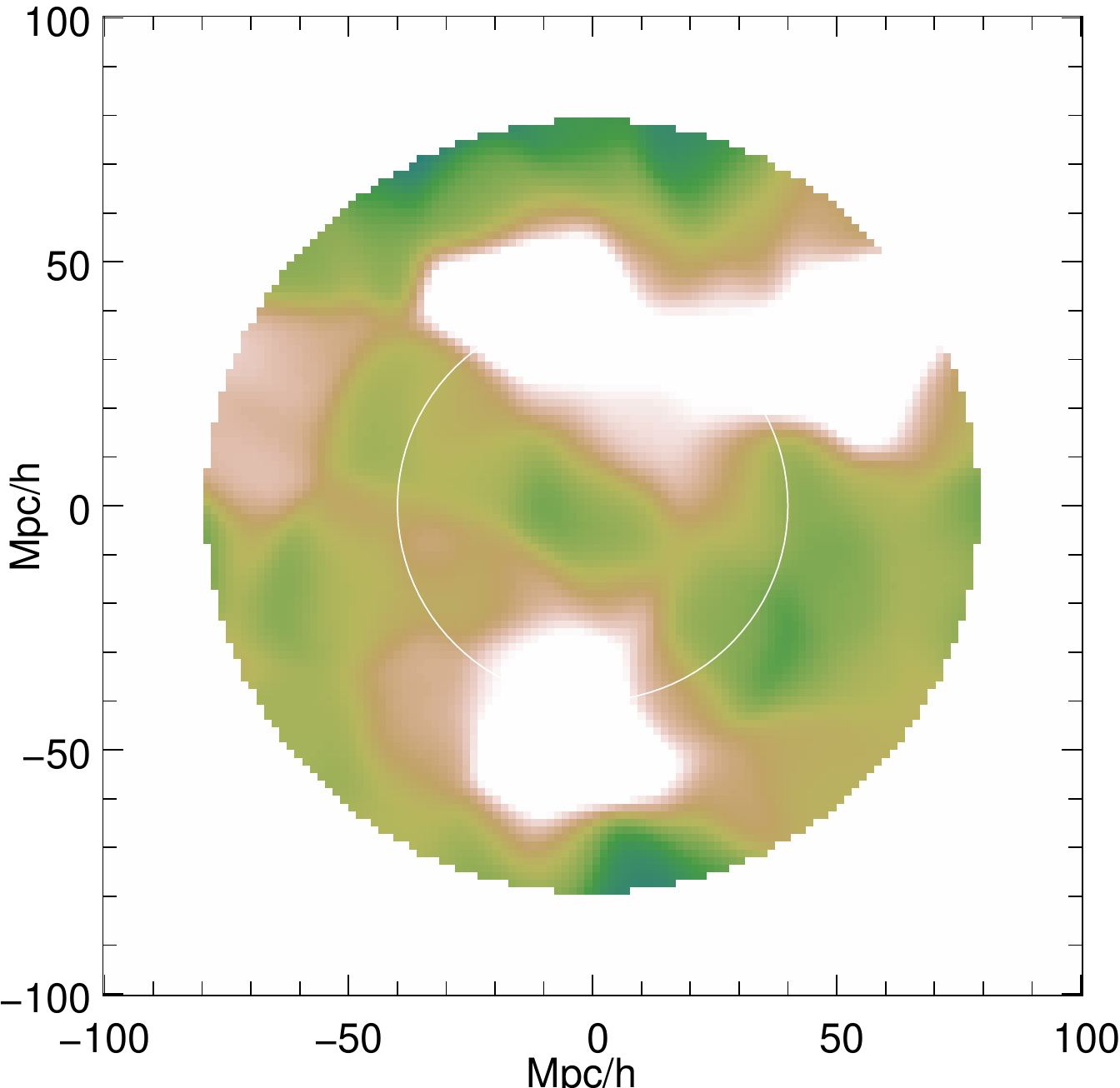}      \\
      \includegraphics[width=.3\linewidth]{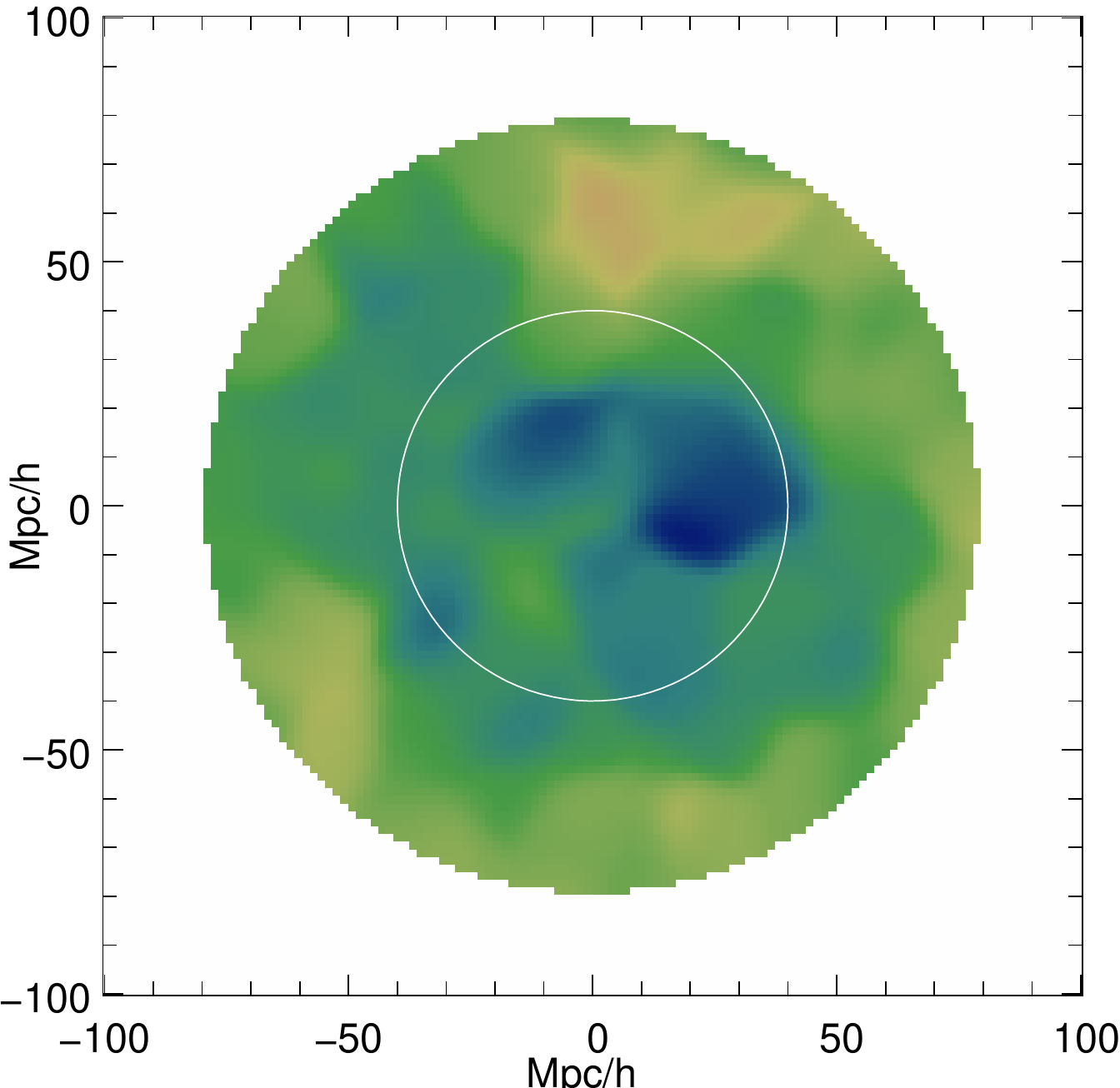} &
      \includegraphics[width=.3\linewidth]{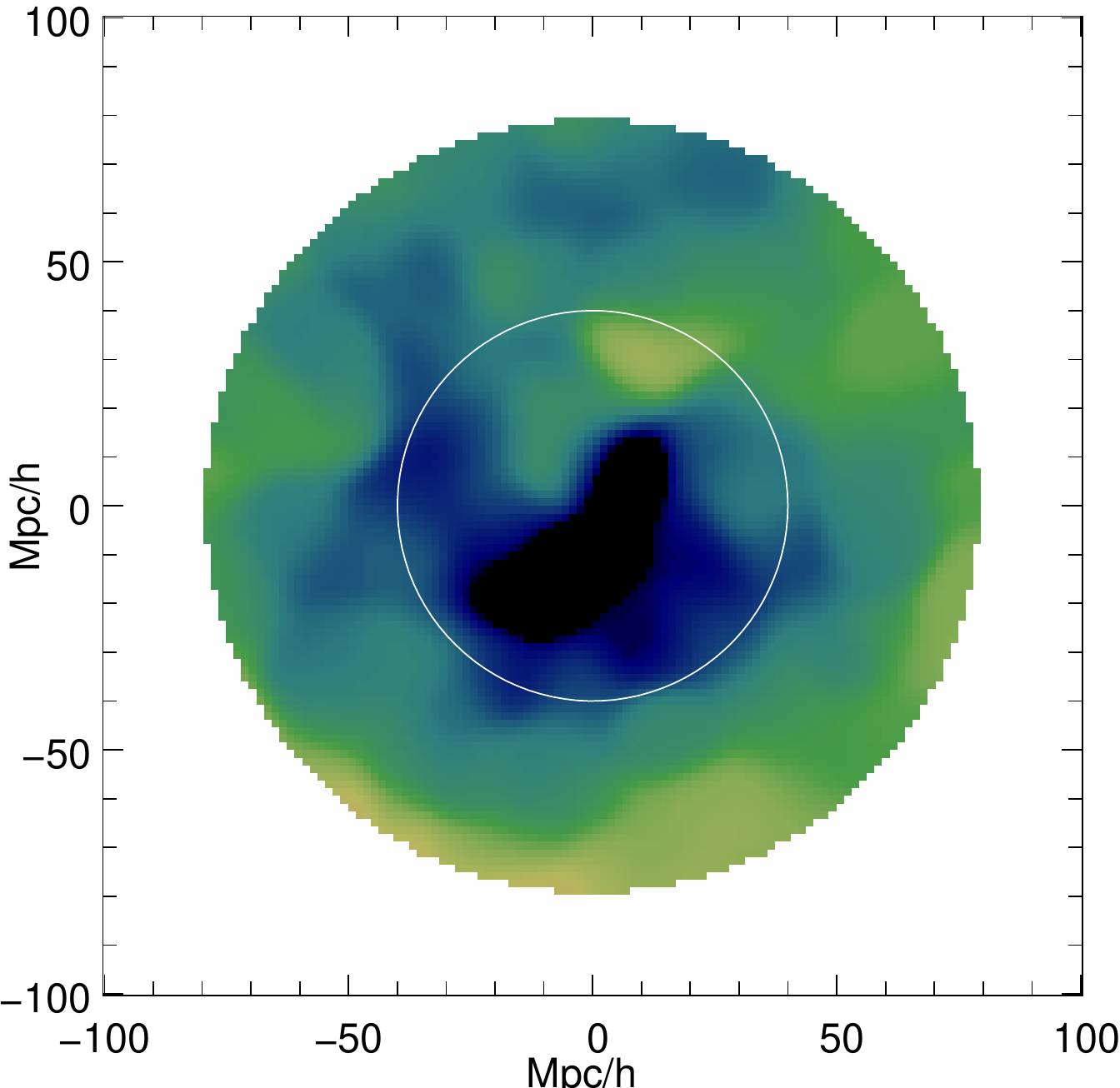} &
      \includegraphics[width=.3\linewidth]{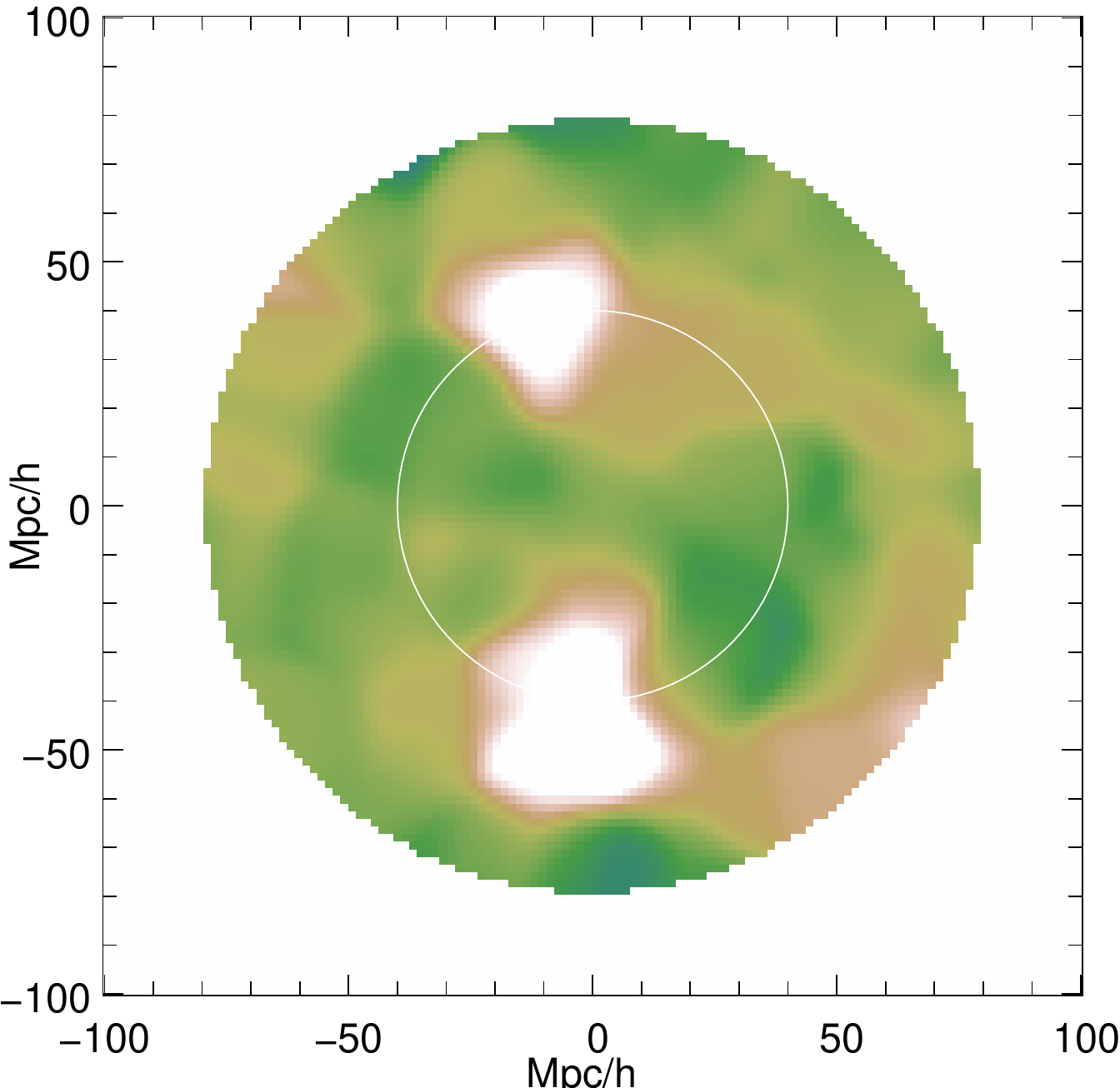} 
      \\
      \includegraphics[width=.3\linewidth]{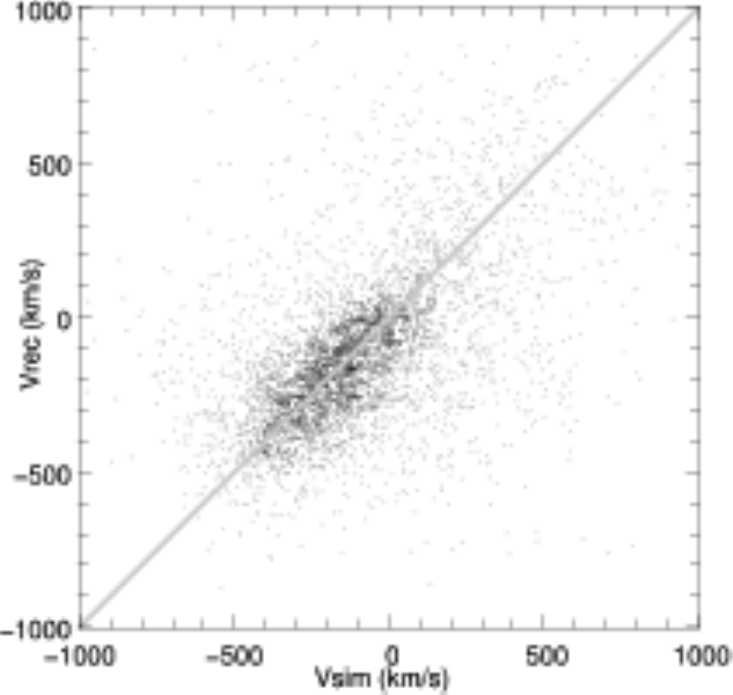} &
      \includegraphics[width=.3\linewidth]{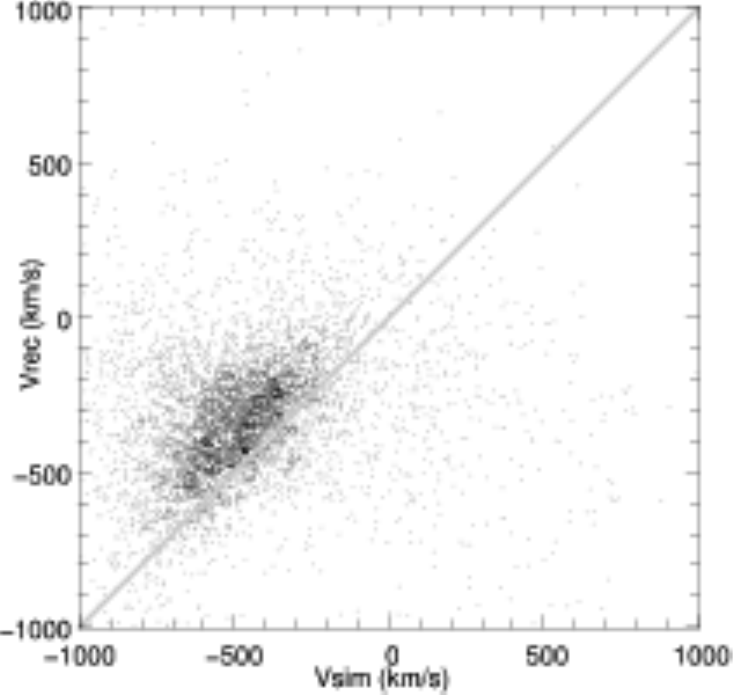} &
      \includegraphics[width=.3\linewidth]{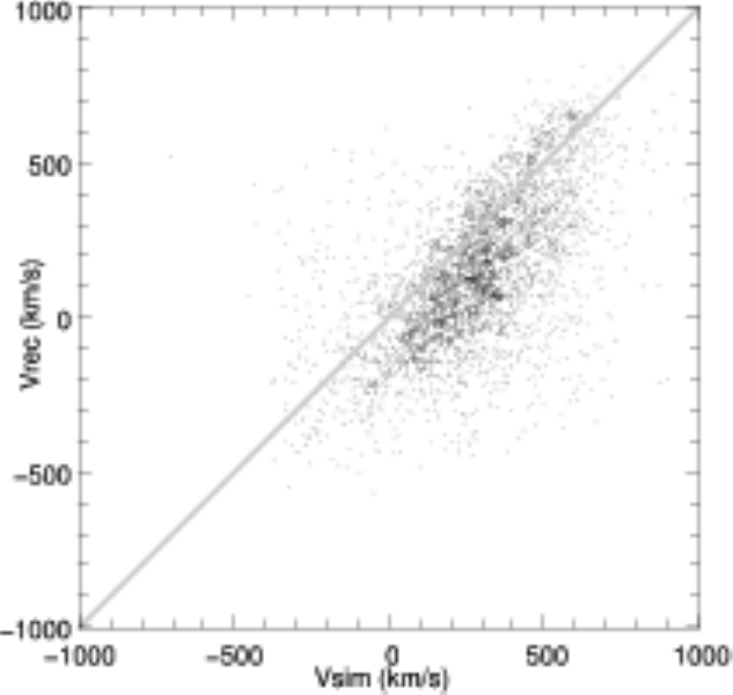}      
    \end{tabular}
  \end{center}
  \vspace*{-.4cm}
  \caption{\label{fig:all_mock_catalogues} {\it Cosmic variance} --
    This figure gives a visual comparison of the three mock catalogues
    used to study cosmic variance effects.
    Top panels: Adaptively smoothed density fields of the considered
    mock catalogues. In each case, we have represented the central
    thin slice that contains the observer. Second row: Simulated
    velocity field, after smoothing with a 5~$h^{-1}$Mpc Gaussian
    window. The white circle gives the limit of the 40~$h^{-1}$Mpc
    volume. Third row: Same as second row, but for the reconstructed
    velocity field. Fourth row: Comparison between reconstructed and simulated
    peculiar velocities. }
\end{figure*}

We generally assume that galaxy catalogues give a fair representation of
the whole universe, but of course we have no \correction{guarantee} that this assumption
is correct. Thus, the result of a MAK reconstruction may be affected by inhomogeneities above the catalogue scale.
For instance, our galaxy may reside in a particularly extreme region
(overdense or underdense), which would produce unusual
peculiar velocities. This effect, known as 
{\it cosmic variance}, can be investigated by our three original basic
mock catalogues: 4k-mock6, 4k-mock7, 4k-mock12 (\S~\ref{sec:mock_building_main}). The
cosmic variance effect is here further enhanced by the finiteness of
the sampled volume. The volume is sufficiently small here to have a
non-zero average line-of-sight velocity. On a 40~$h^{-1}$Mpc
scale, this effect can substantially modify the $\Omega_\text{m}$
measurement (put $\Omega_\text{m,mes}$ in this section) by cutting the
$P(v^s_\text{rec}, v_\text{sim})$ distribution at an inadequate place.

\begin{table*}
  \vspace*{-0.4cm}\caption{\label{table:cosmic_variance} {\it Cosmic
      variance} -- Summary of measurements conducted on the three mock
    catalogues. The reconstruction is either conducted on the basic
    catalogue without any observational effect besides cosmic variance
    (labelled {\it Original}), or on the same catalogue but affected
    by redshift distortion, incompleteness and for which the
    Lagrangian domain is determined using {\it PaddedDom} (reconstruction
    labelled {\it Full}).  The description of the other columns are
    given Table~\ref{tab:ML_errors}. }
  \begin{center}
    \begin{tabular}{|c|c||c|c|c|c|c|c|c|c|c|}
      \hline
       \multirow{2}{1.5cm}{Catalog} & \multirow{2}{2cm}{Reconstruction
         type} & \multicolumn{3}{c||}{Velocity field} &
       \multirow{2}{1.2cm}{$\Omega_{\text{m}}$ ($s$)} &
       \multirow{2}{1.2cm}{$\Omega_{\text{m}}$ ($\mathfrak{L}_\text{min}$)} &
       \multirow{2}{1.2cm}{$\Omega_{\text{m}}$ ($\mathfrak{L}_\text{max}$)} &
       \multirow{2}{1.2cm}{$\Omega_{\text{m}}$ (1.5$\sigma$,$s_\text{med}$)} &
       \multirow{2}{1.2cm}{$\Omega_{\text{m}}$ (1.5$\sigma$,$s_\text{min}$)} &
       \multirow{2}{1.2cm}{$\Omega_{\text{m}}$ (1.5$\sigma$,$s_\text{max}$)} \\
       \cline{3-5}
       & &
       $s$ & $r$ & $\sigma$ &  \\ 
       \hline \hline
       \multirow{2}{2cm}{4k-mock6 ($\Omega_\text{eff} = 0.35$)}
       & Original & 0.80 & 0.76 & 0.65 & 0.313 & 0.28 & 0.38 & 0.35 & 0.28 & 0.43 \\
       &  Full & 0.94 & 0.50 & 0.96 & 0.35  & 0.13 &
       0.31 & 0.31 & 0.16 & 0.70 \\
      \hline
      \multirow{2}{2cm}{4k-mock7 ($\Omega_\text{eff} = 0.5$)} 
      & Original & 0.70 & 0.67 & 0.76 & 0.57 & 0.39 & 0.47 & 0.40 & 0.33 & 0.48   \\
      & Full & 0.88 & 0.11 & 1.33 & 0.43 & 0.41 & 1.62
      & 0.30 & 0.09 & 1.29 \\
      \hline
      \multirow{2}{2cm}{4k-mock12 ($\Omega_\text{eff} = 0.19$)} 
      & Original & 1.12 & 0.81 & 0.66 & 0.24 & 0.235 & 0.27 & 0.24 & 0.22 & 0.26
      \\
      & Full & 1.08 & 0.58 & 1.11 & 0.24 & 0.29 & 0.62
      & 0.15 &  0.08 & 0.31 \\
      \hline
    \end{tabular}
  \end{center}
\end{table*}

The results of the reconstruction on these three mock catalogues are
given in Fig.~\ref{fig:all_mock_catalogues}.
In Table~\ref{table:cosmic_variance}, we give, for each mock catalogue,
the best achievable result (thus highlighting purely the effect of
choosing this mock catalogue) and the results one would obtain through
observation of this piece of the universe. Unknown Lagrangian domain,
redshift distortion  and incompleteness effects are added to the
considered mock catalogue. The problems of mass-to-light assignment
and the zone of avoidance are left apart for the sake of
clarity. Their imprint on the velocities should most likely remain the
same as we have shown in the corresponding previous sections, {\it
  i.e.} biasing for the first and increase of the scatter for the
second. 
Only the cases with the forementioned observational effects are represented in
Fig.~\ref{fig:all_mock_catalogues}. 

Visual inspection of lower scatter plots in
Fig.~\ref{fig:all_mock_catalogues} shows that volume finiteness is
likely making the $\Omega_\text{m,mes}$ measurement sensitive to the
``local'' $\Omega_\text{m}$ ($\Omega_\text{eff}$ in the table).
This assertion is supported by the estimation of $s$ and
$\Omega_\text{m}$ for {\it TrueDom} reconstructions 
given in Table~\ref{table:cosmic_variance}. Moreover, experiments conducted with
the spherical collapse model show that 
$\Omega_\text{m,mes}$ is indeed a weighted average between
$\Omega_\text{eff}$ and $\Omega_\text{m}$.

More specifically, 
reconstructed velocities in {\it 4k-mock7} (including observational
effects) are apparently giving the $\Omega_\text{m}$ of the simulation
but they present a large scatter
rendering the slope estimation dubious. Indeed, doing the same
reconstruction but without observational effects give a measured
$\Omega_\text{m,mes} = 0.40$, which is the exact average between the
simulation $\Omega_\text{m,simu}=0.30$ and
$\Omega_\text{eff} = 0.50$.\footnote{Spherical collapse  rather
  predicts $\Omega_\text{m,mes}=0.35$ for the same setup.} The aforementioned scatter is expected for
this mock catalogue: the velocity field is  
badly reconstructed near the observer in that case (middle panels)
because the local cosmic flow is higher than usual ($\sim$1000~\kms)
and the non-linearities are stronger. Thus the convexity of the
problem is lost on an extended region around the observer when the
reconstruction is conducted in redshift space (see \S~\ref{sec:redshift_distortion}).
A particularly saliant misreconstruction is given by the outflowing ``bubble'' at
the center which disappears in the reconstructed velocity field. The
size of the affected region is about 20~$h^{-1}$Mpc around the
observer in 4k-mock7 and thus limits the
number of objects having  both good reconstructed and observable peculiar
velocities. 

In an opposite way, velocities in {\it 4k-mock12} are reconstructed
with a better correlation, as shown by
Table~\ref{table:cosmic_variance}, but $\Omega_{\text{m}}$ measurement
is strongly weighted toward $\Omega_\text{eff}$. These two ``features'' are largely
due to the huge central void. First, MAK reconstruction and Zel'dovich
approximation are known to work better in low density regions 
and being centered on a void results in inhibiting blueshift
distortion as galaxies are principally going away from the
observer, rendering the reconstruction problem convex in
Eq.~\eqref{eq:zassign}. Second, the low density region  largely affects the 
statistical velocity distribution, which in this case leads to a
measured $\Omega_{\text{m,mes}}$ weighted more strongly towards
the $\Omega_\text{eff}$ of 4k-mock12.\footnote{The spherical collapse
  model would predict a measured $\Omega_\text{m,mes}=0.26$ and this
  is in good agreement with the value measured when no observational effects are
  injected in the mock catalogue.}  This leads
us to a $\Omega_{\text{m,mes}}$ that is nearer $\Omega_\text{m,simu}$ in
4k-mock12 than the mean matter density of the whole simulation. The
volume finiteness also produces an apparent offset between
reconstructed velocities and measured ones. This is expected as
doing a statistical analysis on a finite volume catalogue must
introduce a selection bias effect. 
We have indeed checked that the point set
$\left\{(v_{\text{r},i},\psi^\text{s}_{\text{r},i})\right\}$, obtained through a MAK
reconstruction applied on 4k-mock12, is a subset
of the corresponding set built from a reconstruction on 8k-mock12.
Looking at our ``standard'' {\it 4k-mock6}, one can note that the
simulated velocity distribution is generally more symmetric according to
the null velocity than for the two other mock catalogues, with no
visual bias while comparing reconstructed velocity to simulated
velocity. This supports the initial assertion linking
$\Omega_\text{m,mes}$ to $(\Omega_\text{m,simu},\Omega_\text{eff})$ and
the asymetric distribution of velocities. Potentially, one could recover the
true $\Omega_{\text{m}}$ of the Universe (or here the simulation) from
the measured velocities of any catalogues by predicting how the velocity
distribution asymmetry is linked to local density contrast. However,
the simplest, and more robust solution, would still be to extend the
depth of current catalogues to reach a volume where velocities are
normally distributed.

From a prediction point of view, comparing visually the velocity fields inside
the white circles show that, if we know $\Omega_{\text{m}}$, we reconstruct
plausible velocity fields for the three mock catalogues. Outside the
white circles, the reconstructed velocity field is nearly completely
uncorrelated compared to the simulated one as we have discussed in the
previous section. It must be noted that the velocity field goes
smoothly to zero (green colour) on the edge of all mock catalogues: this is
an expected side effect of
the homogeneous padding which tends to smooth out any fluctuation on
the edge (velocity and density field).

\section{Velocity measurement errors}
\label{sec:malmquist}

\subsection{The need for a likelihood analysis ?}
\label{sec:err_correlated}

\begin{figure*}
  \begin{tabular}{lccc}
    & {\bf\large Cosmic variance} & {\bf \large Lagrangian domain} &
    {\bf\large Redshift distortion} \\
    &\includegraphics[width=.23\linewidth]{fig15j} &
    \includegraphics[width=.23\linewidth]{fig15l} &
    \includegraphics[width=.23\linewidth]{fig16l} \\
    \\    
    {\Large $\mathfrak{L}$} & {\Large $ \Omega_\text{m} = [0.28,0.38]$} &
    {\Large $\Omega_\text{m} = [0.28,0.34]$} &
    {\Large $\Omega_\text{m} = [0.18,0.34]$}  \\
    {\Large 1.5$\sigma$} &  {\Large $ \Omega_\text{m} = [0.28,0.35,0.43]$} & {\Large
      $\Omega_\text{m} = [0.26,0.32,0.38]$} & {\Large $\Omega_\text{m} = [0.15,0.30,0.68]$} \\[1cm]
    & {\bf\large Incompleteness} & 
      \begin{minipage}[c]{3cm} 
        \begin{center}
          {\bf\large Observational errors} ($\correction{\sigma_{\mu}} = 0.17$)          
        \end{center}
      \end{minipage}
         \\[.3cm]
    & \includegraphics[width=.23\linewidth]{fig17j}
    & \includegraphics[width=.23\linewidth]{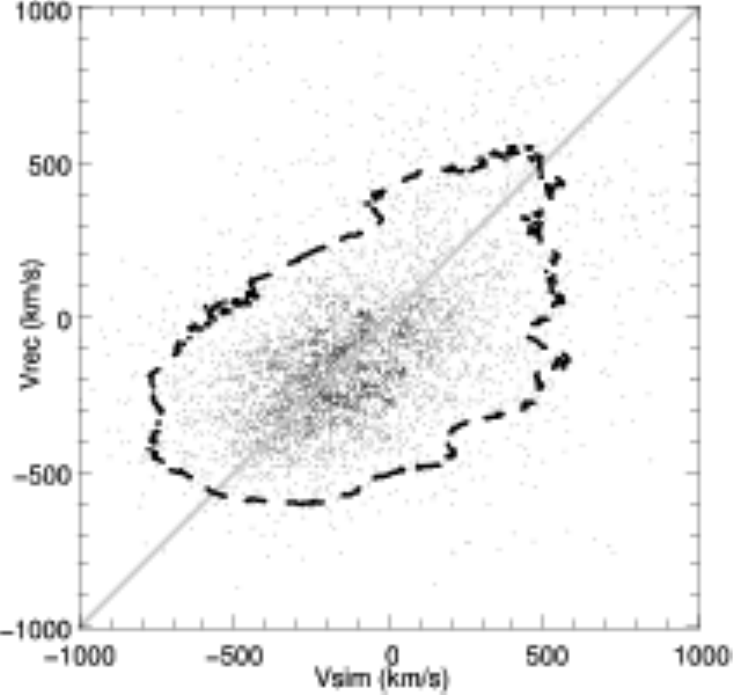} \\[.4cm]
    {\Large $\mathfrak{L}$} & {\Large $\Omega_\text{m} = [0.13,0.31]$} & {\Large $\Omega_\text{m} = [0.10,0.30]$}\\
    {\Large 1.5$\sigma$} & {\Large $\Omega_\text{m} = [0.20,0.38,0.79]$} & {\Large
      $\Omega_\text{m} = [0.08,0.78,1]$} \\[1cm]
  \end{tabular}
  \caption{\label{fig:error_correl} This figure gives the evolution of
  the scatter distribution and of the measurement of $\Omega_\text{m}$
  using it while more and more observational effects are added to 
  8k-mock6 catalogue. All measurements of $\Omega_\text{m}$ are given
  between brackets and are sorted as follows. For measurements obtained
  through the likelihood analysis, labelled by $\mathfrak{L}$, the
  first number corresponds to $\mathfrak{L}_\text{min}$ and the second
  to $\mathfrak{L}_\text{max}$. For measurements obtained using the
  1.5$\sigma$ method, the first number corresponds to $s_\text{min}$,
  then $s_\text{med}$ and finally $s_\text{max}$. The last (lower right) panel uses the full
  likelihood function of Eq.~\eqref{eq:likely_onepoint}. All others
  use a restricted likelihood analysis with $\sigma_0/e = 0$, which is
  nearly equivalent to using Eq.\eqref{eq:restricted_likely} for each
  $(v_{\text{r},i},\psi_{\text{r};i})$ pair. \correction{The
    $1.5\sigma$ isocontour has been plotted with a thick dashed line in the last panel.} }
\end{figure*}

All the effects already described in this paper are present in a redshift
catalogue. Though we expect most of the observational biases should be
independent, some of them may correlate and give worse systematic
errors. We present in Fig.~\ref{fig:error_correl} the progressive
deterioration of the velocity-velocity comparison for {\it 4k-mock6}
based on a reconstruction conducted on {\it 8k-mock6}. The effects are
piled up from left to right. The $\Omega_\text{m}$ measurements for the 1.5$\sigma$ method are indicated below each panel. The obvious conclusion is that the measurements are 
progressively affected but that no extra correlated error seems to
happen when mixing the effects. Another fortunate event is that bias seems to counterbalance themselves to give in the end a nearly unbiased result (last but one panel). Going from {\it TrueDom}/Real to Redshift
tends to decrease $\Omega_\text{m}$ as 
has been seen previously. On
the contrary, injecting incompleteness pushes the measurement to
higher $\Omega_\text{m}$ as we have noticed in
\S~\ref{sec:incompleteness}. The 1.5$\sigma$
 method seems to give the right $\Omega_\text{m}$ value in
all cases, which means that we should be able to use it on galaxy
catalogues provided we have sufficient precision on velocity
measurements. However, looking at the last panel (bottom right) of
Fig.~\ref{fig:error_correl} shows that injecting random velocity
measurement errors (\correction{here we intruduced an optimistic error of \correction{8\%} of the
distance to the object, corresponding to an error on distance
modula of $\sigma_\mu=0.17$}),
renders slope estimation much more
difficult. In that case, the measured $\Omega_\text{m}$ is severely
biased. This is expected as the 1.5$\sigma$ method relies mostly on the
central part of the scatter, which in turn is the one that is the most
affected by random errors. This leads to a circularization of the
1.5$\sigma$ isocontour and thus a completely wrong estimation of the slope. On the
other hand, looking at the global structure of the scatter shows that
the right slope is still hidden in the data, but one should then take
into account the tails of the distribution.
This last test shows the limit of a direct velocity-velocity
comparison in real cases. It might be possible to
recover the original distribution of the scatter by deconvolving from
the noise. However, it seems to be a difficult operation and we prefer
to first try a maximum likelihood approach. Its main advantage would
be to work using distances, thus rendering the
error in measurements more tractable. 

\subsection{Maximum likelihood analysis}
\label{sec:likelihood}
\newcommand{\muset}{\mathfrak{M}}
\newcommand{\redset}{\mathfrak{Z}}
\newcommand{\muerrset}{\mathfrak{S}}
\newcommand{\vset}{\mathcal{V}}
\newcommand{\psiset}{\mathfrak{P}}

Observations of galaxies first give us access to their distances and
not their peculiar velocities. A method based on distances to
make a comparison between a model and observations is potentially less
sensitive to distance measurement errors. Indeed, by comparing directly
distances, one has a small relative error on each measurement instead
of a huge one when peculiar velocities are considered. Below, we
discuss galaxy selection bias and zero-point calibration errors in distance
measurements while keeping the notation of \cite{StraussWillick}. 

{\it Presentation of the Bayesian chain} -- For the Tully-Fisher (TF) relation, one makes an estimate of the absolute
magnitude of a galaxy as a function of its linewidth: the slope between the two
quantities can be biased because the sample is limited in magnitude
\citep{StraussWillick}. This effect which is known as {\it selection
  bias} is purely statistical and if not correctly taken into account
can lead to large systematic errors. Using these absolute magnitudes,
occasionally combined to form groups of galaxies, and the apparent
magnitudes of the same group, one builds the distance modulus
\begin{equation}
  \mu(r) = m(r) - M = 5 \log_{10}\left( \frac{r}{10~pc} \right)
\end{equation}
with $r$ the distance of the considered object (group of galaxies or galaxy).
In addition to the forementioned statistical bias, peculiar
velocity obtained from redshift positions through a Lagrangian
reconstruction, here MAK, are sometimes very noisy, as  shown in
Fig.~\ref{fig:error_correl}. Another more subtle effect is introduced
by the Gaussian distribution of our velocity sample that we are going
to analyze. We need to take care of this ``selection bias'' to avoid
being spoiled by eventual large reconstruction errors present for
objects with a high velocity. Thus we need a Bayesian approach to
account for all these statistical effects. 

\begin{figure*}
  \begin{center}
    \includegraphics[width=.9\hsize]{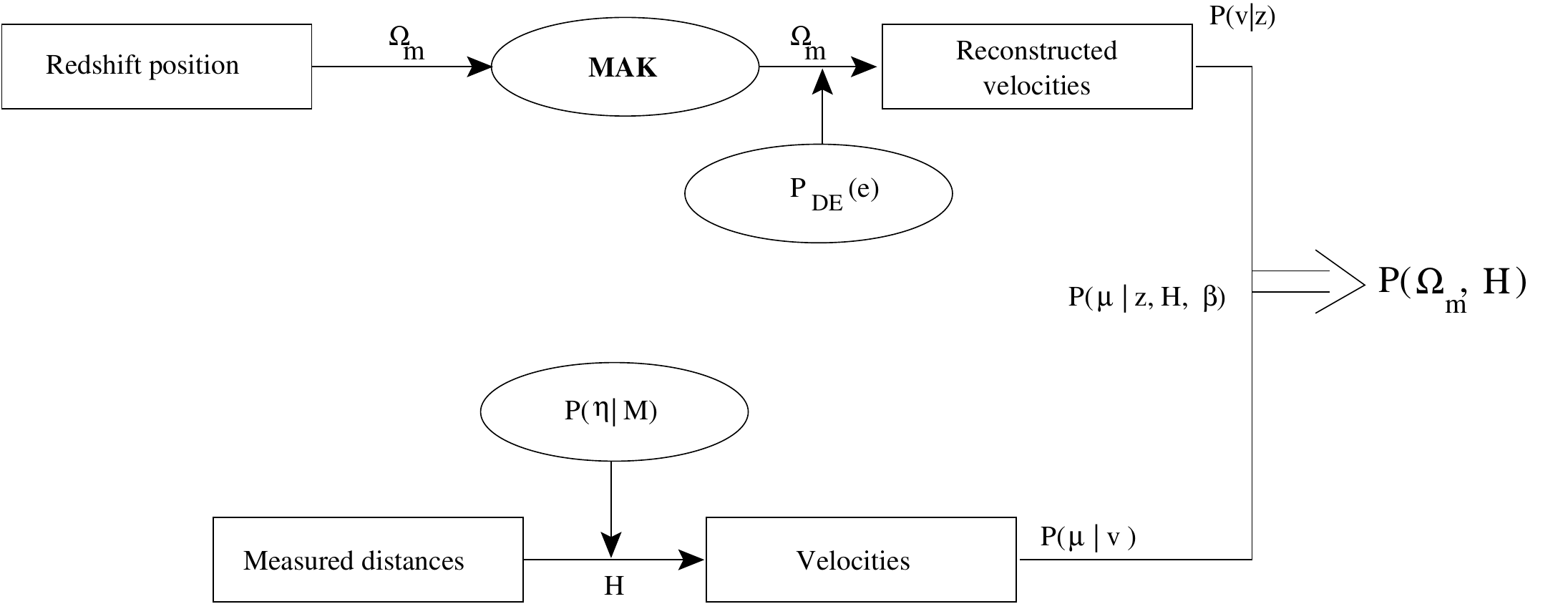}
  \end{center}
  \caption{\label{fig:bayesian_chain} {\it Maximum likelihood
      analysis} -- This sketch illustrates the bayesian chain used
    to establish the likelihood function. The input data are located
    on the left and the output posterior distribution $P(\beta,H)$ on the right.}

\end{figure*}

In principle, the likelihood
function gives a probability for the data, {\it i.e.} here redshift
positions $\redset =\{{\bf z}_i\}$, with $i$ running from 1 to $N$, and
distance moduli $\muset =\{\mu_i\}$, assuming some model described by the vector
parameter $p$. Additionally we assume that we have an estimation of
measurement errors on $\muset$ through the set $\muerrset$. The exact
description of $\muerrset$ will be given in the next paragraph. 
\correction{Typically errors on redshift measurements are of the order of
50-60~\kms. This means that we can consider them as negligible
if we consider objects farther than $R_z=6-10$~\Mpch. The volume enclosed by
the sphere of radius $R_z$ is, in any case, also poorly reconstructed
because of the singularity introduced by redshift distortions near the
observer (\S~\ref{sec:redshift_distortion}). In the following
analysis, we will consider redshift measurements as negligible by
avoiding the objects located at less that 10~\Mpch{} from the observer, thus
we have:
}\footnote{\correction{Though it is in theory possible to avoid this
  hypothesis, it is in practice highly difficult for computational reason as
  one would need to run several MAK reconstructions to evaluate the
  extra integral that would be needed in Eq.~\eqref{eq:likely_basic_relation}.}}
\begin{equation}
  P(\muset,\redset|p,\muerrset) \propto
  P(\muset|\redset,p,\muerrset) = \mathfrak{L}(p) \label{eq:likely_basic_relation}
\end{equation}
The end of this section is devoted to computing the right hand part of
this equation. To achieve this, we will decompose the probability into
small pieces:
\begin{multline}
  P(\muset|\redset,\muerrset,p) = \\
  \iiint\limits_{\muset_r,\vset,D}
  P(D) P(\muset|\muset_r,\muerrset,D,p)\\
  \times P(\muset_r|\vset,\redset,p)
  P(\vset|\redset,p)\;\text{d}\muset_r\text{d}\vset\text{d}D \label{eq:bayesian_chain}
\end{multline}
with $\muset_r=\{\mu_{1,r},\ldots,\mu_{N,r}\}$ representing the ``true'' distance
moduli, with $\mu_{i,R} \in [-\infty,+\infty]$ and
$\vset=\{v_1,\ldots,v_n\}$ the ``true'' object peculiar velocities. 
$P(\muset|\muset_r, \muerrset, p)$ is the probability of measuring the set of distance
moduli $\muset$ given that the real set of distance moduli is
$\muset_r$ and the expected error on the measurement is given by
$\muerrset$. $P(\muset_r|\vset,p)$ is the probability of
obtaining the set of distance moduli $\muset_r$ given the
reconstructed velocities $\vset$.
$P(\vset|\redset,p)$ is the probability the velocities
are well reconstructed from the redshift data $\redset$. The
probability $P(D)$ is going to be introduced in the last paragraph to
account for uncertainty in the calibration of the Tully-Fisher
relation. All those probabilities are computed assuming the model parameters $p$.
We will establish the likelihood
function $\mathfrak{L}(p)$ in three steps:
\begin{itemize}
\item[-] First, the error distributions linked to observations are
  considered to get an unbiased distance estimator for groups. This
  analysis yields the probability $P(\mu_i|\mu,\sigma_{\mu,i},p)$.
\item[-] Second, the errors on reconstructed velocities are considered
  to compute $P(v|\redset,p)$.
\item[-] Last, the two analyses are merged as given above to produce the likelihood
  function which gives the posterior distribution of $\beta$ and the
  Hubble constant $H$.
\end{itemize}
A picture of the above Bayesian chain is given in
Fig.~\ref{fig:bayesian_chain}. 

{\it Distance modulus error distribution} -- To establish the likelihood function comparing the measured distance
to the reconstructed velocity field, we assume the distance catalogues
are obtained using the inverse TF relation \citep{Shay95}, 
\begin{equation}
  \eta^0(M) = -e (M + D)\;,
  \label{eq:inverse_tully}
\end{equation}
where $M$ is the absolute magnitude of the considered galaxy,
$\eta_0(M)$ is its predicted linewidth, $e$ is the slope, and $D$ is
the zero point calibration (the latter two are assumed to be 
known exactly). 
It is known that inverse TF is less sensitive to the selection bias as
compared to forward TF \citep{StraussWillick}. 
Observational data show that the differences between the predicted
linewidth $\eta^0(M)$ and the measured linewidth $\eta$ for an object
of absolute magnitude $M$ are Gaussian distributed\footnote{In fact,
  in writing Eq.\eqref{eq:error_tully}, two effects are mixed: the error
  on the measurement of linewidth, which may reach $10\%$ because of
  the uncertainty in galaxy inclination correction, and the intrinsic modeling errors of the TF
  relation itself.} \citep{Pizagno06,Tully00}. Thus, the probability
of measuring the linewidth $\eta$, given that the object has an
absolute magnitude $M$, and assuming that the TF relation $\eta^0(M)$
is known, is 
\begin{equation}
  P(\eta|M,e,D) = \frac{1}{\sqrt{2\pi}\sigma_\eta(M)} \mathrm{e}^{-\frac{\left(\eta - \eta^0(M)\right)^2}{\sigma_\eta^2(M)}}
  \label{eq:error_tully}
\end{equation}
with $\sigma_\eta(M)$ the linewidth estimation error for the absolute magnitude $M$.
Distance catalogues are composed of estimated distance moduli $\mu_e$
from the inverse TF relation. These estimated distance moduli are
built from the statistics on a single group. Therefore, the joint
probability of having a galaxy in a group with both a linewidth $\eta$
and an absolute magnitude $M$, assuming the TF relation $\eta^0(M)$,
is: 
\begin{equation}
  P(\eta,M|e,D) = F(M) P(\eta|M,e,D)\;,
\end{equation}
where $F(M)$ is the normalized absolute luminosity function of the
group. \footnote{Note that the selection function is assumed to be
  independent of $\eta$ and is hence absorbed in $F(M)$.$F(M)$
  corresponds to $\Phi(M) S(M,\eta)$ in \cite{StraussWillick} notation,
  e.g. eq. (188).} 

The estimator for the distance modulus is given by:
\begin{equation}
  \mu_e = m - M_e(\eta) = M + \mu_0(r) + D' + \frac{\eta}{e'}\;,
\end{equation}
where $D'$ and $e'$ are the estimated inverse TF parameters of
Eq.~\eqref{eq:inverse_tully} and $\mu_0(r)=5\log(\frac{r}{10 pc})$ the
true distance modulus of the considered group. 
The conditional probability that the estimated distance modulus for
the group is $\mu$, assuming that the estimated Tully-Fisher
parameters are $e'$ and $D'$ and that the real parameters for this
group are $e$ and $D$, can be written as 
\begin{multline}
    P(\mu|\mu_0(r), e, e', D, D') =  \left\langle \delta_{\text{D}}\left(\mu - \mu_e\right)\right\rangle_{\text{group}}\\
    \shoveright{ = \left\langle\delta_{\text{D}}\left(\mu - \mu_0(r) + M + D' + \frac{\eta}{e'}\right)\right\rangle_{\text{group}}} \\
    = \int\limits_M \frac{e' F(M)}{\sqrt{2\pi}\sigma_\eta (M)} \mathrm{e}^{-\frac{\left(e'(\mu_0(r)-\mu) + e' D' - e D + (e' - e)M\right)^2}{2\sigma_\eta^2(M)}}~dM\,.
\end{multline}
While working with the inverse TF relation, one can assume that the slope
$e'$ is completely determined and $e'=e$. Since the observed
$\sigma_\eta(M)$ varies little with $M$, it is chosen to be equal to a
constant $\sigma_0$. The previous probability reduces to 
\begin{equation}
  P(\mu|\mu_0(r), e, D, D', \sigma_0) = \frac{e}{\sqrt{2\pi} \sigma_0} \mathrm{e}^{- e^2 \frac{\left(\mu-\mu_0(r) + D'-D\right)^2}{2 \sigma_0^2}}\text{.}
\end{equation}
Though the slope $e'$ is well determined, the zero-point calibration
$D$ may still be affected by non-negligible errors.\footnote{The latest
calibration is given in \cite{TullyVoid07}.} The set
describing errors on distance moduli is thus
$\muerrset=\{\sigma_{0,1}/e,\ldots,\sigma_{0,N}/e\}$\correction{$=\{\sigma_{\mu,1},\ldots,\sigma_{\mu,N}\}$}. The
error on this calibration will affect the distances globally. As a
first approximation we model the error on the zero point by a Gaussian
centered on $D$ with a standard deviation of $\sigma_\text{\sc d}$.  

{\it Linking distance modulus to velocity} -- The second probability
function in Eq.~\eqref{eq:bayesian_chain} is
$P(\muset_r|\vset,\redset,p)$, which is actually a distribution
linking the velocities and redshifts to distance modulus. This
principally corresponds to a change of variable and we give directly
the expression of it, which is inspired by Eq.~\eqref{eq:cz}:
\begin{multline}
  P(\muset_r|\vset,\redset,p) = \\
  \prod\limits_{i=1}^{N}
  H 10^{\mu_{r,i}/5} \delta_\text{D}\left(z_i - v_i - 10\text{ pc}\times H 10^{\mu_{r,i}/5}\right) 
\end{multline}

{\it Reconstructed velocity distribution } -- We are now going to
establish the expression of $P(v|\redset,p)$ with the vector of
parameters of our chosen model
$p=(H,\beta,B_v,\sigma_v,\gamma_{*},e)$ -- $\sigma_v$ and $\gamma_{*}$ are
going to be introduced in the next immediate paragraphs.
One may decompose $P(v|\redset,p)$ that way
\begin{equation}
  P(v|\redset,p) = \int_{\psiset}
  P(v|\psiset,p) P(\psiset|\redset,p)\;\text{d}\psiset\,,
\end{equation}
with $\psiset=\{\psi_{r,i}\}$ the reconstructed displacements. As MAK
reconstruction is deterministic once $\beta$ has been assumed (\S~\ref{sec:redshift_distortion}),
the second probability distribution is simply given in our case by
\begin{equation}
  P(\psiset|\redset,p) = \prod\limits_{i=1}^n \delta_\text{D}\left(\psi_{r,i} - \psi_i\left(\redset,\beta\right) \right)
\end{equation}
with $\psi_i$ representing the MAK reconstructed displacement of the $i$-th object, being a function of all redshift coordinates and $\beta$. 
Thus, studying $P(\vset|\redset,p)$ reduces to examine
$P(\vset|\psiset(p_0),p')$, 
with $p'=(H,\beta',B_v,\sigma_v,\gamma_{*})$,
$p_0=(H,\beta_0,B_v,\sigma_v,\gamma_{*})$, $\beta_0$ being the assumed growth
factor to compute the set $\psiset(p_0)$ using the redshift
reconstruction. $P(\vset|\redset,p)$ and
$P(\vset|\psiset(p_0),p')$ equalizes only if $p=p'=p_0$.
Thus one needs a several redshift
reconstructions to build the probability function
$P(\vset|\redset,p)$. Working with the intermediary set $\psiset$ is easier
than with $\redset$, we thus put the reduced likelihood function:
\begin{multline}
  \mathfrak{L}'_{\beta_0}(p') = \\
  \iint\limits_{\muset_r,\vset}
  P(\muset|\muset_r,\muerrset) P(\muset_r|\vset,\redset,p) P(\vset|\psiset(p_0),p')\;\text{d}\muset_r\text{d}\vset\label{eq:effective_likelihood}
\end{multline}
and we are going to establish the expression of the elementary
probability function $P(v_r|\psi_r,p)$ which will yield
\begin{equation}
  P(\vset|\psiset(p_0),p) = \prod\limits_{i=1}^N P(v_{r,i}|\psi_{r,i},p)
\end{equation}
assuming statistical independance of all $\{v_{r,i},\psi_{r,i}\}$
 duets, and that  $\psiset(p_0)$ is obtained using a redshift
 reconstruction for which $\beta=\beta_0$. $\mathfrak{L}'$ may be
 written in a factorized way:
\begin{multline}
  \mathfrak{L}'_{\beta_0}(p') = \\
    \prod\limits_{i=1}^N\,\, \iint\limits_{\mu_{r},v_{r}}
    P(\mu_{i}|\mu_r,\sigma_{0,i}/e) P(\mu_r|v_r,z_i,p) P(v_r|\psi_{r,i},p)\;\text{d}\mu_r\text{d}v_r\text{.}
\end{multline}
The computation of $\mathfrak{L}'$ is clearly helped using this
factorized form. We may now concentrate on the third probability
function of the above equation.

As has been established in \S~\ref{sec:mak_error}, the distribution of errors
on the reconstructed velocity field is the Lorentzian
\begin{equation}
  P_\text{DE}(e_{\psi v}) \propto \frac{1}{1 + \left(\frac{e_{\psi v}}{B_v}\right)^2}\text{,}
\end{equation}
where $B_v=86$~\kms (redshift reconstruction), with $e_{\psi v}$ the
distance between the reconstructed velocity $\beta\Psi_r$ and the true
velocity $v_r$. This formulation is different from saying
that the reconstructed velocity is affected by error when compared
to the true velocity, and permits  some errors in the MAK
reconstructed displacement field. As has been seen in
\S~\ref{sec:cosmic_var}, the reconstructed velocities may also contain
an extra offset that needs to be removed while measuring
$\beta$. The error distance $e_{\Psi v}$ is thus
\begin{equation}
  e_{\Psi v} = \alpha_{*} v_r - \beta_{*} \Psi_r + \gamma_{*}
\end{equation}
with
\begin{equation}
  \alpha_{*}^2 + \beta_{*}^2=1 \text{ and } \beta=\beta_{*} /
  \alpha_{*}
\end{equation}
and $\gamma_{*}$ to account for a potential spurious offset in
reconstructed velocities. From linear theory \citep{Peebles80}, we
know that the line-of-sight component of the velocity field must be
distributed like a Gaussian function. We now assume that the absolute
probability for an object to have a velocity $v$ is given by a
Gaussian distribution:
\begin{equation}
  P_\text{vel}(v|p) = \frac{1}{\sqrt{2\pi} \sigma_v} \mathrm{e}^{-\frac{v^2}{2 \sigma_v^2}}\text{.}
\end{equation}
It must be noted that it is likely that the observational data does not
encompass a sufficiently large volume so that measured velocities
follow this law. Moreover, this prior is of some importance when we
have to deal with highly scattered data. The shortcomings of such an
approach will be discussed in the next section. One can recover the
standard uniform prior on velocities by taking the limit
$\sigma_v\rightarrow +\infty$ in the next equations.  Assuming
$e_{\psi v}$, as a random variable, is independent of $v_r$ and these two quantities are
themselves statistically independent from $\beta$ and $\gamma_{*}$, we may now write
the joint probability of reconstructing $\psi_r$, having a true
velocity $v_r$:
\begin{multline}
  P(v_r,\psi_r,\beta,\gamma_{*}|B_v,\sigma_v) \\
  =  {\beta_{*} P_\text{DE}(e_{\Psi
    v}|B_v,\sigma_v)\times}\\
\shoveright{P(\beta,\gamma_{*}|B_v,\sigma_v)
P(v_r|B_v,\sigma_v) P(\psi_r|B_v,\sigma_v)} \\  
  = \beta_{*} C(B_v,\sigma_v) P(\beta,\gamma_{*}|B_v,\sigma_v)
  \frac{P(\psi_r|B_v, \sigma_v)\, \mathrm{e}^{-\frac{v_r^2}{2 \sigma^2_v}}}{1 +
      \left(\frac{\beta_{*}\psi_r-\alpha_{*} v_r +
          \gamma_{*}}{B_v}\right)^2}\,,
    \label{eq:proba_main}
\end{multline}
where $C$ is a function eventually depending on $B_v$ and $\sigma_v$. 
The conditional probability that the true velocity is $v_r$ given
the reconstructed displacement $\psi_r$ is now exactly
\begin{multline}
  P(v_r|\psi_r,p) = \\
  \frac{
    \mathrm{e}^{-\frac{v_r^2}{2 \sigma^2_v}} \left(1 +
      \left(\frac{\beta_{*}\psi_r-\alpha_{*} v_r +
          \gamma_{*}}{B_v}\right)^2 \right)^{-1}
    }{
      \int_{v=-\infty}^{+\infty} \mathrm{e}^{-\frac{v^2}{2 \sigma^2_v}} \left(1 +
      \left(\frac{\beta_{*}\psi_r-\alpha_{*} v +
          \gamma_{*}}{B_v}\right)^2 \right)^{-1}\;\text{d}v
    }\;.
  \label{eq:restricted_likely}
\end{multline}
The denominator of the right hand part of this equation must be
computed numerically.\footnote{This function is known as a Voigt
profile.}  It can be shown that, in the limit $\sigma_v\rightarrow
+\infty$, $P(v_r|\psi_r,\beta_{*},\gamma_{*})$ reverts to a pure
Lorentzian form.

{\it Merging the probability distributions} -- 
We may now establish the ``elementary'' conditional probability for an object $i$ to
get a measured distance $\mu_i$  given that its reconstructed
displacement is $\psi_{r,i}$, its redshift is $z_i$, the error on the
linewidth measurement is $\sigma_{0,i}$ and the model parameters are
$p'$ in the notation of this section:
\begin{multline}
  P(\mu_i | \psi_{r,i}(p_0), z_i, \sigma_{0,i}, D', p') \\
   = \iint\limits_{v,\mu_r} P(\mu_{i}|\mu_r,\sigma_{0,i},D',p')
   P(\mu_r|v,z_i,p') \\
     \shoveright{\times P(v|\psi_{r,i}(p_0),p')\;\mathrm{d}\mu_r\mathrm{d}v} \\
   \propto \frac{e H}{\sigma_{0,i}} \int_{\mu_r} 10^{\mu_r/5} \text{e}^{-\frac{e^2}{2
     \sigma^2_{0,i}}\left(\mu_r-\mu_i+D-D'\right)^2} \\
  \times P\left(v=(z_i-10\text{ pc}\times H 10^{\mu_r/5})|\psi_{r,i}(p_0),p'\right)\;\mathrm{d}\mu_r \label{eq:likely_onepoint}\text{,}
\end{multline}
with $p'=(H,\beta',B_v,\sigma_v,\gamma_{*})$, $\psi_{r,i}(p_0)$ being
computed assuming the parameters $p_0$. Looking closely at this
probability, one may notice that changing $D\rightarrow D'=D+\Delta$ is equivalent
to changing $H\rightarrow H'=H\, \text{exp}(\Delta/5)$. Thus the
uncertainty in the zero point calibration translates only in an
uncertainty on $H$ and not on the parameters of the model.

We may now write the full formal expression of $\mathfrak{L}'(p)$, as
already sketched in Eq.~\eqref{eq:effective_likelihood}. As specified
in the discussion we take $P(D')$ to be a Gaussian distribution
centered on $D$ and with a standard deviation
$\sigma_\text{D}$. Now
we may replace and get:
\begin{multline}
  \mathfrak{L}'_{\beta_0}(p') \propto\\
    \int_{D'=-\infty}^{\infty}
    \mathrm{e}^{-\frac{(D-D')^2}{2
        \sigma_\text{D}^2}} \prod_i P(\mu_i | \psi_{r,i}(p_0), z_i, \sigma_{0,i},
    e, p')\;\text{d}D\;,
\end{multline}
with $i$ running on objects of the catalogue. 
Assuming a uniform prior on $\beta$, $H$ and $\gamma_{*}$ and taking care of the
relation between $\mathfrak{L}'_\beta$ and $\mathfrak{L}$ as mentioned above, the
Bayes theorem permits us to write
\begin{equation}
  P(H,\beta,\gamma_{*}|\muset,\muerrset,\redset,e,D',B_v,\sigma_v) 
  \propto \mathfrak{L}(p) = \mathfrak{L}'_\beta (p)\;\text{.}
  \label{eq:posterior}
\end{equation}
We now have access to the posterior distribution of $(H,\beta,\gamma_{*})$.

\subsection{Results}
\label{sec:lik_result}

The results of measuring $\Omega_\text{m}$ using the maximum
likelihood estimator are presented in the tables using the label
$\mathfrak{L}$. 

Except in the case where we consider observational
errors, we use a simplified version of $\mathfrak{L}$ by taking $\sigma_{0,i} = 0$.
While it would have been natural to find the maximum of the likelihood for {\it
all} parameters (including $B_v$, $\sigma_v$, $\gamma_{*}$), we quickly noticed that
it was leading to unacceptably biased measurements and to an
unnecessary increase of the parameter space. Moreover, the results
quite strongly depends on $\sigma_v$ and $\gamma_{*}$, especially when the
reconstruction noise becomes high as in redshift reconstructions (see
Appendix~\ref{sec:stat_bias}). We thus propose to discuss the values
obtained by setting $\gamma_{*}=0$, $B_v=90$~\kms and choosing two
values for $\sigma_v$. First, we use linear
theory to predict the average velocity dispersion of haloes in the
universe, this leads to take $\sigma_v=326$~\kms (the
$\Omega_\text{m}$ measured that way is labelled $\mathfrak{L}_\text{max}$). Second, $\sigma_v=+\infty$ is used to
check the influence of recovering a uniform prior on the velocity distribution (labelled
$\mathfrak{L}_\text{min}$, respectively). 

By looking at all tables of this paper, we noticed that the difference
between the two measured $\Omega_\text{m}$ is mostly 
following the interval defined by $s_\text{min}$ and
$s_\text{max}$. We were expecting such a behaviour ($\sigma_v$ is more
or less controlling the statistical bias of the likelihood function)
but not that it would so clearly follow the other method.  The more
the scatter is important, the more the measurement becomes imprecise as
expected.  It must however be noted that on average the measure
$\mathfrak{L}_\text{max}$ suffers less systematic bias than
$\mathfrak{L}_\text{min}$. This behaviour is supported by the tests
conducted in Appendix~\ref{sec:stat_bias}.

The seemingly well estimated $\Omega_\text{m}$ in the lower right panel of
Fig.~\ref{fig:error_correl} has been computed using the full
likelihood analysis.
Actually, compared to the 1.5$\sigma$ method for
which the measured slope is undefined, $\mathfrak{L}_\text{min}$ and
$\mathfrak{L}_\text{max}$ are basically {\it the same} as when {\it no
observational errors} are introduced. 

The correction based on a Gaussian velocity distribution assumption,
cannot be entirely trusted for 4k-mock7 and 
4k-mock12. As one may note in Fig.~\ref{fig:all_mock_catalogues}, the
velocity distribution is highly non-Gaussian in these cases. This renders incorrect
our distribution modeling in \S~\ref{sec:likelihood}. Looking at
Table~\ref{table:cosmic_variance}, we note that
though the measurements on ``Original'' reconstruction is not strongly affected,
we cannot say the same thing using data obtained from ``Full''
reconstruction. In the first case, the noise is sufficiently low so
that the prior does not have much importance whereas in the second
case the wrong modeling of the velocity distribution leads to a strong
error on the measured $\Omega_\text{m}$. Fortunately, the
scatter distribution presents  different types of properties
that lead to compatible measurements in
Table~\ref{table:cosmic_variance} between the maximum likelihood
($\sigma_v=+\infty$ to remove the Gaussian prior) and the 1.5$\sigma$
method.  For 4k-mock7 and 4k-mock12, the slope estimate is helped by probing
velocities with high magnitudes, leading to  less possibility of
systematic error on the slope. 

One is thus led to use a sufficiently deep distance catalogue to
ensure the velocity distribution is more or less Gaussian to be able
to apply the correction to the likelihood analysis. In this case, one
may rely on the value given by $\mathfrak{L}_\text{max}$. If on the
contrary, the velocity distribution is highly non-Gaussian, one must
use $\mathfrak{L}_\text{min}$. If possible, a visual inspection of the
velocity-velocity scatter plot must be conducted to give a 
check on the amount of statistical biasing.

\section*{Conclusion}
\label{sec:conclusion}

The Monge-Amp\`ere-Kantorovitch method has been applied with success
to reconstruct the velocity field and the density field of simulations
\citep{moh2005}, providing an interesting tool to apply to galaxy
catalogues in order to recover the dynamics of our local
universe. \correction{This method presents the interesting advantage
  of finding the exact solution of an approximated dynamical
  problem written in Lagrangian coordinates.  
  The Lagrangian description presents two major advantages. First, it gives 
  a real estimation of peculiar velocities for each galaxies or
  groups of galaxies, as opposed to a field description which would
  give an average value at a given spatial position (which is also
  possible to build using the Lagrangian description). Second, it
  permits us to use the Zel'dovich approximation, which gives better peculiar velocity prediction
  than linear Eulerian theory applied to the same dark
  matter density field. It means that we expect this method to give
  better results and more spatially resolved than, {\it e.g.}, the
  POTENT method \citep{POTENT} or velocity field reconstruction
  through spherical harmonics \citep{RegosSzalay89}. Now, most previous analyses of
  Lagrangian peculiar velocity reconstruction have been run mostly on
  particle catalogues coming from simulations.} However, galaxy catalogues are not
as simple, and the main problems are as follows:
\begin{itemize}
\item[(i)] Catalogues mostly provide redshift positions of galaxies and
  for a few objects their physical distances from us. 
\item[(ii)] The luminosity is the only known ``dynamical'' quantity for most objects
  in catalogues and so we need extrapolate the M/L relation for known
  objects to the ones that we do not know.
\item[(iii)] Incompleteness effects have to be taken into account: either because of
  magnitude limitation or due to extinction of objects by the galactic
  plane. 
\item[(iv)] The MAK reconstruction also needs the Lagrangian
  domain of the galaxy catalogue.  
\end{itemize}
All these biases and unknown quantities render the reconstruction
problem much more difficult than in simulations. We propose here both to
test the feasibility of such a reconstruction on galaxy catalogues and
the methods to overcome the problems that we have just cited. 
We have tried to address the following problems:
\begin{itemize}
  \item Reducing the introduced systematic errors due to unknown bias between mass and luminosity tracers.
    The dark mass
    can be either put uniformly into the catalogue or put in the
    detected haloes
    (\S~\ref{sec:diffuse_mass}). It appears that there exists an
    optimum way to distribute the mass, as can
    be seen in Fig.~\ref{fig:velocity_random}, which gives {\it unbiased}
    and {\it noiseless} reconstructed velocities, even though the exact location of 63\% of the
    mass in the universe remains unknown.
    In addition to the previous, global, problem, the relative mass
    distribution between objects in the catalogue is also uncertain as we do
    not know their true $M/L$. The induced systematic errors have been
    studied in \S~\ref{sec:ml_ratio} and we show that the naive
    approach corresponding to using
    $M/L=\text{constant}$ inevitably gives a large bias on
    reconstructed velocities. Even a reasonable guess, for instance
    the one proposed by \cite{Marinoni02}, is still significantly biased.
    This suggests some more work must be done on the $M/L$ relation,
    particularly on the high mass end. However, on the positive side, large random
    errors on $M/L$ does not yield any systematic effect and only
    increases the scatter in the velocity-velocity comparison.

  \item We proposed a slightly improved way to correct for incompleteness effects in
    galaxy catalogues and its effect on reconstruction. Though it has
    given good results, we do not expect this method to be completely
    bias-free as it presents the same deficiencies as the previous
    item. However, by enforcing the correction on the {\it mass} distribution,
    we managed to preserve the dynamics in the
    observational data in a better way than would be the case if we had enforced it on
    the {\it luminosity} distribution.

  \item 
    We investigated the eventual systematic errors in redshift
    reconstructions as proposed previously by \cite{MohTu2005} and
    which corresponds to the inverse redshift operator studied by \cite{ValentineTaylor2000}. It
    appears that, though the bias is small, $\Omega_\text{m}$ tends to
    be always underestimated.

  \item Two solutions to overcome the Lagrangian
    volume uncertainty for the case of finite volume catalogues have
    been investigated. The reconstruction method which gives better
    result seems to be {\it PaddedDom}. The other alternative, {\it NaiveDom},
    appears to bias the reconstructed velocities, especially in the
    case of a redshift reconstruction. 

  \item The efficiency of the correction for the zone of avoidance as
    proposed by \cite{Shay95} has been checked (\S~\ref{sec:zoa}). It
    appears that the correction is bias free and only introduce a
    small, but noticeable, additional noise for objects in the
    direction of the zone of avoidance.

  \item We checked that the resulting errors of each effect are
    uncorrelated so they only pile up without producing a strong
    additional bias.  It is fortunate that some
    observational effects produce complementary biases: incompleteness
    effect tends to overestimate $\Omega_\text{m}$ whereas redshift distortion
    underestimates $\Omega_\text{m}$. The resulting bias is thus not so
    important.

  \item We finally tried two estimators to measure $\Omega_{\text{m}}$ from both
    reconstructed displacement and distance measurement
    (\S~\ref{sec:malmquist}): the 1.5$\sigma$ and the maximum
    likelihood estimator.
    However, the first one is not able to work with noisy measured
    velocities, and the second one is badly affected by large
    distribution tails in redshift reconstruction. Adding a prior on
    the distribution of velocities in the catalogue helps to  reduce
    the bias at the cost of having a good measurement of the width of
    this distribution. A good estimate of $\Omega_\text{m}$ is thus
    rendered more problematic though we have shown that it should be feasible
    in principle.
\end{itemize}

We intend to continue this work in the following directions
\begin{itemize}
\item This method can be applied to make a measurement of
  $\Omega_{\text{m}}$ in NBG-8k/NBG-3k catalogues and in the upcoming
  6dFGS redshift and distance catalogues. 
\item A better comparison to the acoustic peaks of the CMB can
  potentially be obtained using the reconstructed displacement field
  \citep{EisensteinSeo06}. 
\item We can apply MAK reconstruction on SDSS and 2MASS catalogues to
  obtain the initial Lagrangian positions and velocities of objects
  in our local universe. This would render the possibility of a
  re-simulation of our local universe for the first time and check the
  MAK prediction and correction schemes on real observations.
\item We want also to improve the reconstruction itself and propose a
  new algorithm to include further gravitational effect during orbit
  reconstructions. This will never give us the internal structure of
  objects but potentially will give better reconstructed velocities
  while keeping the power of the MAK reconstruction.
\end{itemize}

\section*{Acknowledgements}

We are grateful to S.~Prunet, C.~Pichon for useful
discussions and comments on Maximum Likelihood methods. We would like
also to thank D.~Weinberg, M.~Chodorowski for useful discussions.
GL thanks the support and the hospitality of the Institute for
Astronomy (University of Hawaii). This work has been supported by the {\sc horizon}
project (http://www.projet-horizon.fr). 


\begin{appendix}

\section{Construction of a MAK mesh}
\label{app:subsampling}

MAK reconstructions requires a sampling of the matter
distribution with ``particles'' of equal mass corresponding to nodes of an
homogeneous mesh. When considering the simulation, one uses a full
periodic cubic mesh. However, in real galaxy catalogues, the relevant
lagrangian volume is a non-periodic compact subset inscribed in a
larger rectangular mesh. In that case, the assignment is performed only for
``particles'' belonging to this initial volume. Note that the
determination of this initial volume is by itself a great challenge
and a poor guess can have dramatic consequences.

Given a number of ``galaxies'', or tracers, for which the individual
masses $M_i$ are known and a choice of the mass resolution of the
MAK grid, $m_R$, the problem is now to determine how many ``particles'' have
to be assigned to tracer $i$. This number should be $n_{i}=M_i /
m_R$ which is rarely an integer. To address this issue, we construct
an integer function $\tilde{n}_i$ such that the quantity
\begin{equation}
  \chi^2 \equiv \sum_i \left(\frac{\tilde{n}_i m_R - M_i}{M_i}\right)^2
\end{equation}
is minimized given the constrain 
\begin{equation}
  \sum_i \tilde{n}_i = N_\text{MAK} \text{ ,}
\end{equation}
where $N$ is the total number of nodes on the MAK grid, such that $N_\text{MAK}
\times m_R$ is as close as possible to the total mass, $\sum M_i$.
The minimization of $\chi^2$ is performed iteratively until
convergence. Note that the solution of such a minimization is, in general, not
unique due to the possible permutations between objects of the same
mass. Due to this degeneracy, it is needed to shuffle randomly the
tracers prior to the minimization in order to avoid possible 
systematic effects.

Note finally that one must make sure that there is at least a 
few particles per tracer, $n_i \geq \alpha$ with $\alpha > 1$. 
This brings constraints on $m_R$ and
therefore on the size of the MAK mesh.  Unfortunately, it is not always possible to
have $\alpha > 1$ due to the prohibitive CPU cost it would imply
for the MAK reconstruction in the present paper.
To address this problem, we separate the catalogue into groups of
galaxies and field galaxies. For the groups, the
$\chi^2$ minimization is performed as
explained above, with a possible loss of the lightest ones since $n_i$
can still be smaller than unity. For the field galaxies, we use a simpler procedure as
follows. Given the mass $M_i$ of a galaxy $i$, a MAK
tracer is randomly assigned to it with occurence probability $M_i / m_R$.

\section{Tools for error analysis}
\label{app:err_analysis}

To check the accuracy of the reconstructions, we compute the moment of
the joint probability distribution
of the reconstructed velocities ${\bf v}_{\text{rec},i}$ of object $i$ and the simulated velocities of those objects ${\bf v}_{\text{sim},i}$. We write $\langle A\rangle$ the average 
of the quantity $A$
\begin{equation}
  \langle A \rangle =  \frac{1}{N} \sum_{i=1}^N A_i\quad\text{,}
\end{equation}

We define three second moments (after substraction of the average):
\begin{equation}
  \begin{array}{c}
    \begin{array}{lll}
      \sigma^2_\mathrm{r}=\langle v^2_\mathrm{rec}\rangle\text{,} &
      \sigma^2_\mathrm{s} =\langle v^2_\mathrm{sim}\rangle\text{,} &
      \sigma_\mathrm{r,s}=\langle v_\mathrm{rec} v_\mathrm{sim}\rangle
    \end{array}
  \end{array}
\end{equation}
 From these moments we can build the correlation
coefficient:
\begin{equation}
  r = \frac{\sigma_{\text{r,s}}}{\sigma_{\text{r}} \sigma_{\text{s}}}
\end{equation}
and the ratio between the width of the reconstructed field PDF (density or velocity) and
the width of simulated -- mock -- field PDF
\begin{equation}
  s = \frac{\sigma_{r}}{\sigma_{\text{s}}}\text{.}
\end{equation}
For these two quantities the optimum value is $1$.
Alternatively two other ``slope'' estimator of the reconstructed
velocities versus the simulated ones can be built from the above
momenta
\begin{equation}
  \begin{array}{ccc}
    s_\text{min} = \frac{\sigma_\text{r,s}}{\sigma^2_{\text{s}}} = s r & \text{and} & 
    s_\text{max} = \frac{\sigma^2_\text{r}}{\sigma_{\text{r,s}}} = s / r \text{.}
  \end{array}
\end{equation}
These two slopes are interesting when one makes an estimation of
$\Omega_\text{m}$ through $s$ and needs an evaluation of the
uncertainty. The two extra slopes determined using this way should,
ideally, be equal to $s$ but due to the lack of perfect correlation ($r
< 1$), they are actually different from it in realistic cases. In fact,
we have $s_\text{min} < s_\text{med} < s_\text{max}$.

Please note that we can define the relative dispersion
\begin{equation}
  \sigma^2 = \frac{\langle\left( v_{\text{rec}} - v_{\text{sim}}\right)^2\rangle}{\sigma^2_{\text{s}}} = 1 + s^2 - 2 s r\text{,}
\end{equation}
which is a measure of the noise-to-signal ratio: high $\sigma$
corresponds to low signal. Ideally, one wants $\sigma=0$.

\section{Simulating magnitude-limited catalogues}
\label{app:magnitude_limit}

Having only a halo catalogue, we must generate a ``galaxy catalogue''
including incompleteness effects.  The main difficulty in that construction is that the
distribution of galaxies in the universe is a non-trivial, non-linear
functional of the total matter density field.  For instance, bright
galaxies tend to concentrate in massive structures
\citep{Zandivarez06}.  It means that, though most of
the field galaxies are missed, the major groups can still be easily
seen due to the bright galaxies they contain. Thus the galaxy distribution
should mostly trace large haloes at large distances, potentially
introducing a bias in the reconstructed peculiar velocities if
incompleteness corrections are performed unwisely. 
In what follows, we generate mock galaxy catalogues like
NBG-8k/3k. To take properly into account the effects discussed above,
we separate groups of galaxies from field galaxies.
Groups are populated with galaxies following the universal Schechter
form for simplicity, but with a different normalization to
account for their non-trivial $M/L$.

Statistically, NBG-8k/3k catalogues are composed of galaxies measured in the B
band and distributed according to the Schechter form
\begin{equation}
  n(L)\;\text{d}L\simeq n_0  L^{-1} \mathrm{e}^{-L/L_*}\;\text{d}L\text{,}
  \label{eq:lum_schechter}
\end{equation} 
with $L_*\simeq 5.7\times 10^{10}\text{ L}_\odot$ and $n_0 \simeq 0.03\;h^3\text{Mpc}^{-3}$.
Moreover, the NBG-8k catalogue is complete above $3 \times 10^{9}-4 \times10^{9}\text{
  L}_\odot$ inside a sphere of radius $d_\text{comp} = 12\;h^{-1}$Mpc.
As the mean ``galaxy'' (particle) density in the simulation is
$n_\text{sim} = 0.26\;h^3\text{Mpc}^{-3}$ and about $n_\text{cat} =
0.08\;h^3\text{Mpc}^{-3} \simeq 0.30\, n_\text{sim} $ 
in NBG-8k, we must dilute the simulation to get a mock catalogue
similar to NBG-8k. The luminosity $L_\text{G}$
of a detected galaxy at a distance $d$ from the observer must satisfy
the constraint
\begin{equation}
      L_\text{G} > 4 \pi l_\text{cut} d^2
      \label{eq:visibility_cons}
\end{equation}
with $l_\text{cut}$ the minimum flux detectable by the observer.
The fraction of galaxies detected at the distance $d$ in the galaxy
mock catalogue is thus
\begin{equation}
  f_\text{field}(d) =
  \left\{
    \begin{array}{ll}
      0.30 & \text{if } d < d_\text{comp} \\ 
      \frac{\int_{4 \pi l_\text{cut} d^2}^{\infty}
        n(L)\;\text{d}L}{\int_{L_\text{min}}^{\infty}
        n(L)\;\text{d}L} & \text{otherwise}
    \end{array}
  \right.
  \label{eq:frac_schechter_form}
\end{equation}
with $l_\text{cut}$ the minimum flux detectable by the observer. The
fraction is saturated at $0.30$ to follow the dilution constraint
expressed above. We enforce the continuity of $f_\text{field}(d)$ by
choosing $L_\text{min}$ such that $f_\text{field}(d_\text{comp})=0.30$.

The mock galaxy and group of galaxies catalogue is now built:
\begin{itemize}
\item[{\bf I.}] We take a halo $A$ from {\it FullMock} and assume it is a
  group of galaxies. We thus deduce the intrinsic
  luminosity $L_A$ from the mass  $M_A$ of this object using
  Eq.~\eqref{eq:tully_ml}.
\item[{\bf II.}] The observed luminosity $L'_A$ of $A$ is computed assuming that
  its galaxy population follows \eqref{eq:lum_schechter} but
  with a different normalization to achieve the intrinsic luminosity $L_A$.
  If $d_A$ is the distance between the observer and the halo $A$, then
  the galaxies detected in this halo verify \eqref{eq:visibility_cons} for $d=d_A$.
  The total observable luminosity for $A$ is thus
  \begin{equation}
    L'_A = L_A f_L(d_A)
  \end{equation}
  with, assuming $L_\text{min} \ll L_{*}$,
  \begin{equation}
    f_L(d) = \left\{
      \begin{array}{ll}
	\sim 1 & d < d_\text{comp} \\
	\frac{\int_{4 \pi d^2 l_\text{cut}}^\infty L n(L)\,\,\text{d}L}{ \int_{L_\text{min}}^\infty L
	  n(L)\,\,dL} \simeq \mathrm{e}^{-\frac{4 \pi d^2 l_\text{cut}}{L_{*}}} & d \ge d_\text{comp}
      \end{array}\right.      
  \end{equation}
  
\item[{\bf III.}] If $L'_A < 4 \pi d_A^2 l_\text{cut}$ then $A$
  is removed from the catalogue, otherwise it is kept.

\item[{\bf IV.}] This gives us the group component of our magnitude-limited catalogue.

\item[{\bf V.}] The case of the ``field galaxies'' is treated
  separately. Galaxies are identified with dark matter
  particles and their luminosity is assigned following
  \eqref{eq:lum_schechter}. More specifically, we
  choose a shell $S_d$ put at a distance $d$ from the observer. The
  probability of keeping a ``galaxy'' $G$ in $S_d$ is given by
  \eqref{eq:frac_schechter_form}. 
  Inside the shell $S_d$, the selected ``galaxies'' share now a luminosity
  \begin{equation}
    L_\text{f}(d) = \int_{4 \pi d_G^2 l_\text{cut}}^{\infty} L
    n(L)\;\text{d}L
  \end{equation}
  which is distributed evenly among them. Strictly speaking, such a
  repartition should be performed randomly according to
  \eqref{eq:lum_schechter}. That would add a small additional noise
  on the reconstructed velocities. This noise should be
  of insignificant consequence as supported by the discussion of the {\it
    TS-T} case in \S~\ref{sec:ml_ratio}.
\end{itemize}

\begin{figure}
  \begin{center}
    \includegraphics*[width=\linewidth]{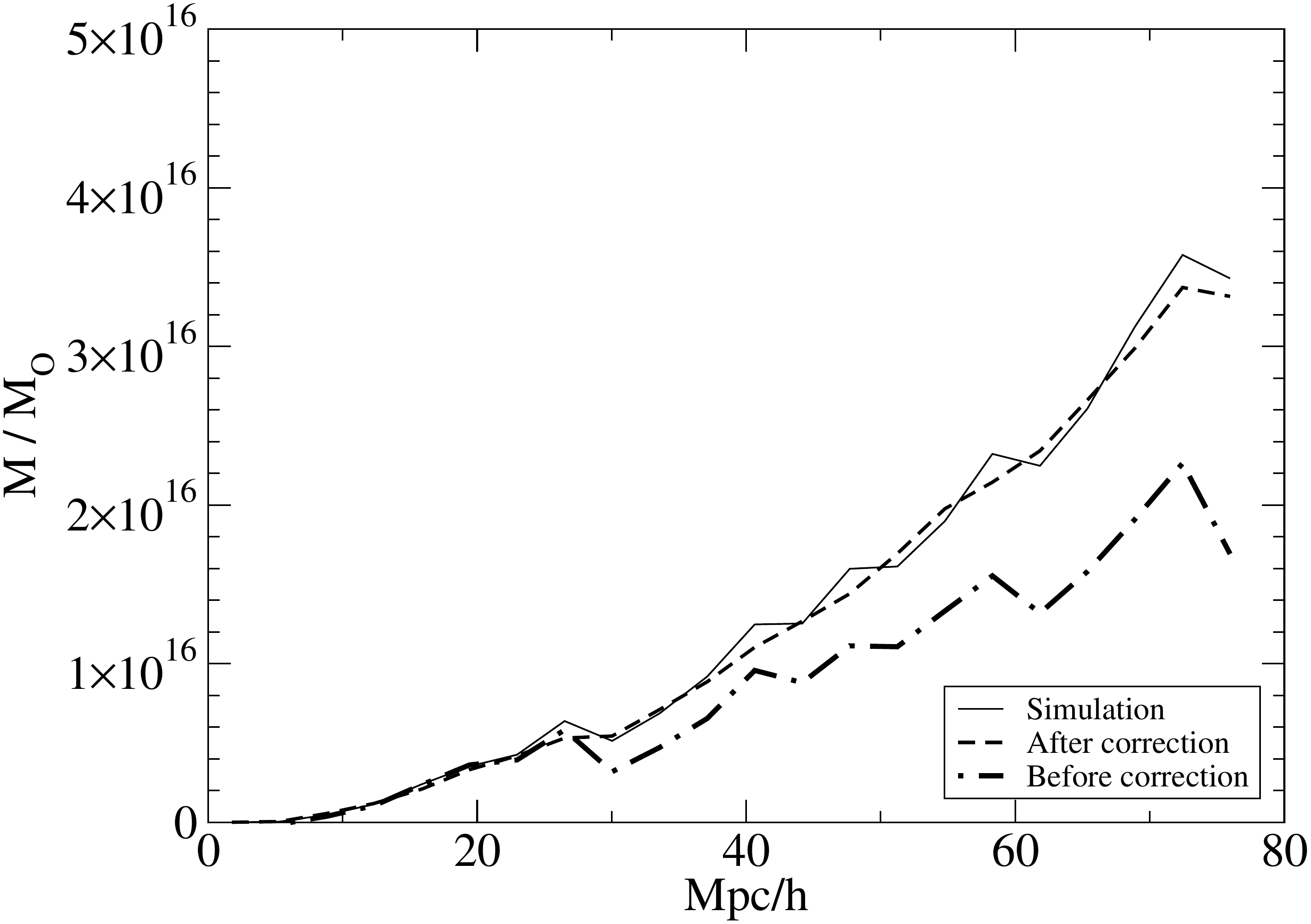}    
  \end{center}
  \caption{\label{fig:mass_check} {\it Magnitude limitation/Filling
      missing mass} -- This plot gives the measured amount of mass in
    a thin shell at different distances from the observer. The solid
    line gives the original mass distribution in the simulation, the
    dot-dashed line the mass distribution after mimicking incompleteness and
    the dashed line the recovered mass distribution after correction for
    incompleteness as described in
    Appendix~\ref{app:magnitude_limit}. }
\end{figure}

We have now a realistic mock catalogue and we try
to account for its incompleteness as we would for NBG-8k:
\begin{itemize}  
\item[{\bf A.}] The missing luminosity in groups is corrected.
  In order to do this, we compute, in a thin shell $S_d$ at some
  distance $d$, the ratio between the expected total luminosity and
  the observed luminosity 
  \begin{equation}
    b(d) = \frac{\int_{0}^{\infty} L n(L)\;\text{d}L}{\int_{4 \pi
    d^2 l_\text{cut}}^{\infty} L n(L)\;\text{d}L} =
    \mathrm{e}^{4 \pi d^2 l_\text{cut} / L_\text{*}} \text{.}
  \end{equation}
  The intrinsic luminosity $L_A$ of a group $A$ in $S_d$ is recovered with
  \begin{equation}
    L_A = L_{\text{obs},A} b(d) \text{.}
  \end{equation}
  The mass $M_A$ of $A$ can then be obtained
  using the {\it non-linear} relation 
  \eqref{eq:tully_ml}.
\item[{\bf B.}] The remaining missing mass in $S_d$ can be written
  \begin{eqnarray}
    \lefteqn{M_{\text{missed},d} = \Upsilon b(d)
      \left(L_{\text{field,obs},d} + L_{\text{group,obs},d}\right) } \nonumber \\
    && \mbox{}
       - M_{\text{field,obs},d} - M_{\text{group,obs},d} \quad\text{,} \label{eq:miss_mass}
  \end{eqnarray}
  with $\Upsilon = 93 \frac{\text{M}_\odot}{\text{L}_\odot}$ the
  average $M/L$,\footnote{Note that a prior assumption on the value of $\Omega_{\text{m}}$ is
  obviously needed to estimate $\Upsilon$.} $L_{\text{group,obs},d}$ the observed
  luminosity of groups, $M_{\text{group,obs},d}$ the masses of groups
  obtained after the above correction, $L_{\text{field,obs},d}$ the
  luminosity of field galaxies. The quantity $M_{\text{missed},d}$
  comes from both missing galaxies and missing group of galaxies. 
  If $M_{\text{missed},d} > 0$ and
  without any further information, the missing mass may either be
  assigned evenly to field galaxies of $S_d$ (our choice, as usually performed in
  the litterature), or distributed uniformly in $S_d$ using
  new random tracers. If $M_{\text{missed},d} \le 0$, the mass distribution in $S_d$ is
  untouched. 

  This procedure is certainly not free from biases. For instance, the
  contrasts between shells are partly smoothed out, as illustrated by
  Fig.~\ref{fig:mass_check}. This is equivalent 
  to reducing the overall magnitude of fluctuations in the density
  field.  As a result, a
  small bias towards larger $\Omega_{\text{m}}$ might occur, as in the lower right panel
  of Fig.~\ref{fig:velocity_random} of \S~\ref{sec:diffuse_mass}. On
  the opposite, if the missing mass is assigned to detected background
  galaxies, the estimated $\Omega_{\text{m}}$ is expected to underestimate the
  true value as discussed in \S~\ref{sec:diffuse_mass}.

\item[{\bf C.}]
  Note that the mass of the ``field galaxies'' is  not the mass of a
  single particle anymore. Procedure explained in
  Appendix~\ref{app:subsampling} is facilitated as follows, for
  the sake of algorithmic simplicity. With $v \in (0;1]$ a uniform random
  variable, a galaxy G of mass $m_\text{G}$ is splitted into
  $n_\text{G}$ subcomponents of mass $m_\text{particle}$ such that:
  \begin{equation}
    n_\text{G} = \left\{
    \begin{array}{ll}
      r_\text{G} & \text{if } \left(\frac{m_\text{G}}{m_\text{particle}} - r_G\right) <
        v \\
      r_\text{G}+1 & \text{ otherwise}
    \end{array}\text{ ,}\right.
  \end{equation}
  with $r_\text{G} = \left\lfloor
  \frac{m_\text{G}}{m_\text{particle}}\right\rfloor$, $\lfloor x \rfloor$
  being the integer part of $x$.
  Each of the subcomponent is now considered as a ``field galaxy'' in
  the procedure explained in Appendix~\ref{app:subsampling}.
\end{itemize}

\section{Statistical bias in the slope estimation}
\label{sec:stat_bias}

The two methods that we used for slope estimation are known to be
biased. A more precise treatment of this bias is \correction{beyond} the scope of
this paper. However we propose here to check the order of magnitude of
the systematic effect of the statistical analysis itself. To achieve
this we produced a set of randomly generated ``velocities'' $v$ and
their ``reconstructed velocities'' $v_\text{R}$ counterpart. The
probability for a point $(v,v_\text{R})$ to have a velocity $v$ is
given by
\begin{equation}
  P_v(v) = \frac{1}{\sqrt{2 \pi} \sigma_v} \text{e}^{-v^2/(2 \sigma_v^2)}\text{,}
\end{equation}
with $\sigma_v = 300$~\kms typically. The probability for it to have a
reconstructed velocity $v_R$ is given by the {\it same} probability law.
We now compute the error $e$ between $v_R$ and $v$, which must be
distributed according to the Lorentzian form
\begin{equation}
  P_\text{DE}(e) = \frac{1}{\pi B} \frac{1}{1 + \left(\frac{e}{B}\right)^2}\text{,}\label{eq:error_law}
\end{equation}
with $B=86$~\kms. The error $e$ is related to $v$ and $v_\text{R}$ by
\begin{equation}
  e = \alpha_* v - \beta_* v_\text{R}\text{.}
\end{equation}
For the rest of the appendix we take
$\alpha_{*} = \beta_{*} = 1/\sqrt{2}$.
The probability of keeping a point $(v,v_R)$ with an
error $e$ is given by, integrating $P_\text{DE}(e')$ between $-e$ and $+e$,
\begin{equation}
  P_\text{keep}(e) = \frac{2}{\pi} \tan^{-1}\left( \frac{e}{B} \right)
\end{equation}

\begin{figure}
  \begin{center}
    \includegraphics[width=.8\linewidth]{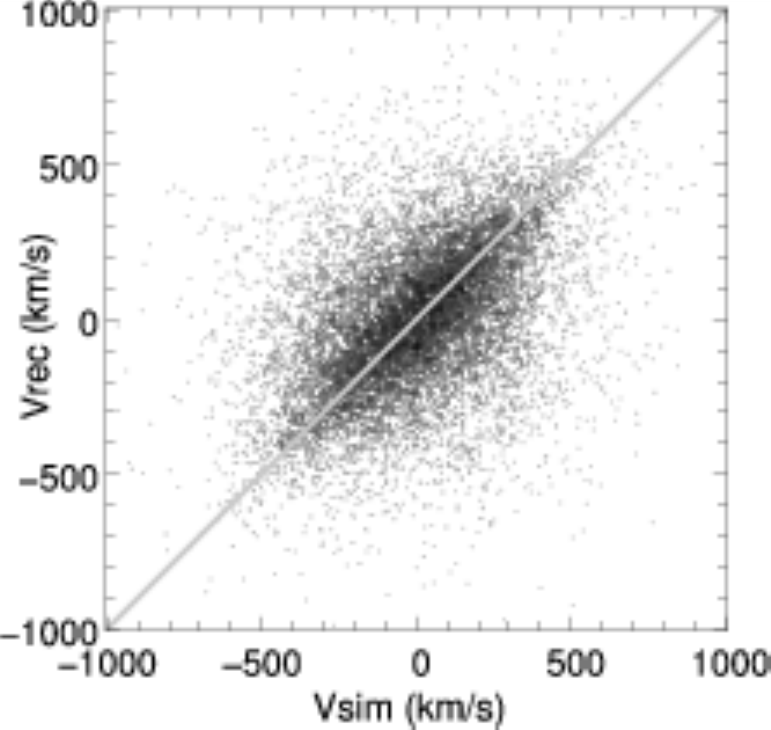}
  \end{center}
  \caption{\label{fig:random_points} {\it Statistical bias} -- Scatter
  plot of 10,000 randomly generated points following the approximated probability
  laws found between reconstructed velocities and simulated velocities.}
\end{figure}

 We represented in
Fig.~\ref{fig:random_points} a scatter plot of 10,000 points generated using this
procedure. As one can see, it does look like a real scatter plot of a
redshift reconstruction.

Conducting a 1.5$\sigma$ analysis on this set of points, we find a
slope $\beta_{*}/\alpha_{*} = 1.0 \pm 0.20$. In our case, this would give
$\Omega_\text{m}=0.30 \pm 0.10$. 
Estimating the slope using the maximum likelihood approach gives, with
$\sigma_v=+\infty$, $\beta_{*}/\alpha_{*} = 0.81 \pm 0.01$
($\Omega_\text{m}=0.20 \pm 0.02$) and with 
$\sigma_v=300$~\kms, $\beta_{*}/\alpha_{*} = 1.074 \pm 0.012$
($\Omega_\text{m}=0.34 \pm 0.02$). Putting $B=40$~\kms, both for
generated data and likelihood function, as for real space
reconstructions, 
reduces the error and gives $\Omega_\text{m}=0.31\pm 0.02$, thus highlighting
the importance of the reconstruction noise for a good estimation of $\Omega_\text{m}$.

Consequently, though one must rely on the likelihood analysis, it may be
strongly biased by the structure of reconstruction errors mixed with
the non-uniform distribution of observables. We tried to make a good
approximate model of the errors, though it seems to quite depends on
the value of $\sigma_v$. Whenever possible, of
course, one must crosscheck the result of the likelihood by a visual
inspection of the scatter plot. 

\end{appendix}

\label{lastpage}

\end{document}